%% file: main.tex
\documentclass[12pt]{article} 
\usepackage{amsmath,amssymb,amsfonts}
\usepackage{setspace}
\usepackage{psfrag}
\usepackage{xcolor}
\usepackage{color}
\definecolor{darkblue}{rgb}{0.1,0.1,.7}
\usepackage[colorlinks, linkcolor=darkblue, citecolor=darkblue, urlcolor=darkblue, linktocpage]{hyperref} 
\usepackage[square, comma, compress,numbers]{natbib}
\usepackage[]{amsmath}
\usepackage[]{graphicx}
\usepackage[]{latexsym}
\usepackage{geometry}

\usepackage{booktabs,multirow}

\usepackage{amscd}
\usepackage[all,cmtip]{xy}
\usepackage{mathrsfs}
\usepackage{bbold}
\usepackage{dsfont}
\usepackage{subfigure}
\usepackage{anyfontsize}

\usepackage{changepage}
\usepackage{mymacros}
\newcommand{\as}[1]{\renewcommand{\arraystretch}{#1}}

\newcommand{\eref}[1]{(\ref{#1})}
\newcommand{\sref}[1]{section~\ref{#1}}
\newcommand{\Figref}[1]{Fig.~\ref{#1}}

\begin{document}

\vspace*{-.6in} \thispagestyle{empty}
\begin{flushright}
\end{flushright}
\vspace{.2in} {\Large
\begin{center}
{\bf Mixed Scalar-Current bootstrap in three dimensions}\\
\end{center}
}
\vspace{.2in}
\begin{center}
{\bf 
Marten Reehorst$^{a}$,
Emilio Trevisani$^{b}$,
Alessandro Vichi$^{a,c}$} 
\\
\vspace{.2in} 
\small
$^a$ {\it Institute of Physics,
\'Ecole Polytechnique F\'ed\'erale de Lausanne (EPFL),\\
Rte de la Sorge, BSP 728, CH-1015 Lausanne, Switzerland}\\
$^b$ {\it Laboratoire de Physique Th\'eorique, \'Ecole Normale Sup\'erieure \& PSL Research University, \\
24 rue Lhomond, 75231 Paris Cedex 05, France\\
Institut des Hautes \'Etudes Scientifiques, Bures-sur-Yvette, France}\\
$^c$ {\it Dipartimento di Fisica dell'Universit\`a di Pisa, Largo Pontecorvo 3, I-56127 Pisa, Italy}
\end{center}

\vspace{.2in}

\begin{abstract}
We study the mixed system of correlation functions involving a scalar field charged under a global $U(1)$ symmetry and the associated conserved spin-1 current $J_\mu$.
Using numerical bootstrap techniques we obtain bounds on new observables not accessible in the usual scalar bootstrap. We then specialize to the $O(2)$ model and extract rigorous bounds on the three-point function coefficient of two currents and the unique relevant scalar singlet, as well as those of two currents and the stress tensor. Using these results, and comparing with a quantum Monte Carlo simulation of the $O(2)$ model conductivity, we give estimates of the thermal one-point function of the relevant singlet and the stress tensor.
We also obtain new bounds on operators in various sectors.
\end{abstract}

\newpage

\tableofcontents

\newpage


\section{Introduction and summary of results}
\input{intro/intro.tex}

\section{Setup}
\input{setup/setup.tex}

\section{Results}
\input{results/results.tex}

\section{Conclusions}
\input{Conclusions/conclusions.tex}

\section*{Acknowledgments}
We thank William Witczak-Krempa and Erik Sorensen for useful discussions about the quantum Monte Carlo simulation. We also thank Joao Penedones and Ning Su for useful comments and discussions.
MR and AV are supported by the Swiss National Science
Foundation under grant no.\ PP00P2-163670. AV is also supported by the
European Research Council Starting Grant under grant no.\ 758903. All
the numerical computations in this paper were run on the EPFL SCITAS cluster.
ET is supported by the Simons Foundation grant 488655 (Simons Collaboration on the Nonperturbative Bootstrap).

\newpage
\appendix 
\input{appendices/appendix_condictivity.tex}

\input{appendices/appendices.tex}

\bibliographystyle{utphys}
\bibliography{references}

\end{document}

%% file: intro/intro.tex
\label{sec:intro}

Three dimensional Conformal Field Theories (CFTs) display a rich variety and range of applications. While some of them were introduced a long time ago in order to describe long known phase transitions in condensed matter and statistical models, in recent years the zoo of renormalization group (RG) fixed points has vastly grown.

The numerical conformal bootstrap represents a powerful tool to shed some light on the intricate world of three dimensional CFTs. After its revival a decade ago \cite{Rattazzi:2008pe,Rychkov:2009ij,Rattazzi:2010gj,Rattazzi:2010yc,Poland:2010wg}, it has been successfully used to extract the most precise prediction of critical exponents in key examples \cite{Kos:2014bka,El-Showk:2014dwa,Kos:2015mba,Simmons-Duffin:2015qma,Kos:2016ysd,Simmons-Duffin:2016wlq,Rong:2018okz}. Moreover, interesting studies also displayed hints of novel (and yet unclassified) CFTs \cite{Kousvos:2018rhl,Stergiou:2019dcv}. Many other great results have been achieved in three dimensions \cite{Kos:2013tga,Iliesiu:2015qra,Chester:2014gqa,Chester:2014fya,Chester:2015lej,Chester:2016wrc,Chester:2017vdh,Iliesiu:2017nrv,Baggio:2017mas,Dymarsky:2017yzx,Li:2017kck,Behan:2018hfx,Agmon:2019imm,Kousvos:2019hgc,Rong:2019qer}.  See  also \cite{Poland:2018epd,Chester:2019wfx} for recent reviews on the subject.

When examining the results obtained in the last few years, it appears evident that  bootstrap methods in presence of a global symmetry seem to be less powerful when compared to simpler systems like the Ising model or its supersymmetric extension. 
One possible argument is that, given that the theory is more involved, one simply needs to consider correlators involving more than two scalars. In particular, relevant scalar operators seem to play a crucial role. 

A second explanation could reside in how the presence of a global symmetry is imposed. In past studies, the existence of a global symmetry was injected by declaring that operators entering a correlation function transform according to  irreducible representations of the global symmetry group. In addition, selection rules were imposed on the operator product expansion (OPE) of these operators.
A complementary approach was also initiated in \cite{Dymarsky:2017xzb}, where the presence of a global symmetry was enforced by studying the correlation function of the associated conserved spin-1 current. The latter method is definitively preferable, but comes at the expense of considering spinning operators and thus complicating the analysis. As a plus side, however, it does not introduce any new parameter to scan over, since conserved currents have fixed dimensions. 
In this work we push this approach one step further, and explore the constraints arising from the mixed system of correlation functions involving one conserved current, associated to a $U(1)$ global symmetry, together with a scalar field charged under it. \\
One should be careful with the latter statement: without further assumptions, including a conserved current in the bootstrap does not give us the right to identify it with the generator of the global symmetry under which the scalar is charged. A trivial counter examples is the tensor product of a generalized free scalar field $\phi$ and a generalized free vector field $J_\mu$. In order to impose that the external scalar and current couple non trivially, one should force the correct global symmetry Ward identity, namely that the three point function $\<\phi \bar\phi J_\mu\>$ is non vanishing.\footnote{In the numerical bootstrap framework this is equivalent to impose a finite current central charge $C_J$, see \sref{sec:3pt}.} In this work we use this assumption in our studies of the $O(2)$ model. We plan to systematically make use of this assumption in more general future explorations.

Among the obvious targets of our investigation one can list the $O(2)$ vector model, the Gross Neveu Yukawa model with $N=2$ fermions,  and QED$_3$, both fermionic or bosonic, where one identifies the global symmetry with the topological $U(1)_T$.
Although in principle our set up could be used to analyze any system possessing a $U(1)$ symmetry, we found that our numerical bounds are subject to the same limitations as the single scalar correlator analysis, namely they loose constraining power as the dimension of the scalar grows. For this reason we mostly focus on the $O(2)$ model where the charge-1 scalar has dimension close to the free value. 
We also explored more general bounds and did not find other evidences of CFTs saturating them.

\subsection{New data for the $O(2)$ model}

In this section we collect the most important constraints obtained in the present work. The interested reader can find all the technical details and proper definitions in the next sections. Additional and more general plots can be found in \sref{sec:results}.

As mentioned in the previous section, we mostly focused on the $O(2)$ model. In this case we identify our scalar $\phi$ with the order parameter of the Landau Ginzburg description of the phase transition, while $J_\mu$ is the current associated to the global $O(2)$ symmetry. According to recent bootstrap results \cite{Kos:2016ysd}, this model is confined to live on a narrow island in the plane $(\Delta_\phi, \Delta_S)$, where $S$ here is the unique neutral relevant scalar operator. Previous bootstrap studies also constrained the dimension of the unique relevant traceless symmetric operator $t_{ij}$, the central charge $C_T$ and the current central charge $C_J$\footnote{These are defined respectively as the normalization of the two-point function of the stress tensor $T_{\mu\nu}$ and the $U(1)$ current $J_\mu$.} \cite{Kos:2013tga,Kos:2015mba}. A few OPE coefficients have also been determined in \cite{Kos:2016ysd}, such as $\lambda_{\phi\bar{\phi}S}$ and $\lambda_{SSS}$.

\begin{figure}[h]
\begin{center}
\subfigure[]{\includegraphics[width=0.45\textwidth]{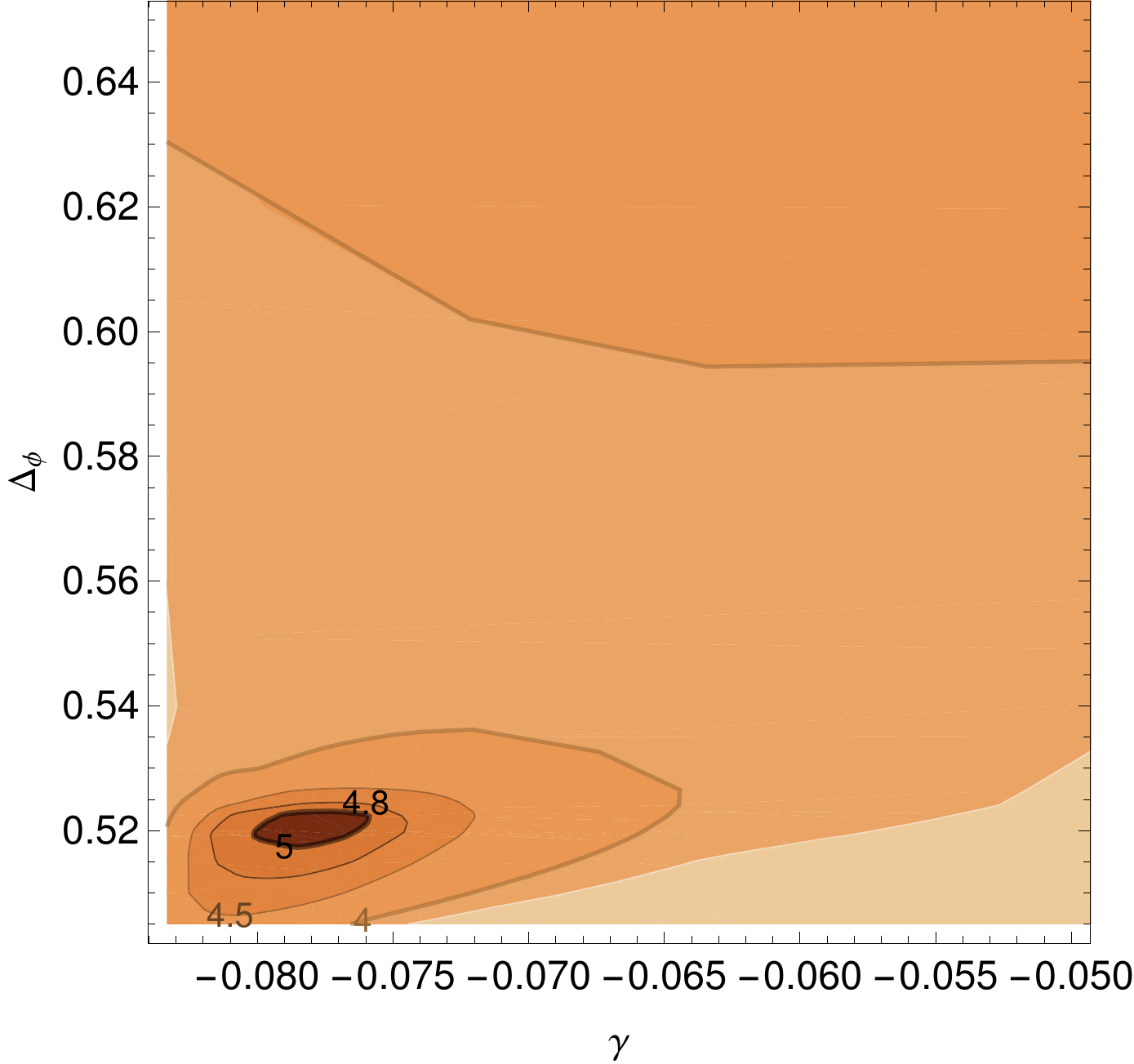}\label{fig:DeltaPhi-gamma}}
\subfigure[]{\includegraphics[width=0.45\textwidth]{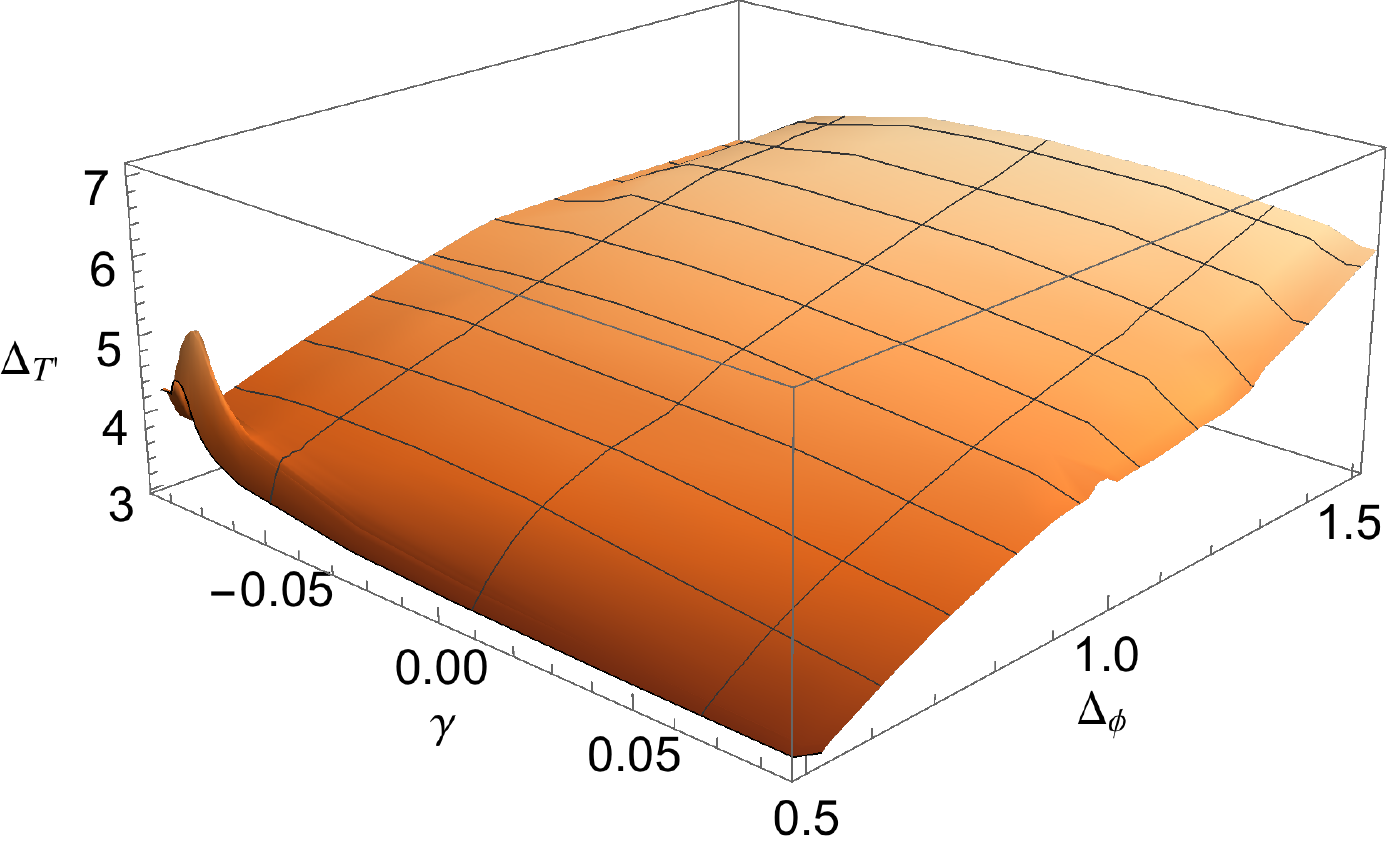}\label{fig:DeltaPhi-gamma3D}}
\caption{On the left: allowed region in the plane $(\Delta_\phi,\gamma)$ assuming that the first spin-2, parity-even and neutral traceless symmetric tensor $T'$ after the conserved stress energy tensor has dimension $\Delta_{T'}\geq 3.8,4,4.5,4.8,5$. As the gap increases the allowed region shrinks to an island. On the right: bound on $\Delta_{T'}$ as a function of $\gamma$ and $\Delta_S$. The bounds have been obtained at $\Lambda=13$.} 
\end{center}
\end{figure}

When bootstrapping a mixed system of scalars, one can impose gaps in various scalar sectors and exploit the existence of few relevant operators to create islands in parameter space. In our setup, however, the same strategy does not work.\footnote{Because of Ward identities, the charge-1 sector does not contain scalars, besides $\phi$ itself. Gaps in the other scalar sectors are not sufficient to create islands.} 
Our strategy will then be to identify a new set of assumptions that allow to create an island and use them to extract constraints on CFT-data that have never been bound before, such as the parameter $\gamma$ and the OPE coefficients $\lambda_{JJS}$.
%
%
While the latter is self explanatory, the former is related to the three point function of two currents and the stress tensor ---see section \ref{sec:3pt}. 
As discussed in \cite{Hofman:2008ar,Buchel:2009sk}, the conformal collider bounds require the parameter $\gamma$ to range between $[-1/12, 1/12]$, with the extremes corresponding to free theories. Numerical evidences of these bounds  were also found by \cite{Dymarsky:2017xzb}. The value of this parameter in the $O(2)$ model was not known, although strong numerical evidences supported a negative value, which was also confirmed by \cite{Dymarsky:2017xzb} under somewhat strong assumptions on the spectrum of the theory.\footnote{In particular we checked that the assumption that all parity-odd operators have twist $\tau=\Delta-\ell\geq2.5$ is inconsistent for the $O(2)$ model. The milder assumption $\tau\geq2$ is still consistent.}

\begin{figure}[t]
\begin{center}
	\includegraphics[width=0.5\textwidth]{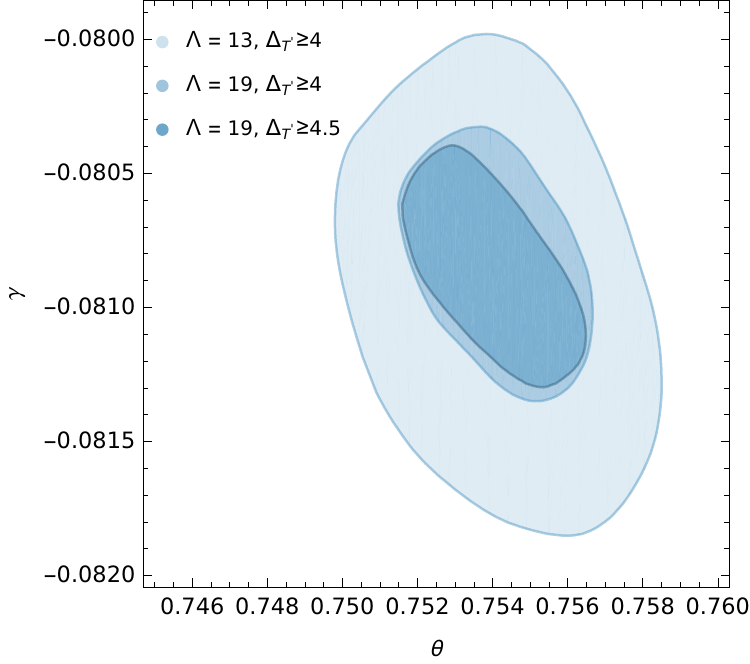}
\caption{Allowed region in the plane $(\gamma, \theta)$ assuming 
the known O(2) constraints shown in Table ~\ref{tab:O2knownconstraints} and $\Delta_{T'}\geq 4,4.5$. The lighter region has been computed at $\Lambda=13$. The two smaller regions instead have been computed at $\Lambda=19$.
} 
\label{fig:gamma_theta}
\end{center}
\end{figure}

In our explorations we found that a discriminant characteristic of the $O(2)$ model is the presence of a rather large gap between the stress tensor and the next spin-2 neutral operator, let us call it $T'$. This property translates in a sharp peak in the bound on $\Delta_{T'}$ as a function of $\Delta_\phi$ and $\gamma$, as shown in  \Figref{fig:DeltaPhi-gamma3D}. Intuitively this happens because fake solutions of crossing or non-local theories do not require a stress tensor but usually posses a spin-2 operator close to the unitarity bound; hence the bound on $T'$ is effectively a bound on the first spin-2 operator and only for local theories (which have a conserved stress tensor) it becomes a bound on the second spin-2 operator. This property was also exploited in \cite{Li:2017kck} to create isolated regions in single correlator bootstrap. \\
In \Figref{fig:DeltaPhi-gamma} we show the allowed region in the plane $(\Delta_\phi,\gamma)$ with increasing gaps on $T'$. 
By raising the gap $\Delta_{T'}$, the allowed region shrinks to a very small island, with a $\Delta_\phi$ value centered around the expected value of the $O(2)$ model. By making the conservative assumption $\Delta_{T'} \geq4$, we are able to create an isolated region, with the parameter $\gamma$ confined close to the lower extreme of its interval.

The above analysis shows that, in order to isolate the $O(2)$ model, we can impose a mild gap between the stress tensor operator and the next operator in the same sector. In order to make this assumption rigorous one could  consider the island created by the mixed correlator bootstrap of scalars as in \cite{Kos:2016ysd} and then derive a rigorous upper and lower bound on $\Delta_{T'}$ by moving inside the island.  In what follows we then use two assumptions to isolate the $O(2)$ model, one more conservative and one more realistic: $\Delta_{T'}\geq 4, 4.5$. A refined analysis \cite{CLLPSDSV} of the $O(2)$ model involving three external scalar operators, $\phi,$ $S$ and the unique relevant charge two scalar  $t$, has found $\Delta_{T'}\geq 4.6$, which is consistent with both our assumptions.

Since in this section we are focusing on the $O(2)$ model, in addition to the gap on $T'$ we will also input information from previous bootstrap analysis and use this assumptions to determine bounds on new quantities. 

Let us begin by $\gamma$ and the OPE coefficient $\lambda_{JJS}$. We remind that, due to our framework, the unique relevant  neutral scalar $S$ appears in two OPEs, schematically:
\bea
J\times J \sim \mathbb 1 + \lambda_{JJS} S + \ldots \, ,\nonumber\\
\phi \times \bar{\phi} \sim 1 + \lambda_{\phi\bar\phi S} S + \ldots \, .
\eea
Let us define the ratio of OPE coefficients,
\be\label{eq:tanTheta}
\tan \theta  = \frac{\lambda_{JJS}}{\lambda_{\phi\bar\phi S}} \, .
\ee
We can then inspect what values of $\gamma$ and the angle $\theta$ are consistent with the $O(2)$ model information we know. \Figref{fig:gamma_theta} shows the allowed region in the $(\gamma,\theta)$ plane once we input the best determination for $\Delta_\phi $ and $\Delta_S$ from \cite{Kos:2016ysd} as well as other known O(2) constraints shown in Table~\ref{tab:O2knownconstraints}.\footnote{We demand positivity for a scan over the allowed intervals for $\Delta_S$ and $\Delta_t$. Instead for $\dphi$ we pick a central value in the allowed island.} Notice that in \Figref{fig:gamma_theta} and in the following plots we fixed the external dimension to a precise value. Given the small size of the allowed range for $\Delta_\phi$ \cite{Kos:2016ysd,CLLPSDSV}, moving this value would not alter the figure in a noticeable way.
 \begin{table*}[h!]
\centering
\as{1.2}
\begin{tabular}{@{}l@{}}
	\toprule
O(2) assumptions\\
	\midrule
	$\Delta_\phi = 0.5191$\\
			$\Delta_S\in [1.509,1.514]$\\
			$\Delta_{S'}>3$\\
			$\Delta_t \in [1.204,1.215] $\\
			$\Delta_{t'}>3$\\
			$\Delta^{Q=0}_{0,-}>3$\\
			$C_J<0.9066 \, C_{J^\text{free}}$\\
	\bottomrule 
\end{tabular}
	\caption{List of assumptions used in our analysis. The bound for $\Delta_S$ is taken from \cite{Kos:2016ysd}. The bound for $\Delta_t$ and  $C_J$ are taken from \cite{Kos:2015mba}.
	$S'$ and $t'$ are respectively the first operators appearing after $S$ and $t$. Evidences for the gap on $\Delta^{Q=0}_{0,-}$ were presented in \cite{Dymarsky:2017xzb}. 
	\label{tab:O2knownconstraints}}
\end{table*}


%
Using the value determined in \cite{Kos:2016ysd}  for $\lambda_{\phi\bar{\phi} S}$ and \eref{eq:tanTheta} we then conclude (for $\Delta_{T'}\geq4$):
\bea\label{eq:lambda_JJS}
&& \gamma = -0.0808(5),\nonumber\\
&& \left|\lambda_{JJS}\right| = 0.645(4)\,.
\eea

\begin{figure}[t]
\begin{center}
\includegraphics[width=0.5\textwidth]{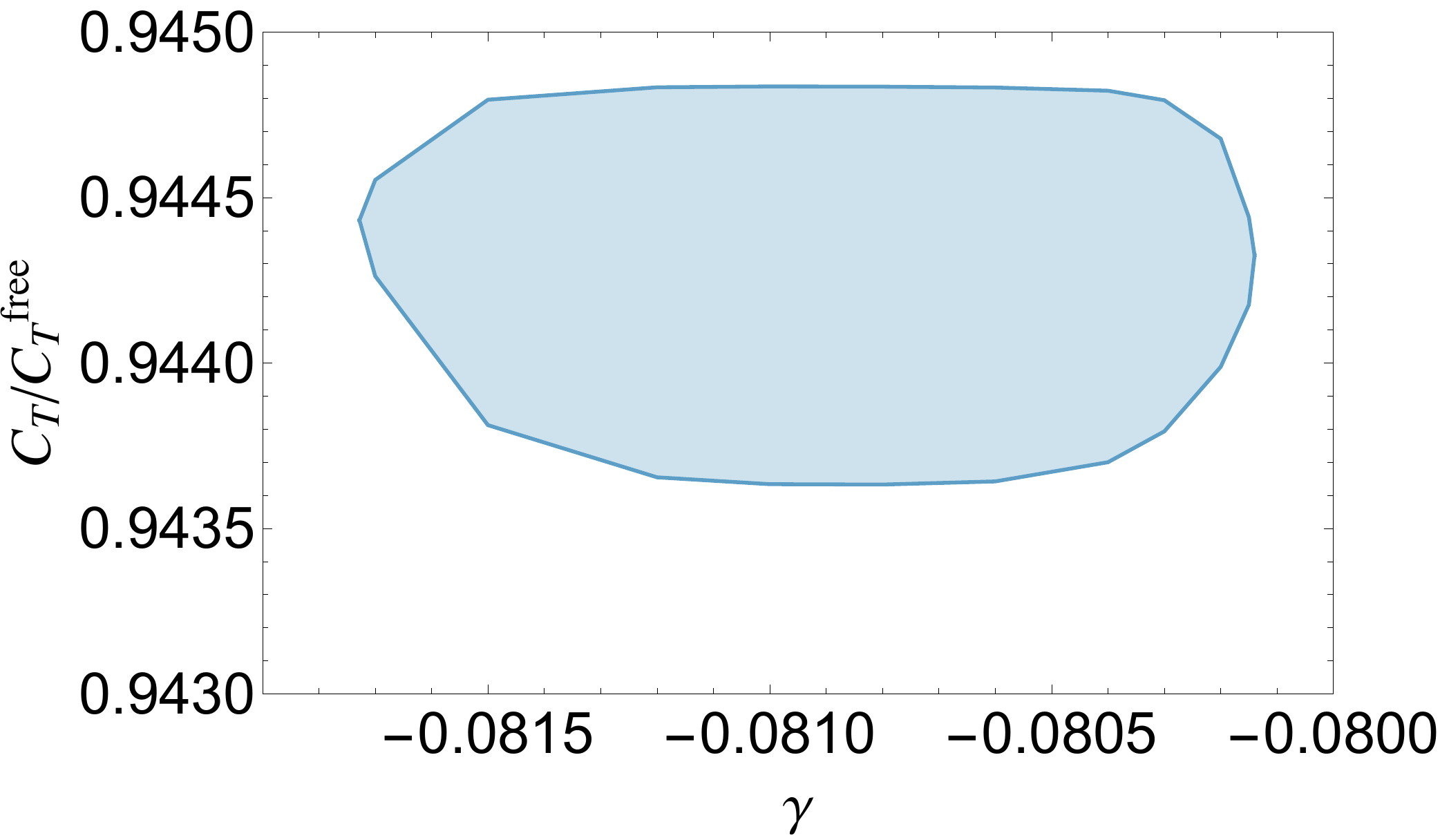}
\caption{Upper and lower bound on the central charge $C_T$ normalized to the free value assuming the constraints shown in Table ~\ref{tab:O2knownconstraints} and $\Delta_{T'}\geq 4.5$. The bounds have been computed at $\Lambda=13$.} 
\label{fig:plotCTO2}
\end{center}
\end{figure}

Similarly, we can extract upper and lower bounds on the central charge $C_T$. These are shown in \Figref{fig:plotCTO2} and allow us to conclude:
\be\label{eq:CTinO2}
\frac{C_T}{C_T^\text{free}}=0.9442(6)\,.
\ee

\begin{figure}[htbp]
\begin{center}
\includegraphics[width=0.45\textwidth]{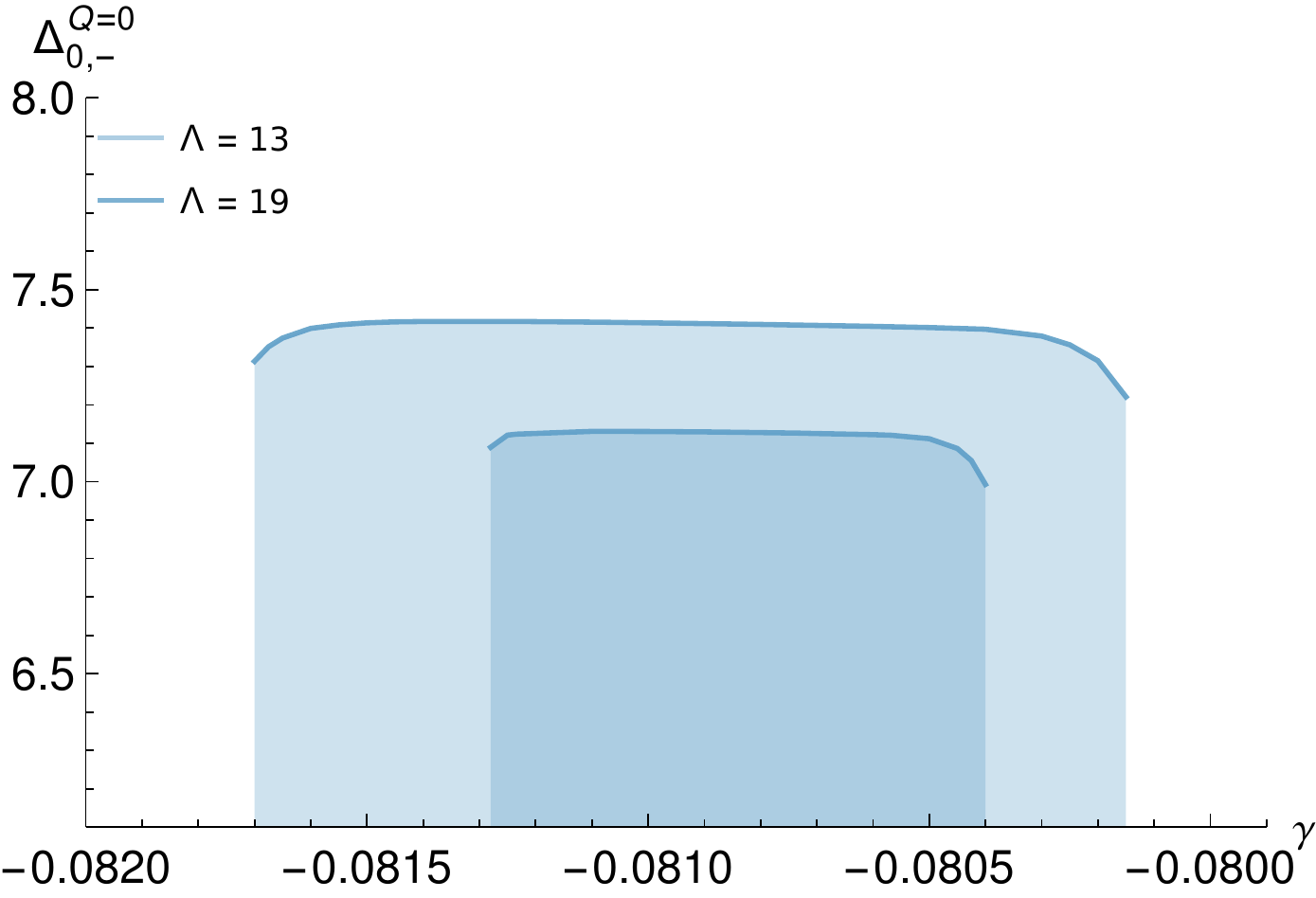}
\caption{Bound on the the dimension of the first neutral parity-odd scalar assuming the known $O(2)$ constraints shown in Table ~\ref{tab:O2knownconstraints} and $\Delta_{T'}\geq 4.5$. The bounds terminate because $\gamma$ is confined in an interval, see \Figref{fig:gamma_theta}.} 
\label{fig:plot0o0O2}
\end{center}
\end{figure}

Finally, using the same set of assumptions, we can extract upper bounds on low lying operators. We stress that these are bona fide upper bounds and are not obtained by the extremal functional method. As an example we show in \Figref{fig:plot0o0O2} the upper bounds on the first neutral parity-odd scalar as a function of $\gamma$ for fixed $\Delta_\phi$. Again changing the value of $\Delta_\phi$ within its allowed range does not affect the results in a noticeable way. Notice that passing from $\Lambda=13$ to $\Lambda=19$ makes the bound stronger by a $5\%$, suggesting that the bound is still not converged. 

We repeated a similar analysis in other channels and we obtained the bounds summarized in Table~\ref{tab:O2operators}.
\begin{table*}[h!]
\centering
\as{1.2}
\begin{tabular}{@{}ccc|cc@{}}
	\toprule
$\ell$ & $P$ & $Q$ &$ \Lambda=13$ & $\Lambda=19$\\
	\midrule
$0$ &  $-$ & $0$ &$7.45$ & $7.13$\\
$1$ &  $-$ & $1$ &$10.14$ & $8.59$\\
$2$ &  $-$ & $1$ &$4.47$ & $4.47$\\
$1$ &  $+$ & $1$ &$2.96$ &$ 2.95$\\
	\bottomrule 
\end{tabular}
\caption{Upper bounds on operators in the $O(2)$ models.}
\label{tab:O2operators}
\end{table*}
\subsection{Conductivity at finite temperature}

CFTs also play an important role in the description of certain quantum critical points. It was observed in  \cite{Katz:2014rla} that transport properties of systems near a quantum critical point can be expressed in terms of CFT-data. There,  the conductivity of a global symmetry current in a (2+1) CFT non-zero temperature was computed in terms of the OPE $J\times J$ and compared with a quantum Monte Carlo simulation (QMC) of the $O(2)$-model in the limit of high frequencies, $w \gg T$. \\
The imaginary frequency conductivity is related to the thermal expectation value of the current two point function by the expression
\be\label{eq:conductivity}
\frac{\sigma(iw)}{\sigma_Q}  = -\frac1{|w|} \<\widetilde J_{\mu=2}(-w) \widetilde J_{\nu=2}(w) \>_T+\text{(contact terms)}\,,
\ee
where $\sigma_Q = e^2/\hbar$ is the conductance quantum unit and $\widetilde J_\mu(w)$ denotes the Fourier transform of the current $J_\mu(x)$.\footnote{The conductivity is defined only on Matsubara frequencies $w_n=2\pi n T$, but can be analytically continued to intermediate values.} \\
When using the OPE, the left-hand side receives contributions to all operators that acquire a thermal expectation value.\footnote{Only primary operators acquire a thermal expectation value.} The leading term comes from the identity exchange and corresponds to a constant value, usually called $\sigma_\infty$, identified with the conductivity at $T=0$. Next, for each primary operator $\mathcal O$ entering the $J\times J$ OPE, the conductivity receives a contribution scaling as $(T/w)^{\Delta_\mathcal{O}}$. As pointed out in \cite{Katz:2014rla}, in the $O(N)$ model, the leading term in the expansion is the unique $O(N)$ singlet relevant scalar, followed by the stress tensor and then irrelevant operators.

In order to compare our bootstrap prediction with the quantum Monte Carlo simulation for the $O(2)$ model, we first need to express the conductivity defined in \eref{eq:conductivity} in terms of the CFT-data. After a brief calculation\footnote{See also \cite{Lucas:2016fju} for a similar expression.}, summarized in appendix~\ref{app:conductivity}, we obtain:
\bea\label{eq:conductivity_us}
\frac{\sigma(iw)}{\sigma_Q}  &=&  \frac{C_J}{32} + \frac{C_J \lambda_{JJS}}{4\pi} \frac{\Gamma(\Delta_S+1)\sin\left(\frac{\pi\Delta_S}2\right)}{2-\Delta_S} \Upsilon^{-1} \left(\frac{T}{w}\right)^{\Delta_S}+ 72\frac{C_J \gamma }{C_T } H_{xx} \left(\frac{T}{w}\right)^3 \ldots \nonumber\\
&=& \sigma_\infty + b_1 \left(\frac{T}{w}\right)^{\Delta_S}+ b_2 \left(\frac{T}{w}\right)^{3}+\ldots \, ,
\eea
where $\Delta_S$ is the dimension of the relevant singlet in the $O(2)$ model, $\Upsilon^{-1}$ measures the normalized  thermal expectation value of $S$, and $H_{xx}$ is the thermal one-point function of the stress tensor $xx$ component,  see appendix~\ref{app:conductivity} for a precise definition. The parameter $\lambda_{JJS}$ is the OPE coefficient determined in \eref{eq:lambda_JJS}. The central charges $C_J$ and $C_T$ measure the normalization of the conserved current $J_\mu$ and the stress tensor $T_{\mu\nu}$. Our conventions are such that in the theory of a single complex scalar one has
 \be
 C_J^\text{free}=2\,, \qquad \qquad C_T^\text{free}=3\,.
 \ee
 
By fitting the quantum Monte Carlo data,  \cite{Katz:2014rla} obtained the values $\sigma_\infty^{QMC}=0.5662(5)$, $b_1^{QMC}=1.43(5)$, $b_2^{QMC}=-0.4(1)$, $\Delta_S^{QMC}=1.526(65)$. In addition, they independently determined $\Upsilon=1.18(13)$, by fitting different observables, namely the one- and two-point function of the operator $S$. Using \eref{eq:conductivity_us} and the bounds on $C_J$ obtained in \cite{Kos:2015mba}, we can test the consistency of the results:
\bea
&\text{Bootstrap + QMC  conductivity fit: } &\Upsilon =  1.257(60)\, , \nonumber\\
&\text{QMC $\Upsilon$ direct fit:} & \Upsilon = 1.18(13) \, . \nonumber
\eea
We see that the two determinations of the parameter $\Upsilon$ are in agreement within their errors. In particular the one using the bootstrap results for $\lambda_{JJS}$ and $\Delta_S$ is more accurate.\footnote{Notice however that the value extracted for $\sigma_\infty^{QMC}$ from the fit of the conductivity is quite off compared to latest bootstrap and Monte Carlo determinations, which could be caused by systematic errors estimated of order 5-10\% in \cite{Witczak-Krempa:2013nua}. The value of $\Delta_S^{QMC}$ has instead larger uncertainties.}

Plugging \eref{eq:lambda_JJS} and \eref{eq:CTinO2} in expression \eref{eq:conductivity_us} we could also extract the value of the stress tensor thermal one-point function. Unfortunately the fit of the conductivity performed in  \cite{Katz:2014rla} is marginally sensitive to the sub-leading terms and the value determined for  $b_2$ has a large uncertainty.\footnote{Notice also that the next correction would come from the second neutral scalar $S'$, which has dimension slightly above 3, and should therefore be treated on equal footing as the stress tensor.} Nevertheless, we can estimate:
\vspace{0.2 cm}
\be
\text{Bootstrap + QMC  conductivity fit: }\qquad H_{xx} =0.105(30) \, . \nonumber \vspace{0.2 cm}
\ee
It would be nice to use the analytic bootstrap at finite temperature \cite{Iliesiu:2018fao,Iliesiu:2018zlz,Manenti:2019wxs} to compute the values of $\Upsilon$ and $H_{xx}$ and compare them with the predictions given in this work.

%% file: setup/setup.tex
\label{sec:setup}
In this section we explain our setup. 
We first discuss which are the possible operators exchanged in the OPEs and we enumerate their associated OPE coefficients. 
In subsection \ref{subsec:crossing} we explain how to write the crossing equations of the mixed $J$-$\phi$ sector (the two sectors with only currents or only scalars were already studied in the literature, \emph{e.g.} \cite{Kos:2013tga, Dymarsky:2017xzb}).
In subsection \ref{sec:CBs} we sketch which are the relevant conformal blocks and how we computed them.
Finally in subsection \ref{subsec:sum_rules} we summarize the full set of bootstrap equations in the form of sum rules.

Before entering the details of the setup, let us introduce the embedding space formalism \cite{Costa:2011mg}, which will be useful to classify conformal invariant tensor structures.
The idea is to uplift each coordinate to a null cone in $\mathbb{R}^{d+1,1}$, namely  $x\in \mathbb{R}^d \rightarrow P\in \mathbb{R}^{d+1,1}$  such that $P^2=0$. 
This is very convenient since the conformal group $SO(d+1,1)$ acts linearly on $\mathbb{R}^{d+1,1}$ thus  trivializing the problem of finding conformal invariants ---in fact the scalar product $P_1 \cdot P_2$ of two embedding points is conformal invariant.
In order to define correlation functions in embedding space we uplift primary operators.
We shall focus on primary operators $\Ocal(x,z)=z_{\m_1}\cdots z_{\m_\ell}\Ocal^{\m_1 \dots \m_\ell}(x)$ in a traceless and symmetric representation of $SO(d)$, which are conveniently contracted with null polarization vectors $z^{\m}$.
Each operator $\Ocal(x,z)$ with dimension $\D$ and spin $\ell$ is associated to a field $\Ocal(P,Z)$, which satisfies the condition
\beq
\Ocal(\lambda P, \alpha Z + \beta P ) = \lambda^{-\Delta} \alpha^{\ell} \Ocal(P,Z)\, ,
\label{eq:EmbField}
\eeq
where $Z\in \mathbb{R}^{d+1,1}$ is an uplifted polarization vector.
In the following we often classify  conformal invariant tensor structures by using the embedding space building blocks introduced in \cite{Costa:2011mg},
%
\begin{align}
H_{ij} &\equiv \frac{(Z_i \cdot Z_j )( P_i\cdot P_j) - (Z_i \cdot P_j )( Z_j\cdot P_i)}{(P_i \cdot P_j )}\,, \nonumber
\\
V_{i,jk} &\equiv \frac{(Z_i \cdot P_j )(P_i\cdot P_k )- (Z_i \cdot P_k )( P_i\cdot P_j)}{\sqrt{-2 (P_i \cdot P_j)(P_i \cdot P_k)(P_j \cdot P_k)}}\,.
\label{eq:HV} 
\end{align} 

For example the two-point function of a primary operator $\Ocal$ with dimension $\D$ and spin $\ell$ is defined as follows
\be
\label{2pt}
\langle  \Ocal(P_1,Z_1)  \Ocal(P_2,Z_2)\rangle = \ \frac{H_{12}^\ell}{P_{12}^{\D}} \ ,
\ee
where $P_{ij}\equiv -2  P_i \cdot P_j$.
The central charges of a theory are defined from the two point functions of canonically normalized currents and stress tensors, 
\be
\label{CJCT}
\langle  J(P_1,Z_1)  J(P_2,Z_2)\rangle =\frac{C_J}{(4\pi)^2} \ \frac{H_{12}}{P_{12}^{d-1}}
\ ,
\qquad
\langle  T(P_1,Z_1)  T(P_2,Z_2)\rangle =\frac{C_J}{(4\pi)^2} \ \frac{H_{12}^2}{P_{12}^{d}}
\ .
\ee
However we keep these operators to be unit normalized according to \eqref{2pt}. Therefore in our conventions $J$ and $T$ are rescaled as follows
 \be
 J \rightarrow J (4\pi)/\sqrt{C_J}\, ,
 \qquad
 T \rightarrow T (4\pi)/\sqrt{C_T} \, .
 \ee

\subsection{3pt functions} 
\label{sec:3pt}
One of the features that makes the scalar-current bootstrap richer and more involved is the presence of various different OPEs:
\begin{align}
J\times J, \qquad J\times \phi, \qquad \phi\times\phi, \qquad  \bar{\phi}\times\phi \ .
\end{align}
Imposing the equality of operators in $J\times J$ and $\phi\times\phi$ and asking for conservation of the currents we can enumerate the allowed OPE tensor structures as indicated in Table \ref{table:OPENUM}.
\begin{table*}[h!]
\centering
\as{1.2}
\begin{tabular}{@{}l l l l l l l@{}}
	\toprule
	 & $JJ\Ocal^{Q=0}_{\ell \, +}$  & $JJ\Ocal^{Q=0}_{\ell\, -}$ & $J \phi \Ocal^{Q=1}_{\ell\, +}$ & $J \phi \Ocal^{Q=1}_{\ell\, - }$ & $\phi \bar{\phi }\Ocal^{Q=0}_{\ell \,+}$ & $ \phi \phi \Ocal^{Q=2}_{\ell \,+}$ \\
	\midrule
	 $\ell=0$ &  $1$  & $1$ & $1$ & $0$ & $1$ & $1$ \\
	$\ell=1$ & $0$  & $0$ & $1$ & $1$ & $1$ & $0$ \\
	$\ell>0$, even \quad  & $2$ & $1$ & $1$ & $1$ & $1$ & $1$ \\
	$\ell>1$, odd  & $0$ & $1$ & $1$ & $1$ & $1$ & $0$ \\
	\bottomrule 
\end{tabular}
\caption{Summary of the number of allowed tensor structures for each three point function in our setup. The labels $\ell,\pm,Q$ respectively correspond to spin, parity and $U(1)$ charge of the exchanged operator. }
\label{table:OPENUM}
\end{table*}
The operators are written in the form $\Ocal^{Q}_{\ell \, p}$, where $\ell$ is the $SO(3)$ spin, $p$ is the parity and $Q$ is the charge under the $U(1)$ global symmetry. In the following we may drop some of these labels for the sake of brevity.

For most of the three-point functions considered in table \ref{table:OPENUM} there exists a unique tensor structure. We will refer to the associated OPE coefficient as $\l$, \emph{i.e.}
 \be
 \label{OPEs}
 \l_{J J \Ocal_{\ell=0 +}} \,, \qquad
 \l_{J J \Ocal_-} \,, \qquad
  \l_{J \phi \Ocal_\pm} \,, \qquad
   \l_{\phi \phi \Ocal_+} \,,  \qquad
   \l_{\phi \phib \Ocal_{+}} \ .
 \ee
Conversely there are two distinct OPE coefficients in the three-point functions of two currents and a parity even operator $\Ocal^{Q=0}_{\ell \, +}$ with even $\ell \neq 0$ which we will define as
\be
\label{lJJ}
 \l_{J J \Ocal_+}^{(1)} \ ,
 \qquad
 \l_{J J \Ocal_+}^{(2)} \ .
\ee
The explicit basis used to define OPE coefficients will not play an important role for the understanding of the results. For this reason we decided to keep this definition implicit in the main text and collect all the conventions in appendix \ref{app:threepoints}.

Next we use Ward identities to relate some OPE coefficients to the central charges $C_J$ and $C_T$ of equation \eqref{CJCT}. Using the Ward identities for $J$, we fix the OPE  $\lambda_{\phi \phib J}$ in terms of $C_J$.
For concreteness, in our normalization this relation takes the form\footnote{We always assume that the external scalar has charge $Q=1$ under the global $U(1)$.}
\begin{equation}\label{eq:phiphiJ}
\lambda_{\phi \phib J}=\frac{4\pi}{\sqrt{C_J}} \ .
\end{equation}

From the Ward identities of $T$  the OPE coefficients $ \lambda _{ \phi  \bar{\phi } T}$ can be fixed in terms of $C_T$ and $\D_{\phi}$. Similarly the OPE coefficients $ \lambda^{(1)} _{J J T},  \lambda^{(2)} _{J J T}$ are fixed in terms of $C_T$ and an extra parameter that we call $\gamma$. In our normalization:
\begin{align}
\lambda _{ \phi  \bar{\phi } T}&=\frac{\sqrt{3}}{ 2} \Delta _{\phi }  \sqrt{\frac{ \CTfree}{ C_T}}  \ , \quad \\
\lambda_{JJT}^{(1)}&=\frac{\sqrt{3}}{8}(1-12\gamma)   \sqrt{\frac{ \CTfree}{ C_T}}  \ , \quad \\
\lambda_{JJT}^{(2)}&=\frac{\sqrt{3}}{4}(5-12\gamma) \sqrt{\frac{ \CTfree}{ C_T}} \ ,
\label{eq:lambdaofgamma}
\end{align}
where $\CTfree \equiv 3 $ is the central charge of a free complex scalar.
The coefficient $\g$  is further constrained by the conformal collider bounds \cite{Hofman:2008ar} to lie in the following interval
\be
-\frac{1}{12} \leq \g \leq  \frac{1}{12} \, .
\ee
The two extremes correspond to  complex free scalar ($\g=-\frac{1}{12}$) and  free fermion theory  ($\g=\frac{1}{12}$).

\subsection{Crossing equations}
\label{subsec:crossing}
In this section we want to obtain all the crossing equations relevant for our setup.
Fortunately a big part of this goal is already solved in previous papers.
For the scalar correlators the situation is the standard one discussed for example in \cite{Kos:2013tga}. 
For the case of four currents we exactly use the same setup detailed in \cite{Dymarsky:2017xzb}.
What is left to discuss is the case of mixed correlators of two scalars and two conserved currents.
In the rest of the section we focus on detailing this case.
%
\vspace{-0.4 cm} 
\paragraph{Tensor structures  \vspace{0.2 cm} \\}
We start by considering four point functions of two scalars $\phi_i$ and two (so far non conserved) vectors $J_i$.
In order to classify the different tensor structures in their four point functions it is convenient to write the correlation functions in embedding space \cite{Costa:2011mg},
\ba 
\label{JsJs_Tensor_structures}
\langle J_1(P_1,Z_1) \phi_1(P_2) J_2(P_3,Z_3) \phi_2(P_4)\rangle
&
\equiv
\Kcal(P_i)
 \sum_{s=1}^{5} f_{s}(u,v) Q^{(f)}_{s}(\{P_i,Z_i\}) \ ,
 \\ 
 \label{JJss_Tensor_structures}
 \langle J_1(P_1,Z_1) J_2(P_2,Z_2)  \phi_1(P_3)  \phi_2(P_4)\rangle
&
\equiv
\Kcal(P_i)
 \sum_{s=1}^{5} g_{s}(u,v) Q^{(g)}_{s}(\{P_i,Z_i\}) \ ,
 \\ 
  \label{sJJs_Tensor_structures}
\langle   \phi_1(P_1) J_1(P_2,Z_2) J_2(P_3,Z_3) \phi_2(P_4)\rangle
&
\equiv
\Kcal(P_i)
 \sum_{s=1}^{5} h_{s}(u,v) Q^{(h)}_{s}(\{P_i,Z_i\}) \ , 
\ea
where $u \equiv P_{12}P_{34}/(P_{13}P_{24})$ and $v\equiv P_{23}P_{14}/(P_{13}P_{24})$ are the usual conformal cross ratios. The function $\Kcal$ is a fixed kinematical factor
\be
\label{convention_new2}
\Kcal(P_i)
\equiv
\k(v)
\frac{ 
\left(\frac{P_{24}}{P_{14}} \right)^{\frac{\D_1-\D_2}{2}}  \left(\frac{P_{14}}{P_{13}} \right)^{\frac{\D_3-\D_4}{2}}}{(P_{12})^{\frac{\D_1+\D_2}{2}}(P_{34})^{\frac{\D_3+\D_4}{2}}} \, ,
\qquad
\k(v) \equiv v^{- \frac{ \D_2 + \D_3}{2}} \ .
\ee
The factor $\k(v)$ is introduced to get nicer definitions for the crossing equations.
The tensor structures $Q_s$ are the $s$-th component of the vectors ${\vec Q}$ defined below
\ba
\label{QsMixed}
\begin{split}
{\vec Q}^{(f)}=  \left\{H_{13},V_{1,23} V_{3,21},V_{1,23} V_{3,41},V_{1,43} V_{3,21},V_{1,43}
   V_{3,41}\right\} \ ,
   \\
   {\vec Q}^{(g)}=\left\{H_{12},V_{1,23} V_{2,14},V_{1,23} V_{2,34},V_{1,43} V_{2,14},V_{1,43}
   V_{2,34}\right\} \ ,
   \\
{\vec Q}^{(h)}=   \left\{H_{23},V_{2,14} V_{3,21},V_{2,14} V_{3,41},V_{2,34} V_{3,21},V_{2,34}
   V_{3,41}\right\} \ ,
   \end{split}
\ea
where the structures $H_{ij}$ and $V_{i,jk}$ are the building blocks of \cite{Costa:2011mg} defined in \eqref{eq:HV}.
So far the structures $Q_s$ are fixed only by scaling. Extra constraints will be imposed in the following by requiring that the two currents $J_i$ are equal and conserved and by imposing that $\Delta_{\phi_1}=\Delta_{\phi_2}$. 
%
\vspace{-0.4 cm} 
\paragraph{Crossing equations\vspace{0.2 cm} \\}
Now that the tensor structures are classified, we are ready to write the crossing equations. 
Crossing equations are obtained by demanding the invariance of the four point functions under the permutations $1 \leftrightarrow 3$ (\emph{i.e.} of the operators inserted at point $P_1$ and $P_3$). This implies relations between different functions $f_{s}$ and relates the functions $g_{s}$ and  $h_{s}$. The resulting equations can be diagonalized by introducing the following change of basis, 
\be
\label{def:hat}
f_s\equiv \tfrac{1}{\sqrt{2}}\sum_{s'=1}^5 (M_f)_{s s'} \hat  f_{s'} 
\ , \qquad
g_s\equiv \tfrac{1}{\sqrt{2}}\sum_{s'=1}^5 (M_g)_{s s'} \hat  g_{s'}
\ , \qquad
h_s\equiv \tfrac{1}{\sqrt{2}}\sum_{s'=1}^5 (M_h)_{s s'} \hat  h_{s'}
\ ,
\ee
where $M_{f,g,h}$ are $5\times 5$ matrices defined as follows
\begingroup
\setlength{\arraycolsep}{3.3pt}
\be
\begin{array}{l}
M_f
\equiv
\left(
\begin{array}{lcccc}
 0 \ & 0 & 0 & \!\!\sqrt{2} & 0 \\
 0 & 1 & 0 & 0 &  \!\scalebox{0.75}[1.0]{\( - \)}1 \\
 1 & 0 & 1 & 0 & 0 \\
 1 & 0 & \!\scalebox{0.75}[1.0]{\( - \)} 1 & 0 & 0 \\
 0 & 1 & 0 & 0 & 1 \\
\end{array}
\right)
\, ,
\quad
M_g
\equiv
\left(
\begin{array}{lcccc}
 0\ & 0 & 0 & \!\!\sqrt{2} & 0 \\
 0 & 1 &  \!\scalebox{0.75}[1.0]{\( - \)}1 & 0 & 0 \\
 1 & 0 & 0 & 0 & 1 \\
 1 & 0 & 0 & 0 &  \!\scalebox{0.75}[1.0]{\( - \)}1 \\
 0 & 1 & 1 & 0 & 0 \\
\end{array}
\right)\, ,
\quad
M_h
\equiv
 \left(
\begin{array}{lcccc}
 0 \ & 0 & 0 & \!\!\sqrt{2} & 0 \\
 1 & 0 & 0 & 0 &  \!\scalebox{0.75}[1.0]{\( - \)}1 \\
 0 & 1 & 1 & 0 & 0 \\
 0 & 1 &  \!\scalebox{0.75}[1.0]{\( - \)}1 & 0 & 0 \\
 1 & 0 & 0 & 0 & 1 \\
\end{array}
\right)\, .
\end{array}
\ee
\endgroup
With these definitions the permutation  $1 \leftrightarrow 3$ in \eqref{JsJs_Tensor_structures} and \eqref{JJss_Tensor_structures} results in the following set of crossing equations
\ba
\begin{split}
\label{crossinghat}
\hat f_s(u,v)&=&\hat f_s(v,u) \ ,
 \quad s=1,2,4,5
\  ,
\qquad \ \ 
\hat f_3(u,v)=- \hat f_3(v,u) \ ,
\\ 
\hat g_s(u,v)&=& \hat h_s(v,u) 
\ ,
 \quad s=1,2,3,4
 \ ,
\qquad \ \ 
\hat g_5(u,v)=- \hat h_5(v,u) \ .
\end{split}
\ea
\paragraph{Equality \vspace{0.2 cm} \\}
When the two vectors and the two scalars are equal (\emph{i.e.} $J_i=J$, $\phi_i=\Ocal$) we can use extra crossing relations (for example a $J\Ocal J\Ocal$ is invariant under $(1,2)\leftrightarrow(3,4)$) 
which constrain the functions $\hat f,\hat g,\hat h$,
\be
\label{solving_conservation}
\hat f_5(u,v)=0 \ , \qquad \hat g_5(u,v)=0 \ , \qquad \hat h_5(u,v)=0 \ .
\ee
However, we are interested in the case of different scalar operators with the same scaling dimension $\Delta_{\phi_1}=\Delta_{\phi_2}$. In this case we are not allowed to use the crossing relation above, however \eqref{solving_conservation} still holds.
Indeed we could show that for $\Delta_{\phi_1}=\Delta_{\phi_2}$ the conformal blocks which decompose the functions $\hat f_5,\hat g_5,\hat h_5$ exactly vanish. Thus, the functions must vanish too. 
\paragraph{Conservation \vspace{0.2 cm} \\}
Conservation of the two operators $J_i$ gives four independent partial differential equations (of the first order) for the functions $\hat f_s$ (similarly for $\hat g_s$ and  $\hat h_s$),
\ba
\label{conservation_fgh}
\begin{split}
\sum_{s=1}^5 [(M^{(f)}_u)_{s' s} \partial_u +(M^{(f)}_v)_{s' s} \partial_v + (M^{(f)}_0)_{s' s}] \hat f_s(u,v)=0 \ , 
\qquad
s'=1,2,3,4, \\
\sum_{s=1}^5 [(M^{(g)}_u)_{s' s} \partial_u +(M^{(g)}_v)_{s' s} \partial_v + (M^{(g)}_0)_{s' s}] \hat g_s(u,v)=0 \ , 
\qquad
s'=1,2,3,4, \\
\sum_{s=1}^5 [(M^{(h)}_u)_{s' s} \partial_u +(M^{(h)}_v)_{s' s} \partial_v + (M^{(h)}_0)_{s' s}] \hat h_s(u,v)=0 \ , 
\qquad
s'=1,2,3,4, 
\end{split}
\ea
where $M_u, M_v, M_0$ are $4\times 5$ matrices which depend on $u,v$. 

Two of the four differential equations in \eqref{conservation_fgh} (for example $s'=3,4$) involve only the fifth functions $\hat f_5$ (similarly for $\hat g_5$ and  $\hat h_5$). These two equations are therefore not important in our setup since, as we argued above, the functions $\hat f_5,\hat g_5,\hat h_5$ must vanish when $J_1=J_2$ and $\Delta_{\phi_1}=\Delta_{\phi_2}$.\footnote{As a curiosity we would like to report that, when the conformal dimensions of the two scalars is $\Delta _{\phi }$ and the one of the currents is $\Delta _{J}$, we could solve these two differential equations, finding
\ba
\hat f_5(u,v) &=&c_1 \ \left(1 - 2 u + (u - v)^2 - 2 v \right)^{-\frac{\Delta _J}{2}} (u v)^{\frac{\Delta _{\phi }}{2}+\frac{\Delta _J}{2}}\ ,
\\
\hat  g_5(u,v) &=&c_2 \ u^{\Delta _J+\frac{1}{2}}  v^{\frac{\Delta _{\phi }}{2}+\frac{\Delta _J}{2}}   \left(1 - 2 u + (u - v)^2 - 2 v\right)^{-\frac{\Delta _J}{2}} \ ,
\\
\hat  h_5(u,v) &=&c_3 \ u^{\frac{\Delta _{\phi }}{2}+\frac{\Delta _J}{2}}  v^{\Delta _J+\frac{1}{2}} \left(1 - 2 u + (u - v)^2 - 2 v\right)^{-\frac{\Delta _J}{2}}
\ ,
\ea
where $c_i$ are constants of integration. Compatibility with the conformal blocks decomposition requires $c_i=0$. 
}

The remaining two differential equations, $s'=1,2$, involve non zero functions. For the case $J_1 \phi_1 J_2  \phi_2$ one can use them to evolve the crossing equations of $\hat f_3(u,v)=- \hat f_3(v,u) $ and $\hat f_4(u,v)=\hat f_4(v,u) $ from the line $u=v$ to the plane. The crossing equation for  $\hat f_4$ is trivially satisfied on the line therefore we do not need to impose extra equations. On the other hand to ensure crossing symmetry for $\hat f_3$ we need to impose the extra condition $\hat f_3(u,u)=0$. 
For the case of $J J  \phi \bar{\phi}$ the conservation equations can be used to evolve the equations
$
\hat g_3(u,v)= \hat h_3(v,u) $ and
$
\hat g_4(u,v)= \hat h_4(v,u) $
from the line $u=v$ to the full plane.
One can in fact explicitly check that the evolution equations for $ \hat g_3(u,v)$ and $\hat g_4(u,v)$ are exactly equal to the ones of $\hat h_3(v,u)$ and $\hat h_4(v,u)$.
In summary the final set of crossing equations for two conserved equal currents and two scalars with equal dimensions are
\be
\label{crossing_Jcons_plane}
\hat f_s(u,v)=\hat f_s(v,u) \ ,
\quad 
\hat g_s(u,v)=\hat h_s(v,u) \ ,
 \qquad (s=1,2)
\ee
with the following constraint on the line
\be
\label{crossing_Jcons_line}
\hat f_3(u,u)=0 \ ,
\quad
\hat g_3(u,u)=\hat h_3(u,u) \ ,
\quad
\hat g_4(u,u)=\hat h_4(u,u) \ .
\ee
\subsection{Conformal Blocks}
\label{sec:CBs}
In the previous section we explained how to write the crossing equations. 
The basic idea of the bootstrap is to require the compatibility of the crossing equations with the conformal block decomposition. In this section we explain which are the relevant conformal blocks for our setup.

Let us consider a four point functions $\langle \Ocal_1 \Ocal_2\Ocal_3 \Ocal_4\rangle $ of operators $\Ocal_i$ with dimensions $\D_i$ and spin $\ell_i$. By taking the OPE $\Ocal_1\times \Ocal_2$ and  $\Ocal_3\times \Ocal_4$ we obtain the following conformal block decomposition
\be
\label{CB_decomposition}
\langle \Ocal_1 \Ocal_2\Ocal_3 \Ocal_4\rangle = \Kcal \sum_{p,q} \l^{(p)}_{\Ocal_1\Ocal_2\Ocal}\l^{(q)}_{\Ocal_3\Ocal_4\Ocal}\sum_{s} g^{(p,q) \, \Ocal_1 \Ocal_2 \Ocal_3 \Ocal_4}_{\Ocal, s}(u,v) Q_s \, ,
\ee
where $\Kcal$ is the prefactor defined in \eqref{convention_new2}, $Q_s$ are the four-point function conformal invariant tensor structures (E.g. \eqref{QsMixed}) and $\l^{(p)},\l^{(q)}$ are the left and right OPE coefficients.
The conformal blocks $g^{(p,q) \, \Ocal_1 \Ocal_2 \Ocal_3 \Ocal_4}_{\Ocal, s}(u,v)$ are functions of the cross ratios $u$ and $v$, built out of the insertions of the four operators. They depend on the representation of the exchanged operator $\Ocal$ which is labelled by $\D$, $\ell$ and $\pm$.
The dependence on the external operators $ \Ocal_i$ is twofold. Firstly, they depend on their conformal dimension $\D_i$, through the combinations $\D_{12}$ and $\D_{34}$, where $\Delta_{ij}\equiv \D_i-\D_j$. 
Most importantly they depend on the spins $\ell_i$ of $ \Ocal_i$ which are responsible for the presence of different tensor structures both for the OPE and for the four point function. This affects the possible values of the conformal block labels $p$, $q$ and $s$.

There are different strategies to compute conformal blocks. In this paper we mostly used a recurrence relation \cite{Kos:2013tga, Penedones:2015aga} which builds the blocks as a power series in the radial coordinates $r\equiv |\rho|, \eta\equiv (\rho+\bar \rho)/(2 |\rho|)$ of \cite{Hogervorst:2013sma},
where
\be \label{radcoord}
\rho=\frac{z}{(1+\sqrt{1-z})^2} \ , \qquad \bar\rho=\frac{\bar z}{(1+\sqrt{1-\bar z})^2} \ ,
\ee
and $u=z\bar{z}$ and $v=(1-z)(1-\bar{z})$.
The recurrence relation 
is defined by studying the analytic structure of the conformal blocks as functions of $\D$.
It takes the form
\be
\label{recurrence_rel}
h^{(p,q) \, \Ocal_1 \Ocal_2 \Ocal_3 \Ocal_4}_{\Delta \ell, s}(r,\eta)=
h^{(p,q) \, \Ocal_1 \Ocal_2 \Ocal_3 \Ocal_4}_{\infty \ell, s}(r,\eta)+
\sum_{A} (4r)^{n_A}
\frac{(R_{A})_{p p' q q'}}{\D-\D_A^\star}
h^{(p,q) \, \Ocal_1 \Ocal_2 \Ocal_3 \Ocal_4}_{\Delta_A \ell_A, s}(r,\eta) \ ,
\ee 
where $h^{(p,q)}_{\Delta \ell, s}(r,\eta)\equiv (4r)^{-\D} g^{(p,q)}_{\Delta \ell, s}(r,\eta)$.
There are a few ingredients that enter this formula: $h_{\infty}$, $R_{A}$ and the labels $\D_A^\star,\Delta_A, \ell_A, n_A$. The latter are known from representation theory for any conformal block in generic dimensions, while $h_{\infty}$, $R_{A}$ can be computed by some standard computations \cite{Penedones:2015aga, Costa:2016xah, Costa:2016hju}. Moreover, recently the paper \cite{Erramilli:2019njx} appeared with a closed form solution for $h_{\infty}$ and $R_{A}$ for any conformal block in $d=3$. This will be a very useful tool to implement the conformal bootstrap in more complicated situations involving mixed correlators with spinning operators.
 
For our setup we need to compute five different kinds of conformal blocks
\be
g^{\phi \phi \phi \phi}_{\Ocal}(u,v) \, , \qquad
g^{ J \phi J \phi}_{\Ocal, s}(u,v)\, , \qquad
g^{ \phi J  J \phi}_{\Ocal, s}(u,v)\, , \qquad
g^{(p) \,J  J\phi \phi}_{\Ocal, s}(u,v)\, , \qquad
g^{(p,q) \, J J  J J}_{\Ocal, s}(u,v)\, ,
\ee
where $\phi$ here stands for any scalar operator of dimension $\D_{\phi}$.
In the following we discuss how we computed these conformal blocks and some of their features.
\begin{itemize}
	\item[$\phi \phi \phi \phi$:] 
	The scalar block is a single function of the cross ratios which we computed, as customary, by means of the recurrence relation \eqref{recurrence_rel}.
		\item[$J\phi J \phi$:] 
	The mixed blocks $g^{ J \phi J \phi}_{\Ocal, s}(u,v)$ were computed in \cite{Costa:2016xah} using the recurrence relation \eqref{recurrence_rel}. We used the ancillary file that was included in the publication.
Notice that in \cite{Costa:2016xah} the blocks are defined for generic spacetime dimension $d$ and also for generic non conserved vectors $J_1,J_2$ and different scalars $\phi_1,\phi_2$. The package generates  $g^{(p,q) \, J \phi J \phi}_{\Ocal, s}(u,v)$ for $s=1,\dots 5$ and for $\Ocal$ belonging both to traceless symmetric representation of spin $\ell$ (in this case $p,q=1,2$)  and to the mixed symmetric representation $(\ell,1)$ of $SO(d)$  (in this case $p,q=1$). 
For our setup we need to consider $\phi_1=\phi_2$ and $J_1=J_2$ conserved in dimension $d=3$  ---their normalization is discussed in appendix \ref{app:CBJsJs}.
This implies that we do not need to compute the $5$ structures labelled by $s$ but only the three combinations  useful to decompose \eqref{crossing_Jcons_plane} and \eqref{crossing_Jcons_line} (one of these combinations is computed only in the $u=v$ line). 
Finally, we stress that the $(\ell,1)$ representation of $SO(d)$ is identified as a parity odd spin $\ell$ representation of $SO(3)$.
\item[$\phi JJ \phi$:] 
The blocks  $g^{  \phi JJ \phi}_{\Ocal, s}(u,v)$ have the same exact features as $g^{ J \phi J \phi}_{\Ocal, s}(u,v)$.
Indeed they can be computed from the latter by using permutation of the operators $1 \leftrightarrow 2$ as explained in appendix \ref{app:CBsJJs}. In particular one can compute the blocks $g^{  \phi JJ \phi}_{\Ocal, s}(u,v)$ at some order in the $r$ expansion by knowing the blocks $g^{ J \phi J \phi}_{\Ocal, s}(u,v)$ at the same order (provided that the complete dependence in the variable $\eta$ is known at that order). However we decided to compute these blocks by using the differential operators of \cite{Costa:2011dw} to test if this algorithm was as effective as the recurrence relation. 
In our implementations, the recurrence relation was  faster.
	\item[$J J \phi \phi$:] 
We computed $g^{ \phi J  J \phi}_{\Ocal, s}(u,v)$ using the recurrence relation \eqref{recurrence_rel} as we detail in appendix \ref{app:CBJJss}.
Our program works in arbitrary dimensions and for generic vector and scalar operators.
For a generic setup $p$ and $s$ take values from $1$ to $5$.  In our case, due to conservation, $p$ only runs over $1$ and $2$ and we only require $s=1,2,3,4$  (two on the $u,v$ plane plus two at the $u=v$ line) which are enough to expand the crossing equations \eqref{crossing_Jcons_plane} and \eqref{crossing_Jcons_line}. 
\item[$J J J J$:] 
The $g^{(p,q) \, J J  J J}_{\Ocal, s}(u,v)$ blocks were computed in \cite{Dymarsky:2017xzb} using a recurrence relation valid for generic vectors in $d=3$. 
Of the $41$ values of $s$, only $11$ combinations are useful to expand the crossing equation for conserved equal currents ($5$ on the full $u,v$ plane, $5$ on the $u=v$ line and $1$ at the point $u=v=1/4$). Moreover, for the conserved blocks, the values of $p,q$ again run only from $1$ to $2$.
In this work we re-used the blocks generated for the paper \cite{Dymarsky:2017xzb}. 
   \end{itemize}
Generating the five ingredients is not trivial. 
The $JJJJ$ blocks are the hardest task which was already done. However, computing the blocks $\phi JJ \phi$ and $J\phi J \phi$ is also very expensive because they both depend on the value of $\D_{\phi}$. We decided to generate these functions at low derivative order with an explicit dependence on $\D_{\phi}$. This enabled us to perform some exploratory scans in the dimension of $\phi$. We then computed them at higher derivatives for some fixed values of $\D_{\phi}$ compatible with the $O(2)$ model.
Finally, the computation of the $JJ \phi \phi$ and $\phi \phi\phi \phi$ blocks is reasonably fast.
\subsection{Sum rules}
\label{subsec:sum_rules}
The bootstrap equations are obtained by combining the crossing equations of subsection \ref{subsec:crossing} with the conformal block decomposition  of subsection \ref{sec:CBs}.
They take the form of sum rules for some functions $F^{[\pm]}$ which are defined in terms of combinations of conformal blocks,
\ba
\label{Fpm}
F^{[\pm] \Ocal_1 \Ocal_2 \Ocal_3\Ocal_4} &\equiv&  \k(v) g^{ \Ocal_1 \Ocal_2 \Ocal_3\Ocal_4}_{\Ocal}(u,v) \pm (u \leftrightarrow v) \ , 
\qquad 
\k(v) \equiv v^{- \frac{ \D_2 + \D_3}{2}} \ .
\ea
In what follows we write down the sum rules for all the considered correlators.
The goal is to reach a single vectorial bootstrap equation that can be analyzed by means of semidefinite programming.\vspace{-0.4 cm} 
\paragraph{The scalar sector\vspace{0.2 cm} \\}
Let us start by reviewing how the scalar bootstrap equations arise.
By equating the OPE channels $(12)(34)=(13)(24)$ of the correlation function  $\langle \phi  \phib \phi \phib \rangle$, we get the following equation
\be
\sum_{\Ocal^{Q=0}_{\D\, \ell \,  + } } (-1)^\ell |\l_{\phi \bar \phi \Ocal}|^2 F ^{[-]\phi \bar \phi\phi \bar \phi}_{\Ocal}(u,v)=0 \, .
\ee
The same strategy applied to $\langle  \phib \phi \phi \phib \rangle$ generates two equations
\be
\label{sbsssb}
 \sum_{\Ocal^{Q=0}_{\D\, \ell \,  + } } |\l_{\phi \bar \phi \Ocal}|^2 F^{[\pm]\bar \phi \phi\phi \bar \phi}_{\Ocal}(u,v) 
 \mp
\sum_{
\Ocal^{Q=2}_{\D\, \ell \,  + } } |\l_{\phi \phi  \Ocal}|^2 F^{[\pm]\phi  \phi \bar \phi \bar \phi}_{\Ocal}(u,v)=
0 \, ,
\ee
parametrized by the label  $\pm$. \vspace{-0.4 cm} 
\paragraph{The mixed sector \vspace{0.2 cm} \\}
As we explain in subsection \ref{subsec:crossing}, the crossing equations for the mixed correlators take a  simpler form in the hatted basis \eqref{def:hat}.
It is therefore convenient to define new functions $F^{[\pm] }$ which are rotated accordingly,
\be
\begin{array}{ccl}
F^{[\pm] J \phi J \phib}_{\Ocal,  s}(u,v) &\equiv& \k(v)^{-1} \sum_{s'=1}^5 (M_f^{-1})_{s s'} g^{[J \phi J \phib]}_{\Ocal, s'}(u,v) \pm (u \leftrightarrow v) \ , 
\\
F^{(q)[\pm] J  J \phi \phib }_{\Ocal,  s}(u,v) &\equiv& \k(v)^{-1} \sum_{s'=1}^5 (M_g^{-1})_{s s'} g^{(q)[J  J\phi  \phib]}_{\Ocal, s'}(u,v) \pm (u \leftrightarrow v) \ ,
\\
F^{[\pm] \phi J J \phib}_{\Ocal,  s}(u,v) &\equiv& \k(v)^{-1} \sum_{s'=1}^5 (M_h^{-1})_{s s'} g^{[\phi J   J \phib]}_{\Ocal, s'}(u,v) \pm (u \leftrightarrow v) \ ,
\end{array}
\ee
In this notation it is easy to write the bootstrap equations.
From $\langle J \phi J \bar \phi \rangle$ we get two equations on the plane and one on a line
\ba
\sum_{
\Ocal^{Q=1}_{\D\, \ell \,  \pm } 
} \sigma_\Ocal  |\l_{J \phi \Ocal}|^2F^{[-]  J \phi J \phib }_{\Ocal,  s}(u,v) =0 \, ,&
\; s=1,2\, ,\\
\sum_{
\Ocal^{Q=1}_{\D\, \ell \,  \pm } 
}  \sigma_\Ocal  |\l_{J \phi \Ocal}|^2F^{[+]  J \phi J \phib }_{\Ocal,  s}(u,u) =0 \ ,&
\; s=3\, .
\ea
Here $\sigma$ is a sign which depends on the normalization of the three point functions.\footnote{This is due to the fact that we need rewrite the OPE coefficients in a positive combination,
\ba
\l_{ \phi J  \bar \phi } \l_{\phi J \bar \phi } = \l_{J \phi  \bar \phi } \l_{\phi J \bar \phi } &=& |\l_{J \phi  \bar \phi }|^2 \, , \\
\l_{\phi J  \Ocal} \l_{ \bar \Ocal J \bar \phi } =\l_{J \phi  \Ocal} \l_{ \bar \Ocal J \bar \phi } &=&(-1)^{\ell+p+1} |\l_{J \phi  \Ocal}|^2 \qquad (l>0)\, .
\ea
} In our case
\be
\label{sigma_sign}
\sigma_\Ocal =
\left\{
\begin{array}{ll}
1 & \mbox{if } \Ocal=\bar \phi \\
(-1)^{\ell+p+1} & \mbox{if }  \Ocal \neq \bar \phi
\end{array}
\right.
\ ,
\ee
where $p=0,1$ and $\ell$ are respectively the parity and the spin of the exchanged operator $\Ocal$.

From $\langle  \phi  JJ \bar \phi \rangle$ we get four equations on a plane (labelled by $s=1,2$ and $[\pm]$) and two on the line ($s=3,4$),
\ba
\sum_{
\Ocal^{Q=0}_{\D\, \ell \,  + } 
}\sum_{q=1}^2 \l^{(q)}_{J J \Ocal} \l_{ \phi \phib \Ocal} F^{(q)[\pm] J J \phi \phib}_{\Ocal,  s}(u,v) \mp \sum_{
\Ocal^{Q=1}_{\D\, \ell \,  \pm } 
} \sigma_\Ocal  |\l_{J \phi  \Ocal}|^2 F^{[\pm]   \phi JJ \phib}_{\Ocal,  s}(u,v) =0 \, ,&
\; s=1,2\, , 
\\
\sum_{
\Ocal^{Q=0}_{\D\, \ell \,  + } 
}\sum_{q=1}^2 \l^{(q)}_{J J \Ocal} \l_{\phi \phib \Ocal} F^{(q)[+] J J \phi \phib}_{\Ocal,  s}(u,u) - \sum_{
\Ocal^{Q=1}_{\D\, \ell \,  \pm } 
} \sigma_\Ocal  |\l_{J \phi  \Ocal}|^2 F^{[+]   \phi JJ \phib}_{\Ocal,  s}(u,u) =0 \, ,&
\; s=3,4\, .
\ea
\paragraph{The current sector \vspace{0.2 cm} \\}
Finally we review the sum rules for $\langle JJJJ \rangle$  as obtained in \cite{Dymarsky:2017xzb}.
In that case conservation and equality of the currents produced a set of $5$ crossing equations valid on the plane, $5$ on a line and a single one at a point. 
In terms of opportune functions $F^{[\pm]}$,\footnote{The exact meaning of the functions $F^{[\pm]}$ is defined in \cite{Dymarsky:2017xzb}, where it is used the notation $F^{[+]}\to \tilde H$ and  $F^{[-]}\to \tilde F$.} the equations are casted into the following sum rules 
\be
\nonumber
\sum_{p=1}^{2} \sum_{q=1}^{2} \sum_{
\Ocal^{Q=0}_{\D\, \ell \,  + } 
} \l^{(p)}_{J J \Ocal_{+}}\l^{(q)}_{J J \Ocal_{+}} F^{[-](p,q)JJJJ}_{\Ocal_{+},s}(u,v)+
 \sum_{\Ocal^{Q=0}_{\D\, \ell \,  - } } | \l_{J J \Ocal_{-}}|^2  F^{[-]JJJJ}_{\Ocal_{-},s}(u,v)
=0 \ ,\quad s=13,15,16,17
\ee
\be
\nonumber
\sum_{p=1}^{2} \sum_{q=1}^{2} \sum_{
\Ocal^{Q=0}_{\D\, \ell \,  + } 
} \l^{(p)}_{J J \Ocal_{+}}\l^{(q)}_{J J \Ocal_{+}} F^{[+](p,q)JJJJ}_{\Ocal_{+},s}(u,v)+
 \sum_{\Ocal^{Q=0}_{\D\, \ell \,  - } } | \l_{J J \Ocal_{-}}|^2  F^{[+]JJJJ}_{\Ocal_{-},s}(u,v)=0 
 \ ,
\ \ \ \ \ \ \ \  \ \ \ \ \ \quad s=7
\ee
\be
\nonumber
\sum_{p=1}^{2} \sum_{q=1}^{2} \sum_{\Ocal^{Q=0}_{\D\, \ell \,  + } } \l^{(p)}_{J J \Ocal_{+}}\l^{(q)}_{J J \Ocal_{+}} F^{[+](p,q)JJJJ}_{\Ocal_{+},s}(u,u)+
 \sum_{\Ocal^{Q=0}_{\D\, \ell \,  - } } | \l_{J J \Ocal_{-}}|^2  F^{[+]JJJJ}_{\Ocal_{-},s}(u,u)
=0
 \ ,
 \quad s=1,2,4,5,6
\ee
\be
\nonumber
\sum_{p=1}^{2} \sum_{q=1}^{2} \sum_{\Ocal^{Q=0}_{\D\, \ell \,  + } } \l^{(p)}_{J J \Ocal_{+}}\l^{(q)}_{J J \Ocal_+} F^{[+](p,q)JJJJ}_{\Ocal_{+},s}(\tfrac{1}{4},\tfrac{1}{4})+
 \sum_{\Ocal^{Q=0}_{\D\, \ell \,  - } } | \l_{J J \Ocal_{-}}|^2  F^{[+]JJJJ}_{\Ocal_{-},s}(\tfrac{1}{4},\tfrac{1}{4})
=0
 \ ,
 \ \ \ \ \ \ \ \  \ \ \ \ \
 \quad s=3
\ee
In the equations above we explicitly show the parity of the exchanged operator since the number of OPE coefficients depends on this quantum number. \vspace{-0.4 cm} 
%
\paragraph{The bootstrap equation\vspace{0.2 cm} \\}
Since  $ \l^{(p)}_{J J \Ocal_{+}}\l^{(q)}_{J J \Ocal_{+}}$ and $\l^{(p)}_{J J \Ocal} \l_{\phi \bar \phi \Ocal} $ are not ensured to be positive quantities, it is necessary to rearrange the equations into a single expression that can be studied using the standard semidefinite programming techniques,
\ba
\label{BOOTSTRAP_EQNS}
\begin{split}
& \vec{\lambda}_\mathbb{1}^T \vec V_{0,0,+}^{Q=0} \vec{\lambda}_\mathbb{1} +
\sum_{ 
\Ocal_{\ell=0,+} ^{Q=0}
\atop 
{}
 } 
\left(
\begin{array}{c}
\l_{ \phi  \bar \phi \Ocal_{+}}
\\
\\
\l^{(1)}_{ J  J  \Ocal_{+}}
\end{array}
\right)^{\! \!T}
\!
\!
\!
\cdot
\!
\vec
V^{Q=0}_{\D,0,+}
\!
\cdot
\!
\left(
\begin{array}{c}
\l_{ \phi  \bar \phi \Ocal_{+}}
\\
\\
\l^{(1)}_{ J  J  \Ocal_{+}}
\end{array}
\right)
+
\sum_{
\Ocal_{\ell>0 \textrm{ even},+} ^{Q=0}
 } 
\left(
\begin{array}{c}
\l_{ \phi  \bar \phi \Ocal_{+}}
\\
\l^{(1)}_{ J  J  \Ocal_{+}}
\\
\l^{(2)}_{ J  J  \Ocal_{+}}
\end{array}
\right)^{\! \! T}
\!\!\!
\cdot
\!
\vec
V^{Q=0}_{\D,\ell,+}
\!
\cdot
\!
\left(
\begin{array}{c}
\l_{ \phi  \bar \phi \Ocal_{+}}
\\
\l^{(1)}_{ J  J  \Ocal_{+}}
\\
\l^{(2)}_{ J  J  \Ocal_{+}}
\end{array}
\right)
\\
+&
\sum_{
\Ocal_{\ell \textrm{ odd},+} ^{Q=0}
} 
|\l_{ \phi  \bar \phi \Ocal}|^2 
\vec V^{Q=0}_{\D,\ell,+}
 +
\sum_{
\Ocal_{\ell \neq 1,-} ^{Q=0}
} 
|\l_{J J \Ocal_{-}}|^2
\vec V^{Q=0}_{\D,\ell,-}
+
\sum_{
\Ocal_{ \ell \geq1,+} ^{Q=1}
 }
|\l_{J \phi \Ocal}|^2
\vec V^{Q=1}_{\D,\ell,+}
+
\sum_{
\Ocal_{ \ell \geq 1,-} ^{Q=1}
 }
|\l_{J  \phi \Ocal}|^2
\vec
V^{Q=1}_{\D,\ell,-}
\\
+&
\sum_{
\Ocal_{\ell \textrm{ even} ,+} ^{Q=2}
}
|\l_{\phi \phi  \Ocal}|^2
\vec
V^{Q=2}_{\D,\ell,+}+ |\l_{J \phi  \bar \phi }|^2 
 \
\vec V^{Q=1}_{\D_\phi,0,+} 
=
0
\, .
\end{split}
\ea
where $\vec{\lambda}_\mathbb{1}=(1,1)$.
Here we have separated the case of $\Ocal_{\ell=0,+} ^{Q=0}$ from the other $\ell>0$, since in former case there is no OPE coefficient $\l_{JJ\Ocal}^{(2)}$. In appendix \ref{vectors_bootstrap} we write explicitly all the vectors $\vec V^{Q}_{\D,\ell,\pm}$, where $\D$ is the conformal dimension, $\ell$ is the spin, $\pm$ is the parity and $Q=0,1,2$ is the charge of the exchanged operator. By construction all the vectors $\vec V$ are 23 dimensional. The 23 components of the vector $\vec V^{Q=0}_{\D,\ell,+}$ are $3 \times 3$ matrices for $\ell>0$, and $2 \times 2$ matrices for $\ell=0$. The components of all the other vectors  $\vec V$ do not have any matrix structure.

In the next sections we use the following convention to denote the gaps of the exchanged operators
\be
\label{def:gaps}
\D_{\ell,\pm}^{ Q}
\equiv
\mbox{Gap for operators with spin $\ell$, parity $\pm$ and charge $Q$} \ .
\ee
In this notation the bootstrap equations \eqref{BOOTSTRAP_EQNS} depend on the following five infinite families of gaps,
\be
\label{5knobs}
\D_{\ell,+}^{ Q=0} 
\, ,
\qquad
\qquad
\D_{\ell,-}^{ Q=0} 
\, ,
\qquad
\qquad
\D_{\ell,+}^{ Q=1} 
\, ,
\qquad
\qquad
\D_{\ell,-}^{ Q=1} 
\, ,
\qquad
\qquad
\D_{\ell,+}^{ Q=2} 
\, .
\ee
For brevity we sometimes refer to the gap of important operators by their name, \emph{e.g.}
 $\D_S=\D_{\ell=0,+}^{ Q=0}$, $\D_{\phi^2}=\D_{\ell=0,+}^{ Q=2}$, 
$\D_{T}=\D_{\ell=2,+}^{ Q=0}$, and so on.
We assume that the CFT is unitarity, namely that all gaps in \eqref{BOOTSTRAP_EQNS} are consistent with the unitarity bounds,
\be
\D_{\ell=0,\pm}^{ Q} \geq \frac{d}{2}-1 \ ,
\qquad
\qquad
\qquad
\D_{\ell>0,\pm}^{ Q}\geq \ell+d-2  \ .
\ee
If we increase enough some of the gaps $\D_{\ell,\pm}^{ Q} $, we may find that the equations \eqref{BOOTSTRAP_EQNS} cannot be satisfied.
In this case we say that the corresponding CFT is excluded.
We can thus think of $\D_\phi$ and the gaps \eqref{5knobs} as the knobs which can turn to generate bounds. 
Equation \eqref{BOOTSTRAP_EQNS} can also be used to compute upper bounds on OPE coefficients.
In the following we show some interesting bounds obtained in this setup. All semi-definite problems have been solved using SDPB \cite{Simmons-Duffin:2015qma} with parameters as in \cite{Dymarsky:2017xzb}.

%% file: results/results.tex
\label{sec:results}

%
%

\subsection{Bounds on operator dimensions}

\subsubsection*{Scalar operators}

\begin{figure}[h!]
\begin{center}
\includegraphics[width=0.45\textwidth]{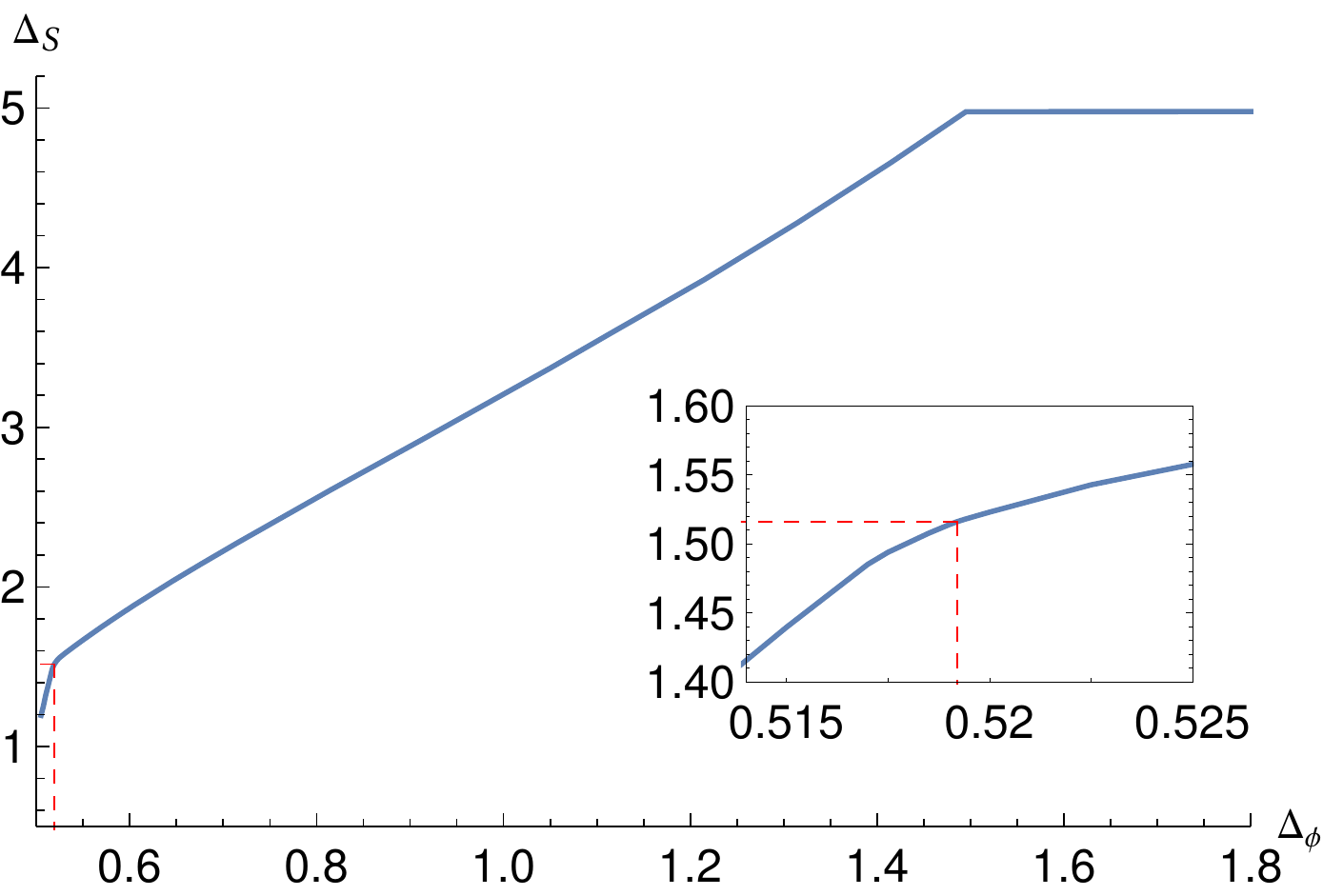}
\caption{Bound on the dimension of the first neutral parity-even scalar operator as a function of $\Delta_\phi$. For $\Delta_\phi\lesssim 1.5$ the bound is driven by the scalar bound. The plateau for larger values of $\Delta_\phi$ corresponds to the bound from the current bootstrap. The bound displays a kink corresponding to the $O(2)$ model. The bounds have been obtained at $\Lambda=13$.} 
\label{fig:DeltaPhi-DeltaS}
\end{center}
\end{figure}

We begin by studying the bound on the first parity-even scalar, neutral under the global $U(1)$ symmetry. We denote its dimension by $\Delta_S$. As shown in \Figref{fig:DeltaPhi-DeltaS} 
the bound coincides with the constraint one would get by bootstrapping only the scalar correlator $\langle\phi\phib\phi\phib\rangle$, until it reaches the maximal value allowed by the current bound \cite{Dymarsky:2017xzb}. At that point the bound becomes flat and independent on the external dimension. Although no new interesting features appears, this bound represents a validation of our methods and shows how the different crossing relations interplay.

\begin{figure}[t]
\begin{center}
\includegraphics[width=0.45\textwidth]{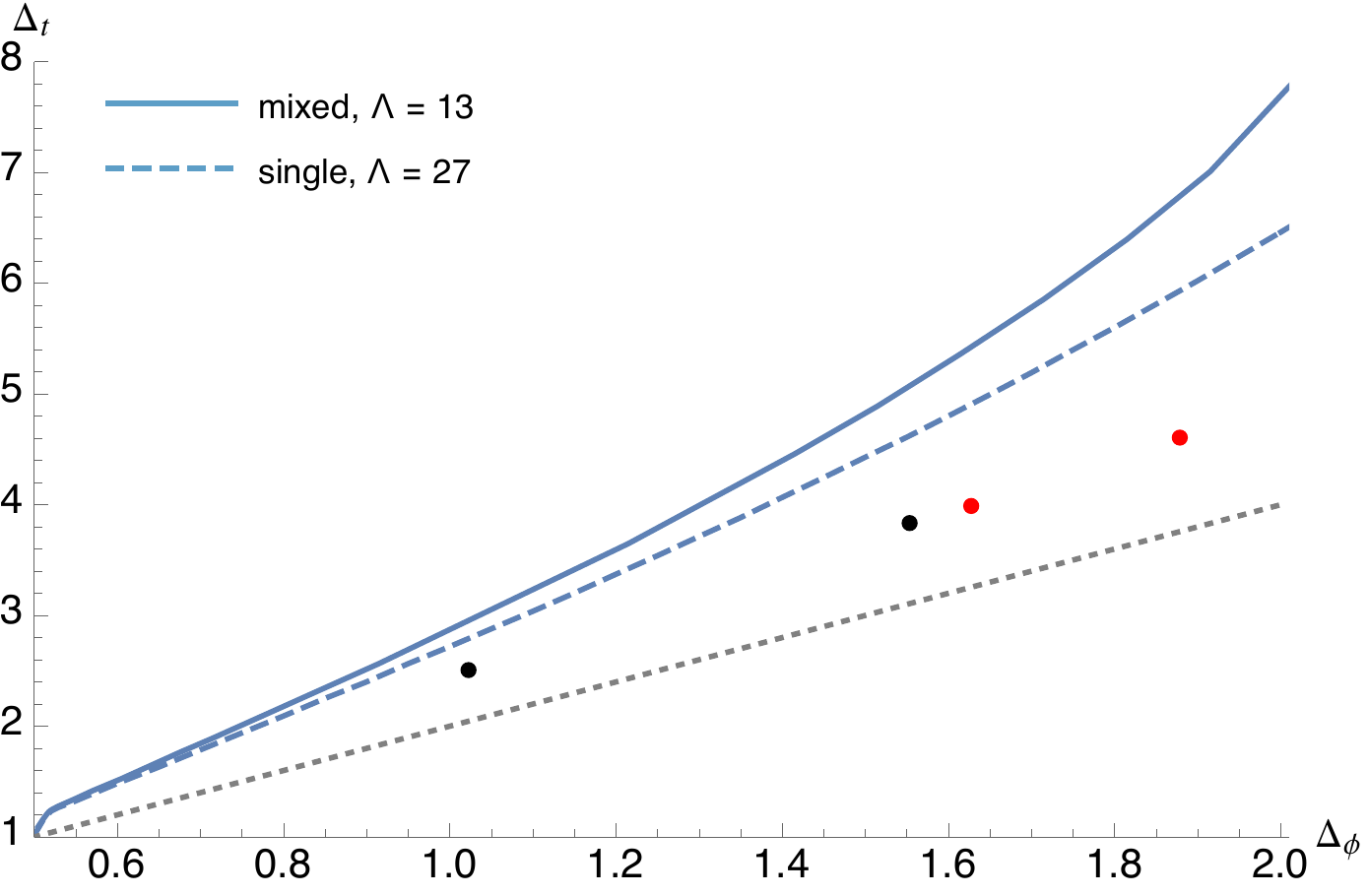}
\caption{Bound on the dimension of the first charge-2 parity-even scalar operator as a function of $\Delta_\phi$. The bound displays a kink in corresponding to the $O(2)$ model. The continuous line was obtained using the mixed system of bootstrap equations at $\Lambda=13$; the blue dashed line only uses the scalar correlator at $\Lambda=27$. The grey dotted line is the generalized free theory line. The red (black) dots correspond to the dimension of the monopoles with charge $q=1/2$ and $q=1$ in bosonic (fermionic) $\text{QED}_3$ computed in large-$N$ expansion  \cite{Dyer:2015zha, Dyer:2013fja}. Here we show respectively $N=10,12$ and  $N=4,6$.} 
\label{fig:DeltaPhi-Delta}
\end{center}
\end{figure}

Next we consider a bound on the dimension of the first parity-even charge-2 scalar $t$. We denote its dimension by $\Delta_t$. This operator only appears in the $\phi\times\phi$ OPE, thus it is natural to expect that the bound is completely driven by the scalar crossing equations only. We show this plot in \Figref{fig:DeltaPhi-Delta}, together with the same bound obtained using only scalar correlators at higher $\Lambda$.\footnote{Given the numerical complexity of our setup we could not push the mixed correlator analysis to the same value of $\Lambda$.} The bound only shows a kink in corresponding to the $O(2)$ model, nevertheless it allows to make contact with another set of CFTs that must obey our exclusion plots. \\
The infrared fixed point of fermionic and bosonic $\text{QED}_3$ contains a topological global $U(1)$ symmetry: we can then interpret $\phi$ as a scalar monopole operator with topological charge $q=1/2$ under this symmetry and identify $J_\mu$ with the associated current; then the bound on $\Delta_t$ is interpreted as the bound on the smaller monopole with charge $q=1$. Interestingly the dimension of these operators have been computed in a large $N$ expansion \cite{Dyer:2013fja}, where $N$ is the number of copies of fermions or bosons in the gauge theory. The predictions are shown in \Figref{fig:DeltaPhi-Delta}: although they do not saturate the bound, they seem to get close for small values of $N$ (where however the large-$N$ expansion is not accurate).

The only other sector containing scalars in the mixed system of $J$ and $\phi$ is the neutral parity-odd one. We do not show its bound here, since it coincides exactly with the one obtained in \cite{Dymarsky:2017xzb}, except that it has a termination point dictated by the maximal value of $\Delta_S$ allowed as a function of $\Delta_\phi$.

\subsubsection*{Operators with spin}
\begin{figure}[h!]
\begin{center}
\includegraphics[width=0.49\textwidth]{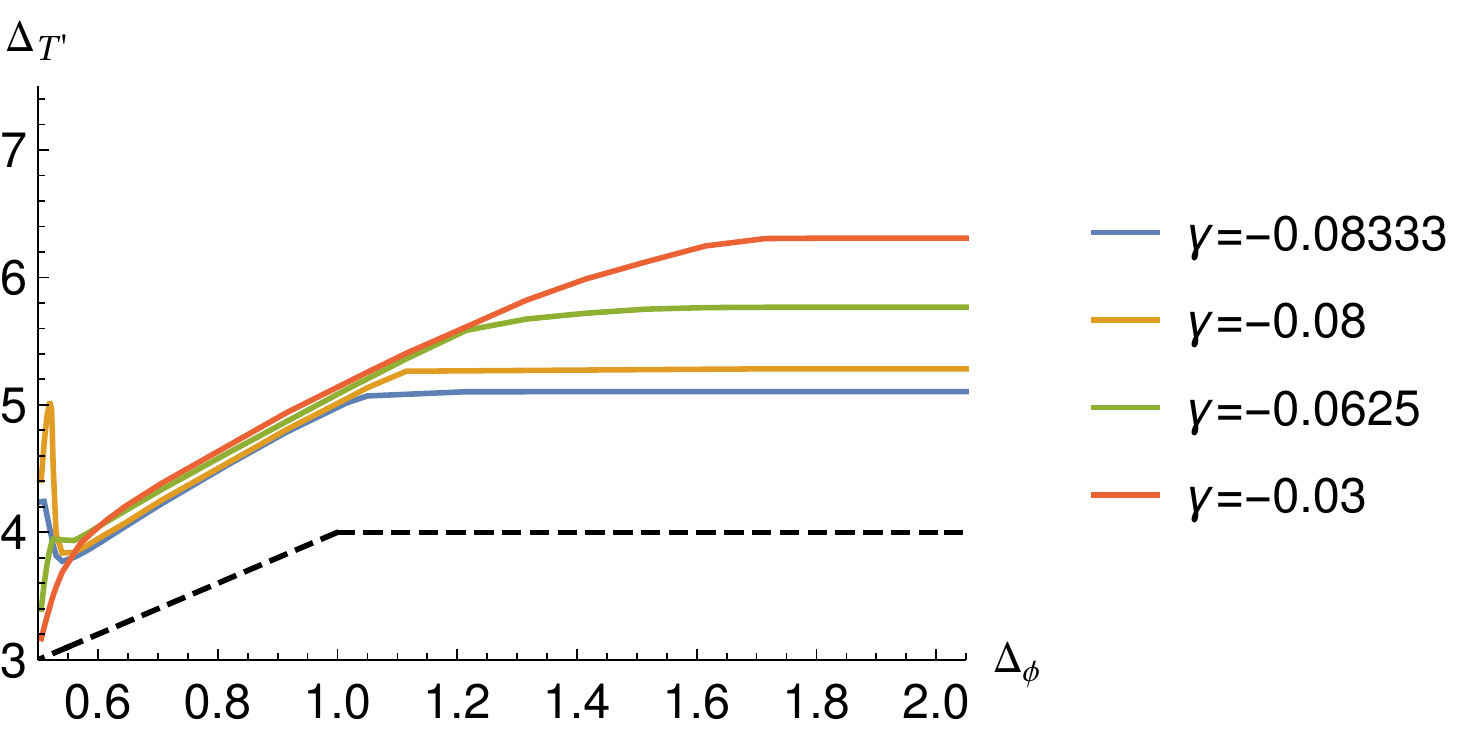}\hspace{10pt}\includegraphics[width=0.49\textwidth]{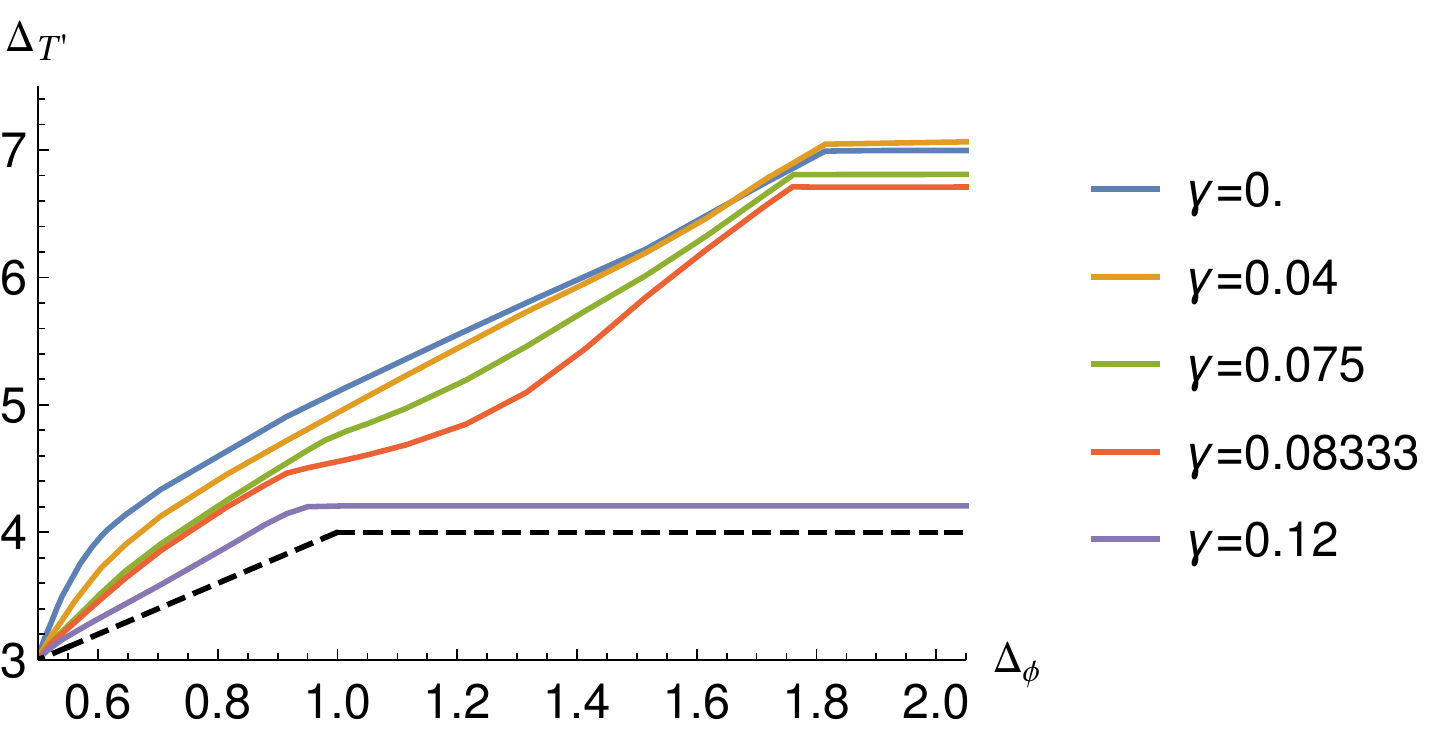}
\caption{Bound on the dimension of the first neutral parity-even spin-2 operator after the stress tensor as a function of $\Delta_\phi$. Different curves corresponds to different values of the parameter $\gamma$ defined in \eref{eq:lambdaofgamma}. The dashed curves correspond to the function $\text{min}(4,2\Delta_\phi+2)$. The bounds have been obtained at $\Lambda=13$.} 
\label{fig:DeltaPhi-DeltaTprime}
\end{center}
\end{figure}

We now move to bounds on operators with spin. We have already pointed out in \sref{sec:setup} the advantages of bounding the dimension of $T_{\mu\nu}'$ ---the first neutral parity-even spin-2 operator after the stress tensor--- to pinpoint the $O(2)$ model. Let us now review this statement by exploring the bound on $\Delta_{T'}$ on a broader range of parameters. In \Figref{fig:DeltaPhi-DeltaTprime} we show its upper bound as a function of the external dimension $\Delta_\phi$ and the parameter $\gamma\in [-1/12,1/12]$ defined in \eref{eq:lambdaofgamma}. 
%

We observe two interesting features at the extremes of the $\gamma$ interval. Close to the $\gamma\sim-1/12$ the upper bound on $\Delta_{T'}$ develops a sharp peak corresponding to a somewhat large gap in the spin-2 sector. We interpret this gap as the signal of the existence of a local CFT. Non-local theories or fake solutions of the crossing equations do not require a conserved spin-2 primary. As a result for those theories the bound on $\Delta_{T'}$ is actually a bound on the first spin-2 operator. In \sref{sec:intro} we exploited this peak to create an island in the $(\Delta_\phi,\gamma)$ plane and provide the first precise determination of $\gamma$ for the $O(2)$ model.\footnote{One can show the peak persist once additional information about the $O(2)$ model is injected, such as the presence of a single relevant neutral scalar.} 

In the proximity of the other extreme the bounds shows instead a clear kink around $\Delta_\phi\sim0.91$. Although it would be nice to interpret this feature as an existing CFT, we are not aware of obvious candidates. The value $\gamma\sim1/12$ suggests that the putative CFT should admit a description in terms of fermions: in that case the scalar $\phi$ could be a fermion bilinear with a large anomalous dimension. Another possibility is that $\phi$ is a monopole operator of a $\text{QED}_3$-like theory.\footnote{For instance large-$N$ computation and bootstrap studies suggest that the smallest monopole in fermionic $\text{QED}_3$ with 4 flavours has dimension $\Delta_M\sim1.034$. Bosonic $\text{QED}_3$ is instead believed not to have a fixed point at small $N$. In principle $U(1)$ Chern-Simons theories with non vanishing $\kappa$ must also obey our bounds. } Unfortunately the parameter $\gamma$ for the topological current has never been computed.  We leave the investigation of this kink for future studies. 
It is also plausible that this is a reminiscence of the trivial solution in which $\phi$ is a generalised free field and $J$ is a decoupled conserved current. In this case one has $\Delta_{T'}=\text{min}(4,2\Delta_\phi+2)$. This solution is shown by a black dashed curve in \Figref{fig:DeltaPhi-DeltaTprime}. We observe indeed that for values of $\gamma$ outside the conformal collider interval the bounds approaches this solution.

We also notice that all the curves in \Figref{fig:DeltaPhi-DeltaTprime} eventually reach a plateau. We checked that at this point the $\dphi$-independent constraints from $\langle JJJJ\rangle$ take over. 

\begin{figure}[t]
\begin{center}
\subfigure[]{\includegraphics[width=0.4\textwidth]{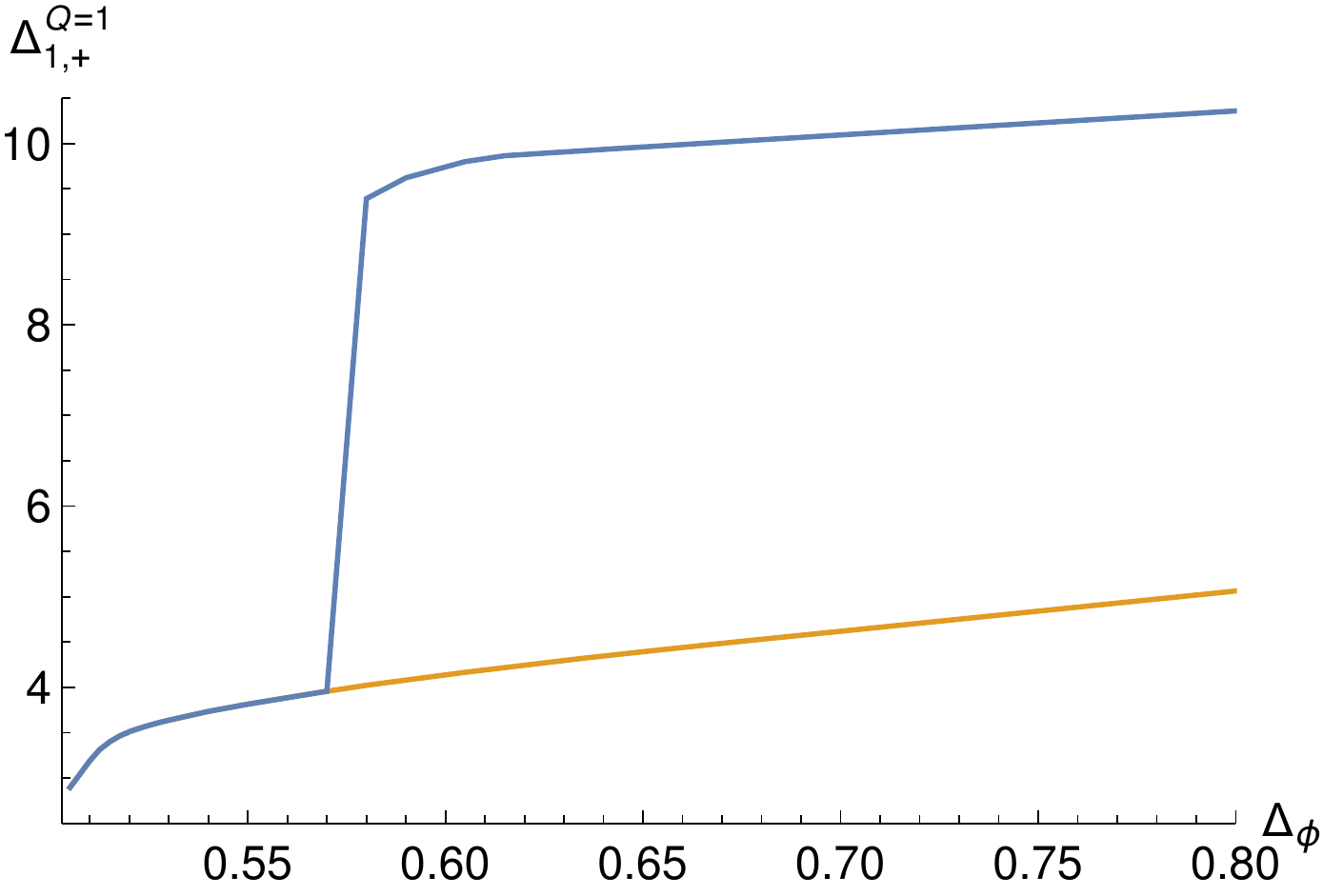}\label{fig:DeltaPhi-DeltaVcharged}}\qquad \ \ \ \ \
\subfigure[]{\includegraphics[width=0.4\textwidth]{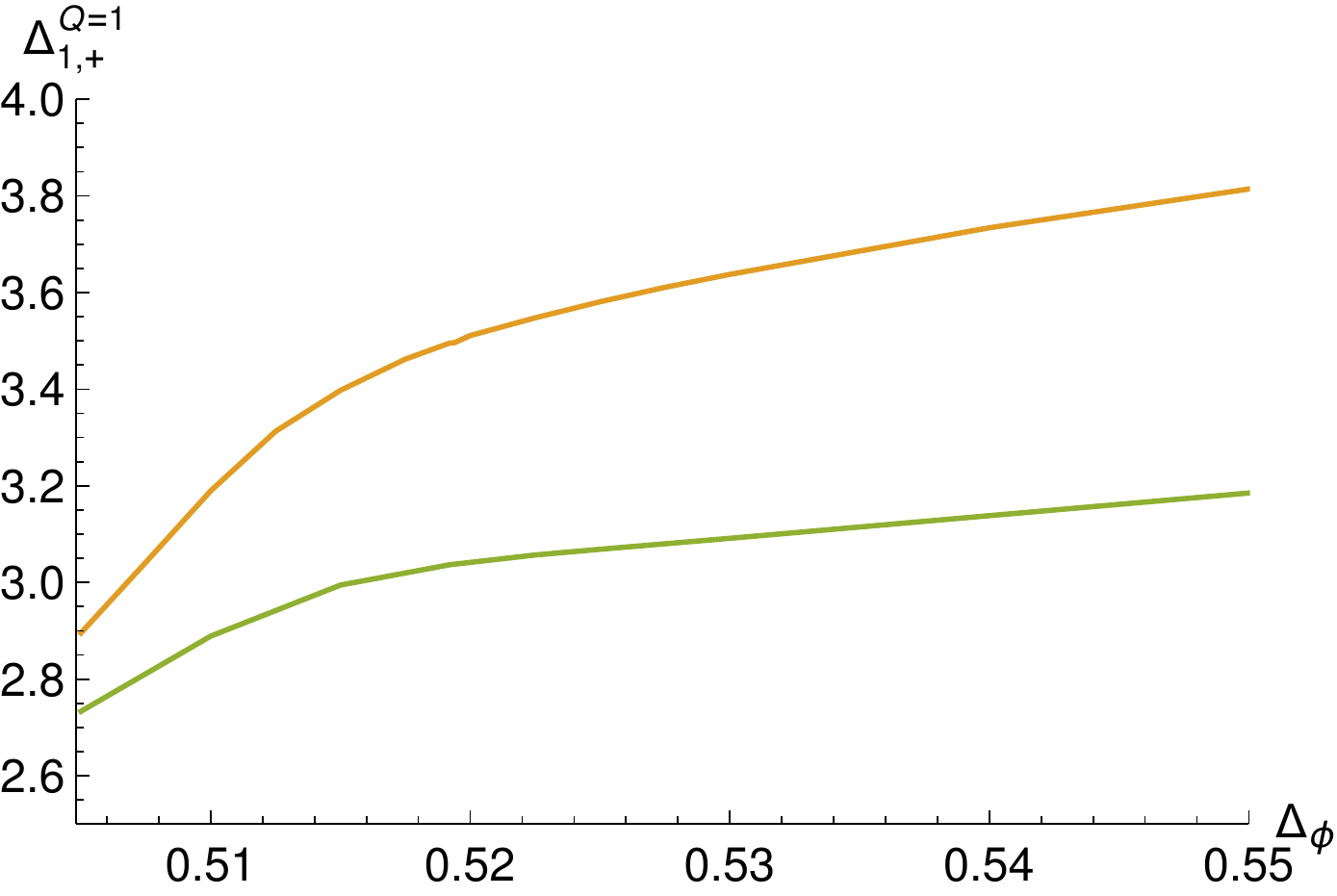}\label{fig:DeltaPhi-DeltaVcharged-assumptions}}
\caption{On the left: bound on the dimension of the first charge-1 parity-even spin-1 operator as a function of $\Delta_\phi$. Assuming a gap of 4 in the charge-1 parity-even spin-2 sector removes the fake primary effect. On the right: same bound with and without the assumption of no relevant neutral vectors  besides $J$. The bounds have been obtained at $\Lambda=13$.} 
\end{center}
\end{figure}

So far we have considered bounds on operators that were accessible both using the scalar correlator alone or the four current correlator alone. Let us now move to operators in the $Q=1$ sector, i.e. appearing in the OPE $J\times \phi$. We recall that, due to conservation, the only scalar allowed in the OPE is $\phi$ itself. Moving to spin-1 operators, we find parity-even and parity-odd charge-1 vectors. \\
In \Figref{fig:DeltaPhi-DeltaVcharged} we plot the bound on the dimension of the first parity-even vector charged under the global $U(1)$. With no additional assumption the bound displays the characteristic fake-primary effect discussed in \cite{Karateev:2019pvw} due to a contamination from charge-1 spin-2 operators at threshold. By imposing a gap in the latter sector, the fake-primary effect is removed. In addition to the jump, the bound also displays a kink approximatively in correspondence with the $O(2)$ model. However we observed that, injecting additional information, the height of the kink changes substantially.\footnote{It could be that the kink observed without further assumptions corresponds to another CFT.} For instance, by imposing the existence of a single relevant spin-1 neutral current, the bound drops as shown in \Figref{fig:DeltaPhi-DeltaVcharged-assumptions}. We checked that imposing extra assumptions does not substantially improve the bound further.\footnote{See also the upper bound presented in Table~\ref{fig:gamma_theta} found under additional constraints.}

\begin{figure}[h!]
\begin{center}
\subfigure[]{\includegraphics[width=0.43\textwidth]{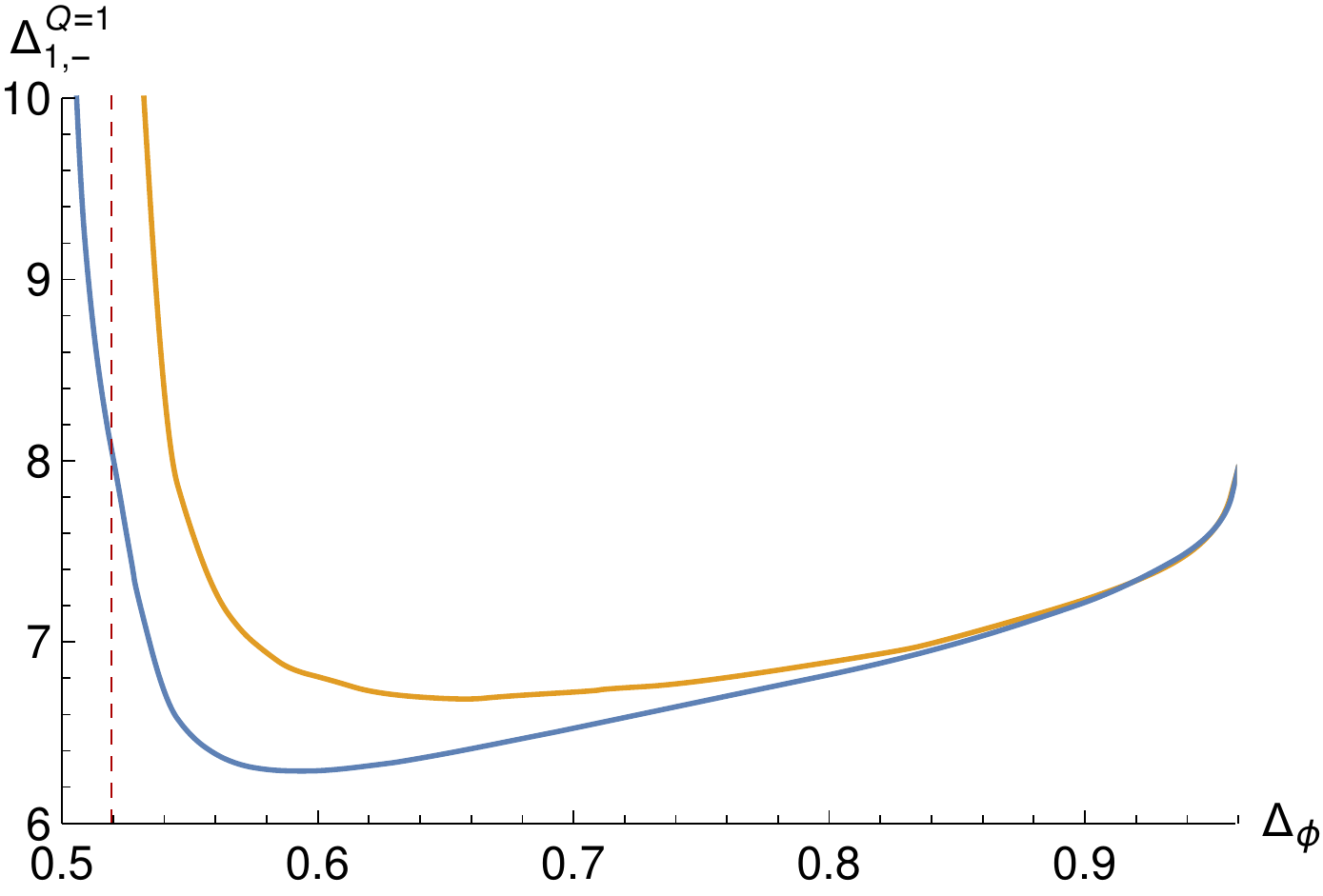}\label{fig:DeltaPhi-DeltaVchargedPodd}}\qquad \ \
\subfigure[]{\includegraphics[width=0.43\textwidth]{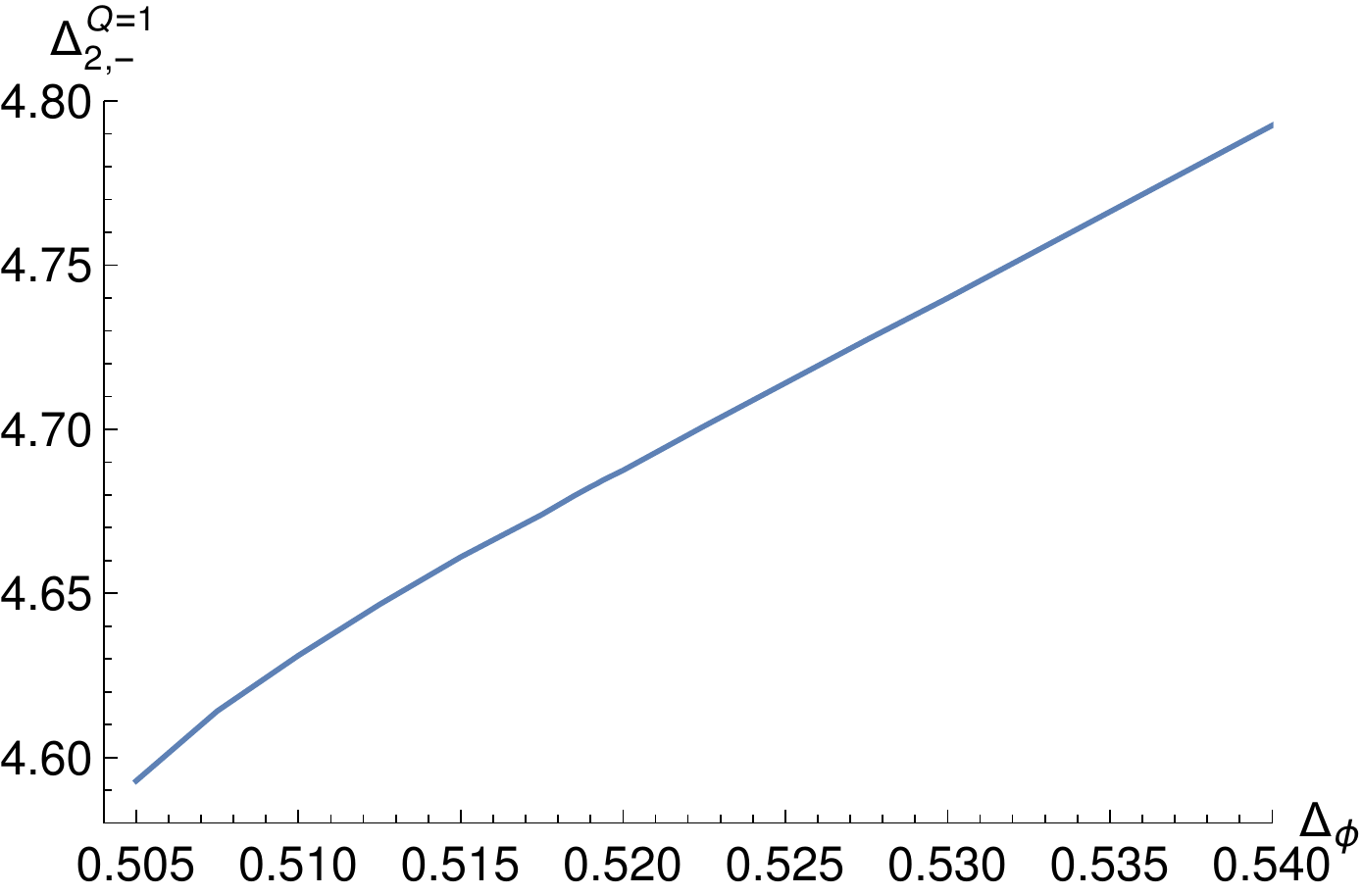}\label{fig:DeltaPhi-DeltaTchargedPodd}}
\caption{On the left: bound on the dimension of the first charge-1 parity-odd spin-1 operator as a function of $\Delta_\phi$. The dashed red line corresponds to the value of $\D_{\phi}$ of the $O(2)$ model. The two lines corresponds to different gaps in the charge-1 spin-2 parity-odd sector to remove the fake primary effect. On the right: bound on the dimension of the first charge-1 spin-2 parity-even operator as a function of $\Delta_\phi$. When the bound reaches 5 it jumps to a much higher value. The bounds have been obtained at $\Lambda=13$.
}
\end{center}
\end{figure}

We conclude the section by presenting in \Figref{fig:DeltaPhi-DeltaVchargedPodd} and \Figref{fig:DeltaPhi-DeltaTchargedPodd}  bounds on the first parity-odd charge-1 vector and tensor. Also in this case we must remove the fake primary effect by imposing a finite gap in the spin-2 and spin-3 charge-1 parity-odd sector. In the former case, however, the bound turns out to be heavily dependent on the gap. With a gap smaller that $4.1$ the bound seems to diverge when approaching the $O(2)$ model, however increasing the gap to 4.5, changes drastically the shape of the bound. We should point out that further investigations show that a gap in the spin-2 charged parity-odd sector of 4.5 is inconsistent with additional assumptions about the $O(2)$ model.

Notice also that the bound in \Figref{fig:DeltaPhi-DeltaVchargedPodd} stops existing as $\Delta_\phi$ approaches 0.96. A similar phenomenon was observed in  \cite{Li:2017ddj}. Also in our case, the point where the bound stops existing shifts as we increase $\Lambda$. We believe this is a numerical artifact which could be cured by imposing ad hoc gaps in the spectrum. 

\subsection{Bounds on central charges}
Among the OPE coefficients appearing in the conformal block decomposition of our correlation functions, the one associated to the exchange of the stress tensor plays a special role. It is indeed related to the central charge $C_T$ by conformal  Ward identities. As shown in \eref{eq:lambdaofgamma}, the precise relation involves the parameter $\gamma$, which due to the collider bounds is constrained in the interval $[-1/12,1/12]$. Using the bootstrap, we can then place a lower bound on the central charge as a function of the external dimension $\Delta_\phi$ and the parameter $\gamma$. 
\begin{figure}[t]
\centering
 \subfigure[]{\includegraphics[width=0.37\textwidth]{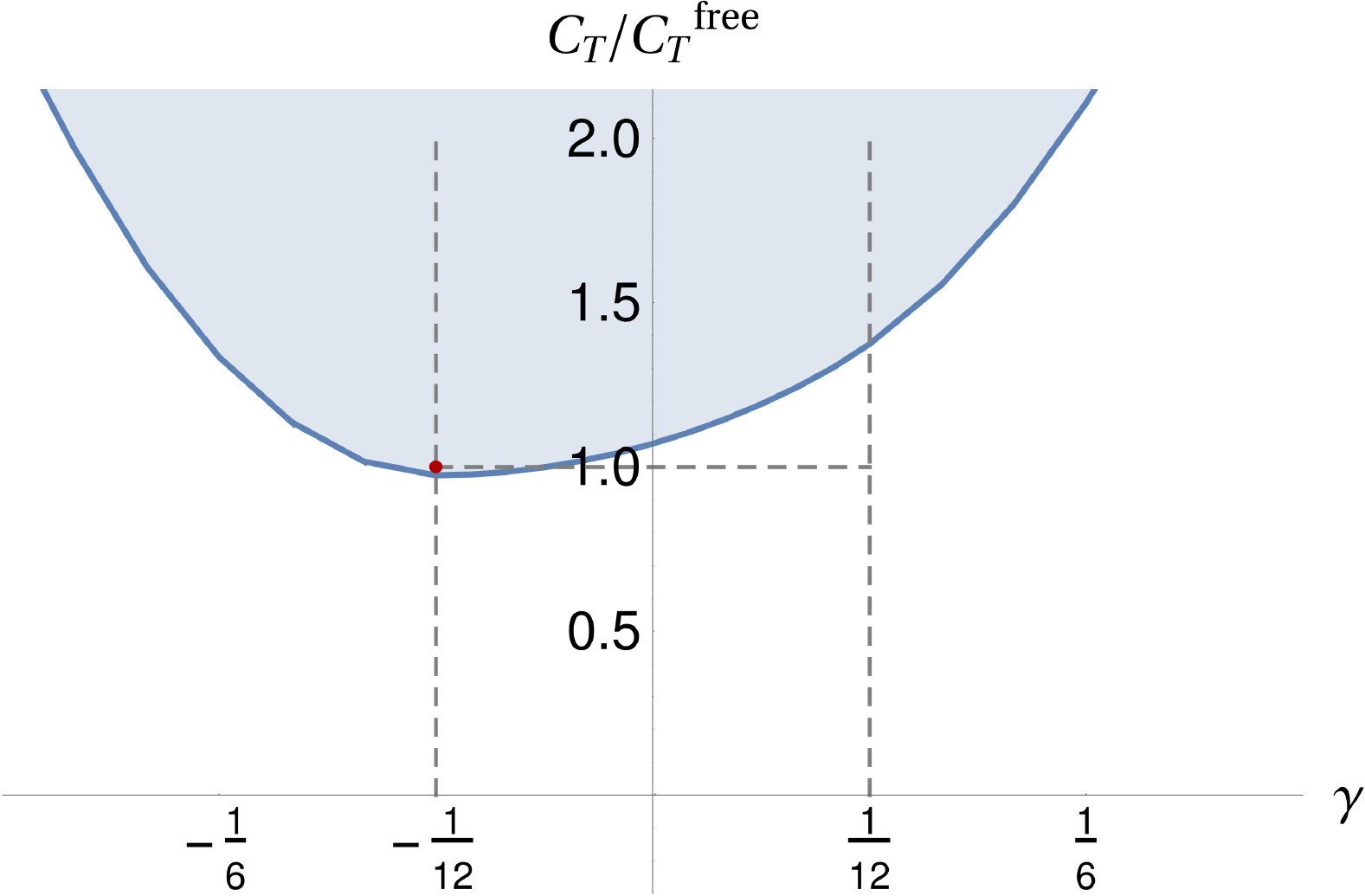}\label{fig:CTatphi0_505}}
 \ \ \ \ \ \ \
 \subfigure[]{\includegraphics[width=0.37\textwidth]{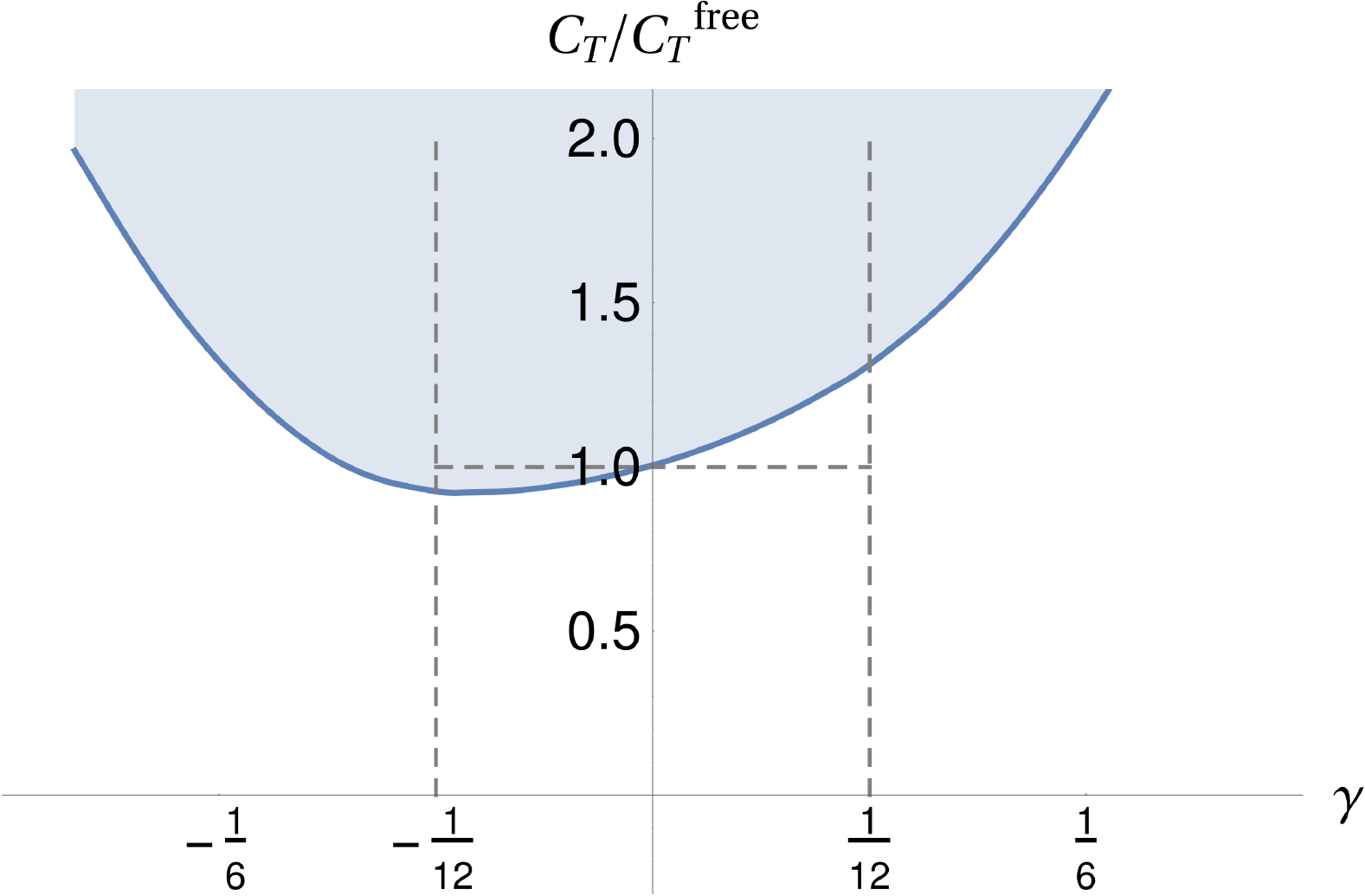}\label{fig:CTatphi0_5192}}
 \\
 \subfigure[]{\includegraphics[width=0.37\textwidth]{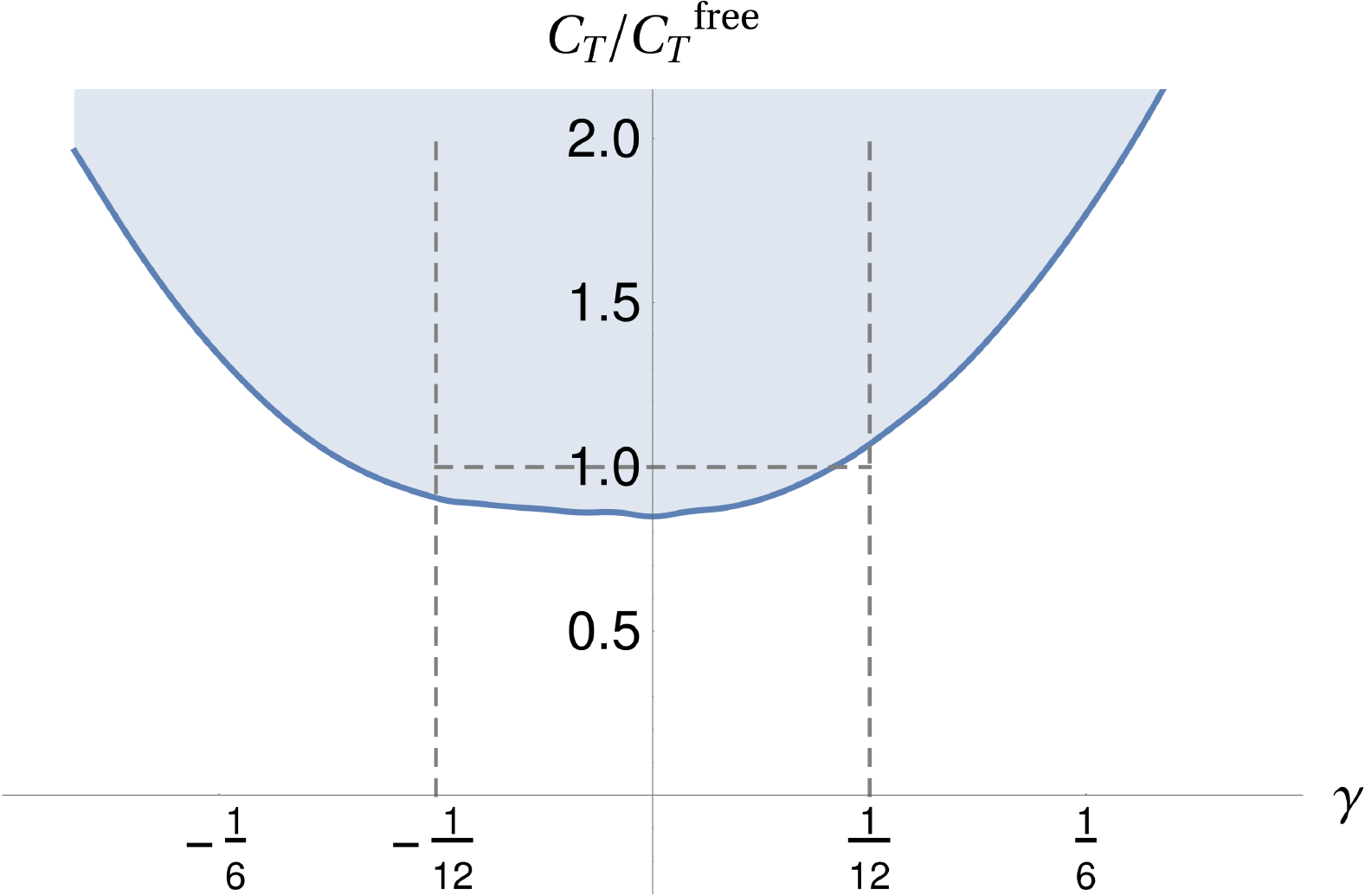}\label{fig:CTatphi0_605}}
  \ \ \ \ \ \ \
  \subfigure[]{\includegraphics[width=0.37\textwidth]{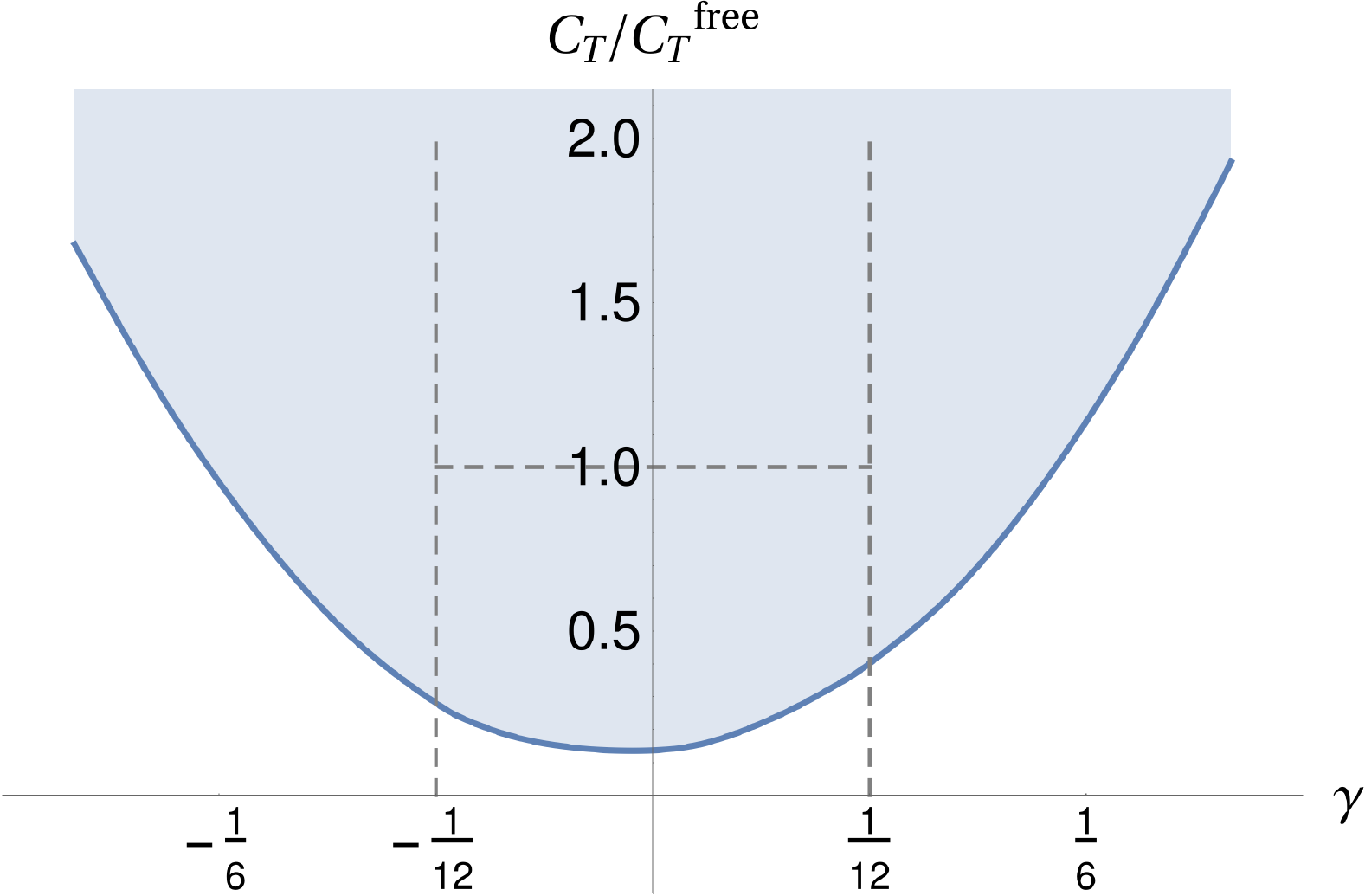}\label{fig:CTatphi1_05}}
 \caption{Lower bound on the central charge $C_T/C_{T^\textrm{free}}$ as a function of $\g$ for $\D_\phi=0.505, 0.5192,0.605,1.05$. The shaded region is allowed. The bounds have been obtained at $\Lambda=13$.}
\label{fig:CTvsgammaphi}
\end{figure}
Lower bounds on the central charge for theories with $O(2)$ symmetry have also been computed using the scalar correlator \cite{Kos:2013tga} or the current correlator only \cite{Dymarsky:2017xzb}. In the former case the bound decreases with $\Delta_\phi$ and is always weaker than the free theory value $C_T^{\text{free}}$, with a change of slope in proximity of the $O(2)$ model.\footnote{The discontinuity appeared to be slightly off in $\Delta_\phi$.} In the latter case the bound remains below the free theory value for the allowed range of $\gamma$ and rapidly increases 
outside.\footnote{Assuming a mild gap after the stress tensor make the collider bounds more manifest and the bounds rapidly grows for $|\gamma|>1/12$.}

In \Figref{fig:CTvsgammaphi} we show the results of our analysis. For values of $\Delta_\phi$ close to unitarity, the bound displays a minimum in correspondence with the free scalar theory values of $\gamma$ and $C_T$ (the red dot in the picture). Increasing $\Delta_\phi$ to $0.5192$ the bound gets slightly weaker to accommodate a smaller central charge, as expected in the $O(2)$ model (dashed line in \Figref{fig:CTatphi0_5192}). Interestingly one can already observe that $C_T\leq C_T^\text{free}$ requires a negative $\gamma$. Increasing further the value of $\Delta_\phi$ makes the bound relax to the bounds obtained using currents alone, \Figref{fig:CTatphi1_05}.

\begin{figure}[t]
\centering
\includegraphics[width=0.45\textwidth]{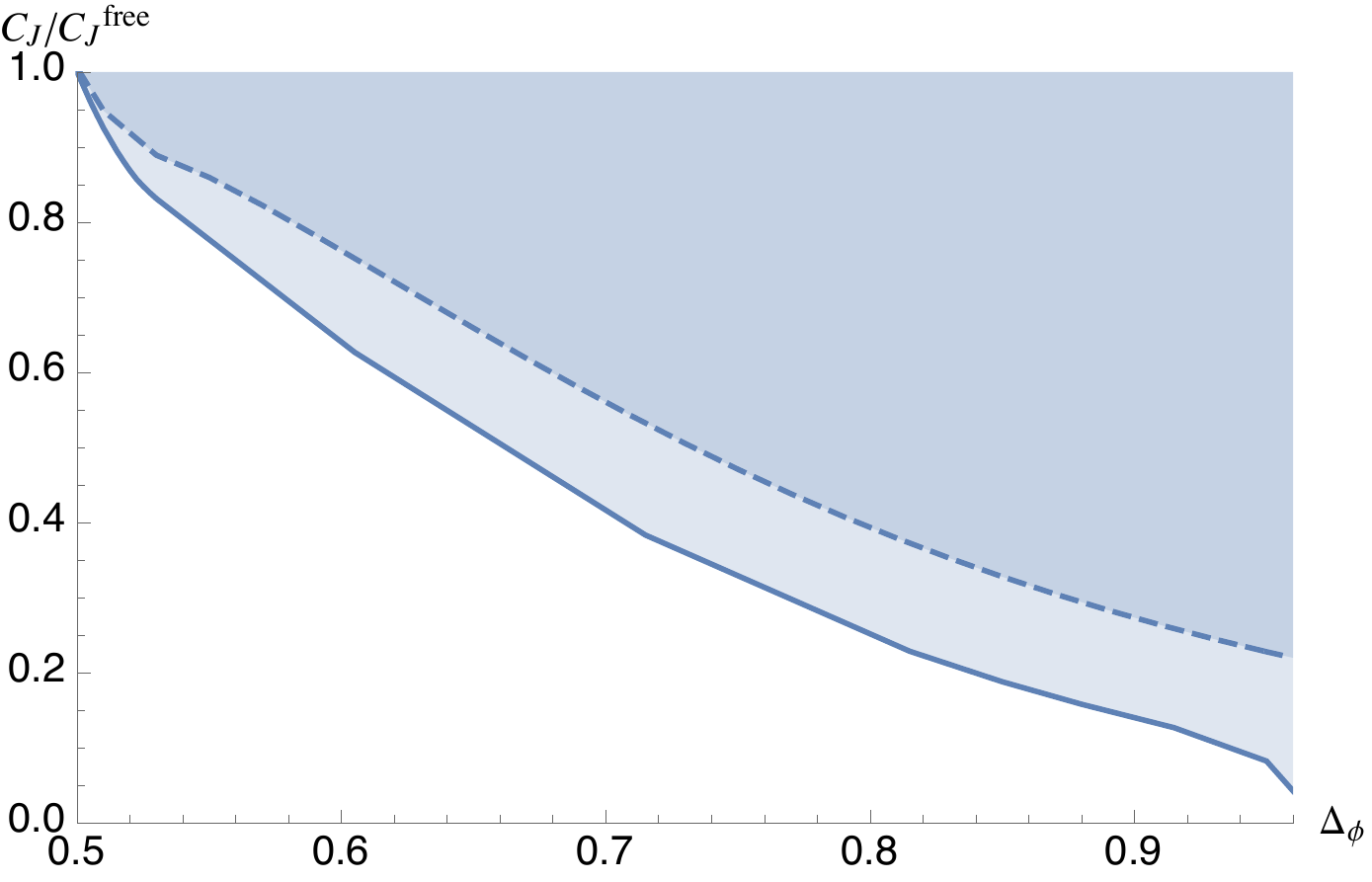}
 \caption{Lower bound on $C_J$ as a function of $\D_\phi$. The solid line is computed at $\Lambda=13$ using the mixed system of $J$-$\phi$ correlators, while the dashed line is computed at $\Lambda=27$ using only the scalar correlator. Both lines show a feature corresponding to the $O(2)$ model.\label{fig:CJmixedANDscalar}}
\end{figure}

We conclude this section by studying the constraints imposed on the central charge $C_J$. Due to Ward identities, this quantity is related to the inverse of the OPE coefficient $\lambda_{\phi\bar{\phi} J}$ according to \eref{eq:phiphiJ}. Notice that the latter OPE coefficient appears both in the scalar correlator and in the mixed channel, schematically:
\begin{align}
&\phi \times \bar{\phi} \sim \mathbb 1 + \lambda_{\phi\bar{\phi} J}\, J + \ldots \,,\nonumber\\ 
& J \times \phi \sim  \lambda_{\phi J \bar{\phi}} \, \phi + \ldots\,.
\end{align}
There is however an important difference between the above expressions: in the first line the block associated to the exchange of a conserved current is continuously connected to non conserved spin-1 blocks; on the contrary, in the mixed channel, the block associated to the exchange of $\bar{\phi}$ itself plays a special role and is, in fact, isolated. In practice this means that this block cannot be mimicked by an operator arbitrarily close in dimension and one can hope to place also an upper bound on $C_J$ under suitable assumptions. We will come back to this shortly.

Let us begin by exploring lower bounds on $C_J$. This is shown in \Figref{fig:CJmixedANDscalar} as a function of $\Delta_\phi$. By comparison we also show the bound obtained using the scalar correlator with higher numerical power. The shape is substantially similar and the only distinguishable feature is in correspondence with the $O(2)$ model, as already observed in \cite{Nakayama:2014yia}.\footnote{The fact that the bound decreases for large external dimensions is expected: if one interpret $J$ as a topological $U(1)$ current in $\text{QED}_3$ and $\phi$ as a monopole operator then one has the asymptotic behavior \cite{Dyer:2013fja,Giombi:2016fct}:
\begin{align}
\begin{split}
&\Delta_\phi \simeq 0.265 N_f - 0.0383+ O\left(\frac{1}{N_f}\right) \,,
\qquad  \frac{C_J}{C_J^\text{free}} \simeq \frac{3.2423}{N_f} \left(1 - \frac{0.1423}{N_f} + O\left(\frac{1}{N_f^2}\right)\right)\, , 
\end{split}
\end{align}
where $N_f$ is the number of fermions in the UV theory. Unfortunately our bound is still far from these values.}

\begin{figure}[h]
\centering
\includegraphics[width=0.5\textwidth]{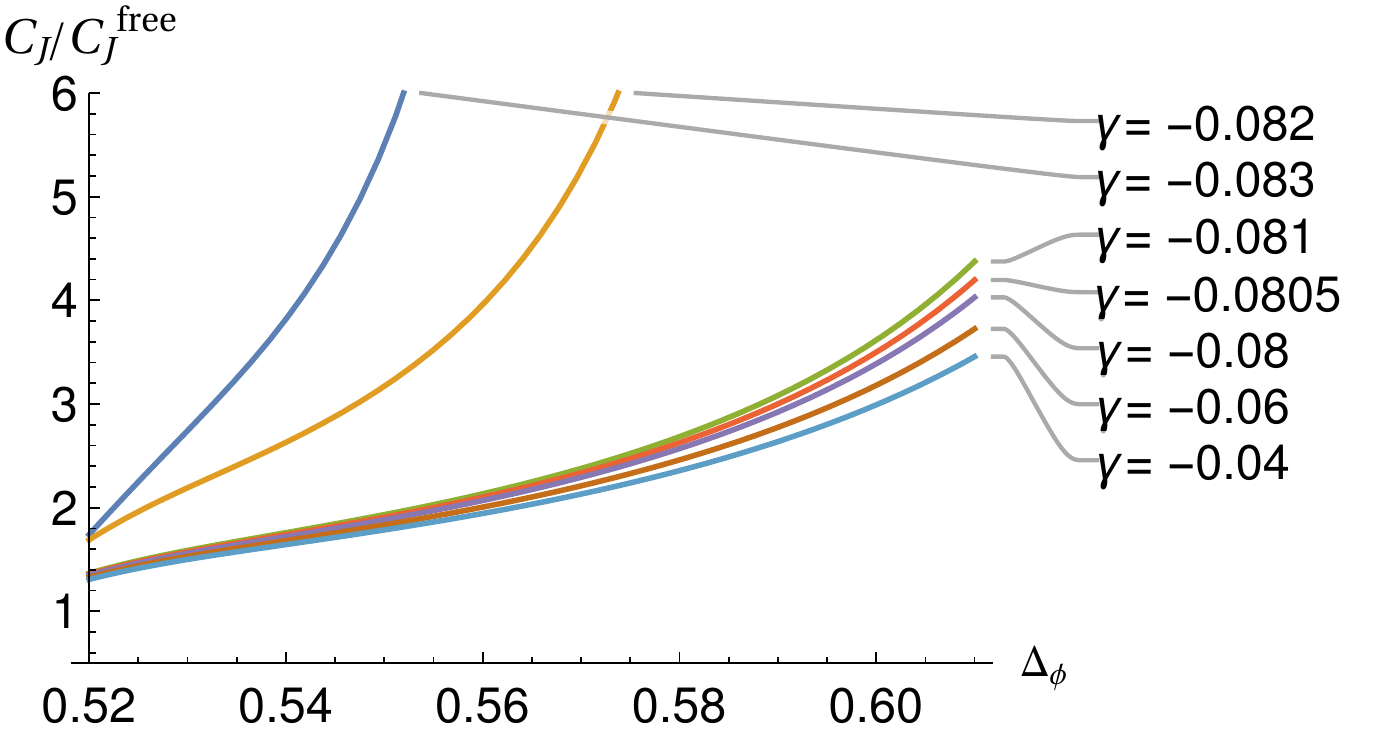}
 \caption{Upper bound on $C_J$ normalized to the central charge of a free complex boson as a function  of $\D_\phi$   at fixed values of the parameter $\gamma$ assuming $C_T\leq 0.95 C_T^\text{free}$. The bounds have been obtained at $\Lambda=13$. \label{fig:CJmaxWithCTsmaller1}}
\end{figure}

As mentioned earlier, in a pure scalar bootstrap setup, extracting bounds on $C_J$ would require to isolate the current conformal block by assuming a gap on the next spin-1 operator.
In the present framework, however, the isolated nature of the $\phi$-conformal block in the mixed channel can be exploited to compute such a bound. Notice that a finite value of $C_J$ implies that the scalar is indeed charged under the external current $J$. Despite the fact that we would like to focus on those cases, it is perfectly legitimate to have a correlation function of a conserved current associated to a $U(1)$ under which the complex scalar $\phi$ is neutral.\footnote{The simplest case is a tensor product of a generalized free vector field and a generalized free scalar theory. Alternatively one can consider for instance the $O(4)$-vector model identify $\phi\equiv \phi_1+i\phi_2$ and $J$ as the generator of rotations in the 3-4 direction.} Thus, we do not expect an upper bound to exist without further assumptions. \\
In our investigations we found that assuming a small value of the central charge $C_T$ forces a finite value of $C_J$. This is shown in \Figref{fig:CJmaxWithCTsmaller1}, where we computed an upper bound on $C_J$ as a function of $\Delta_\phi$ for several values of $\gamma$. This is the first numerical evidence that the existence of a local stress tensor, together with a set of selection rules, implies the presence of conserved current in the scalar OPE.

\begin{figure}[t]
\centering
\includegraphics[width=0.5\textwidth]{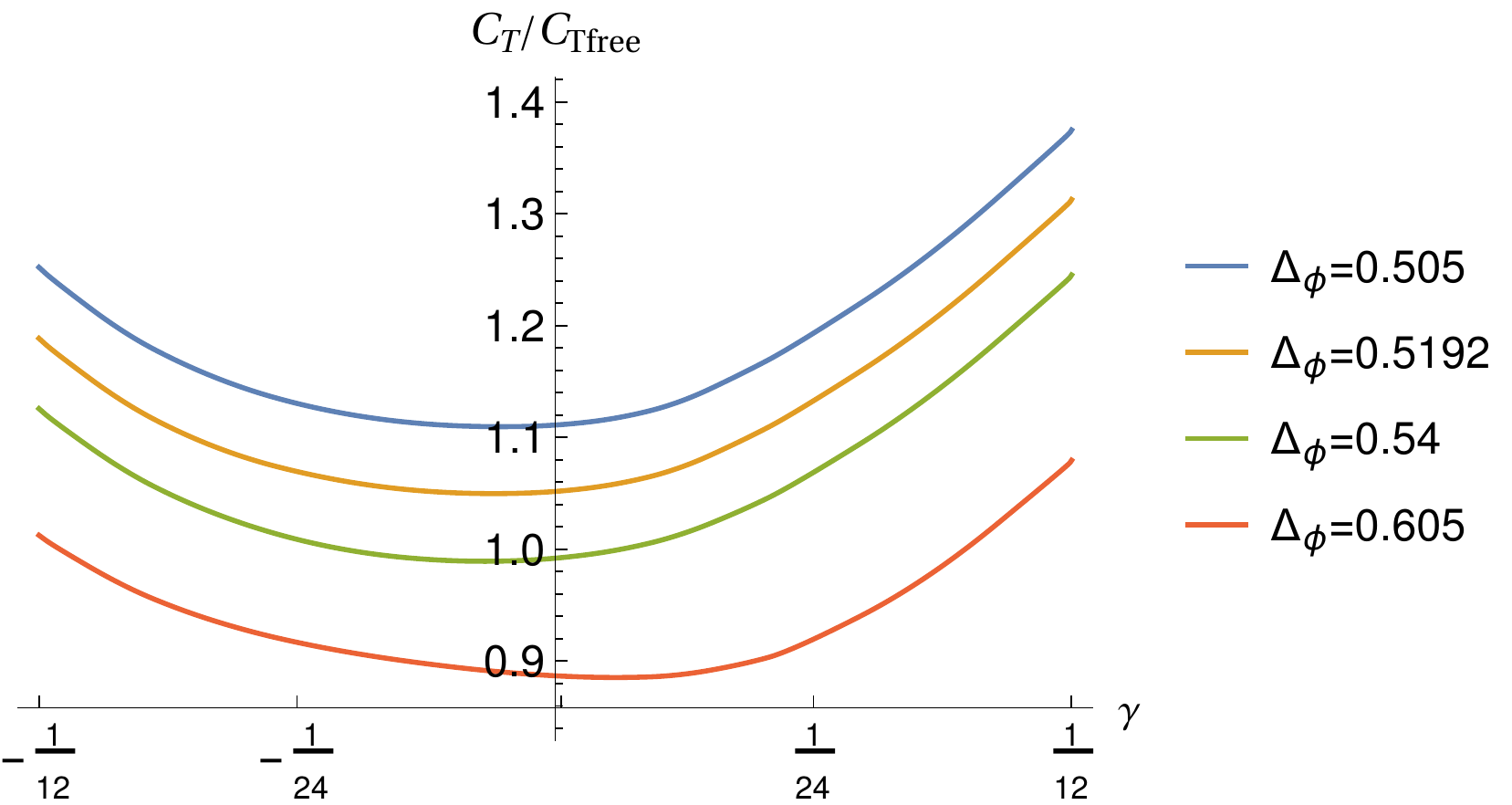}
 \caption{Lower bound on the central charge normalized to the central charge of a free complex boson as a function of the parameter $\gamma$  at fixed values of $\D_\phi$ assuming that the scalar $\phi$ is neutral under the symmetry generated by the current $J$. The bounds have been obtained at $\Lambda=13$. \label{fig:CTphiNeutral}}
\end{figure}

We could go one step further and ask for what values of the central charge such a bound exists. This question can also be recast as a lower bound on $C_T$, assuming $C_J\rightarrow\infty$. 
\Figref{fig:CTphiNeutral} answer precisely this question. We observe that if the central charge is below the bound for given $\gamma$ and $\Delta_\phi$, the scalar must be charged under the external current $J$. Intuitively this result can be restated as:  in order to have an extended symmetry, one needs enough degrees of freedom. While this statement is obvious in free theories, it interesting to show that it can be  extended to interacting CFTs, although at present the bound on $C_J$ only exists for small $C_T$ and $\Delta_\phi$ close to the unitarity bound.\footnote{Indeed for larger values  $\Delta_\phi$, the bounds of \Figref{fig:CTphiNeutral} are on top of the bounds with no assumptions on $C_J$ shown in \Figref{fig:CTvsgammaphi}.}

%% file: Conclusions/conclusions.tex
\label{sec:conclusions}

In this work we studied the impact of considering correlation functions involving a spin-1 conserved current and a scalar operator charged under the associated global $U(1)$. Using numerical bootstrap techniques we have explored the space of constraints. We found that only the $O(2)$ model seems to stand out, appearing as kinks in several operators bounds and as a sharp peak in the bound on the first spin-2 operator after the stress tensor. By using these features we manage to constrain several observables that are not accessible by the scalar bootstrap, such as dimensions of certain operators and three-point functions coefficients involving two currents and a third operator. In particular, we determined with some accuracy the OPE coefficient $\lambda_{JJS}$, where $S$ is the unique relevant neutral deformation in the $O(2)$ model. This parameter controls the leading correction to the conductivity $\sigma$ in the $O(2)$ model at finite temperature and high frequencies. We expressed the dependence of $\sigma$ in terms of the CFT data and compared it to the QMC simulations of \cite{Katz:2014rla} to extract the expectation value of the operator $S$ at finite temperature. Our determination agrees with the direct QMC determination and is more accurate.  

We also accurately determined two more quantities, $\gamma$ and the central charge $C_T$, appearing in the next-to-leading correction of $\sigma$. Their knowledge allowed to extract the thermal expectation value of the stress tensor from the fit of the conductivity. In order to improve the sensitivity to sub-leading corrections, the precision of the QMC simulation should be increased, with particular attention to systematic errors. \\
Recently a QMC study of the Gross-Neveu model was performed in \cite{liu2019designer}, together with a fit of the conductivity. It would be interesting to repeat our bootstrap analysis and extract the relevant CFT-data to compare with those results.\footnote{We thank William Witczak-Krempa for bringing this work to our attention.} In order to focus on the Gross-Neveu model one should presumably consider external fermions.

Part of the motivation of this work was to establish whether it is worthwhile to include conserved currents in the bootstrap. For certain questions we observed that the presence of the spin-1 current was not determinant. On the other hand, when scanning over parameters such as $\gamma$ and $\theta$, we observed interesting interplay of the crossing equations.
To make a conclusive statement one should consider an even more complicated system and include the neutral scalar $S$ as an external operator. That analysis, in conjunction with new algorithms to cheaply scan over the OPE parameter space \cite{CLLPSDSV} could represent the correct approach to deal with CFTs with global symmetries.

%% file: appendices/appendix_condictivity.tex

\section{Conductivity in terms of CFT data}
\label{app:conductivity}

We begin by defining the two point function of the $U(1)$ current:
\be
\label{eq:jjtwopoint}
\< J(x_1) J(x_2) \> \equiv \< J(x_1,z_1) J(x_2,z_2) \>= \frac{C_J}{(4\pi)^2}\frac{1}{x_{12}^4}\left[z_1\cdot z_2 -2 \frac{(x_{12}\cdot z_1)(x_{12}\cdot z_2)}{|x_{12}|^2} \right].
\ee
In the above expression all polarizations $z_i$ and ccordinates are three dimensional and, as usual $x_{12}^\mu = (x_1-x_2)^\mu$.
With this normalization the current $J^\mu$ satisfies the global symmetry Ward identity and in the case of a free scalar field $C_J^\text{free}=2$.

We are interested in the leading terms in the OPE expansion of $J^{\mu} \times J^\nu$, with a particular interest in the contribution of the smallest dimension scalar operator, let us call it $S$, which is normalized according to 
\be
\< S(x)S(0)\> = \frac{A}{|x|^{2\Delta_S}} \,.
\ee
 This can be obtained by matching with the leading term in the  $x_1\rightarrow x_2$ expansion of the three-point function \cite{Dymarsky:2017xzb}:
\be
\<J(x_1) J(x_2) S(x_3)\> = \frac{C_J\sqrt{A}}{(4\pi)^2}  \, \hat\lambda_{JJS}\, \frac{(\Delta_S-2) \hat H_{12}+\Delta_S \hat V_{1,23} \hat V_{2,31}}{|x_{12}|^{4-\Delta_S}|x_{13}|^{\Delta_S}|x_{23}|^{\Delta_S}} \, ,
\ee
where as usual we are working in three dimensions. The prefactor $C_J\sqrt{A}/(4\pi)^2 $ has been added so that the three-point function coefficient $\hat\lambda_{JJS}$ is defined for unit-normalized current and scalar. 
The three point function coefficient is related to the one used in the main text and appendix \ref{app:threepoints} simply by
\be
\hat\lambda_{JJS} = - \Delta_S \lambda_{JJS}\,.
\ee
The structures $\hat H$ and $\hat V$ ---which correspond to the physical space projection of the $H$ and $V$ of \eqref{eq:HV}--- are written as follows,
\bea
&\hat H_{12} = z_1\cdot z_2 -2 \frac{(x_{12}\cdot z_1)(x_{12}\cdot z_2)}{|x_{12}|^2} \, , \nonumber\\
&\hat V_{1,23} = \frac{(x_{12}\cdot z_1) |x_{13}|}{|x_{12}||x_{23}|} - \frac{(x_{13}\cdot z_1) |x_{12}|}{|x_{13}||x_{23}|}\, ,\\
&\hat V_{2,31} = \frac{(x_{23}\cdot z_2) |x_{12}|}{|x_{23}||x_{13}|} + \frac{(x_{12}\cdot z_2) |x_{23}|}{|x_{12}||x_{13}|}\, .\nonumber
\eea
By matching with the OPE expansion we get:
\bea\label{eq:JJSOPE}
\begin{split}
J(x_1) \times J(x_2) &\sim  \frac{C_J}{(4\pi)^2} \bigg[\hat H_{12} \mathbb1 +  \frac{\hat \lambda_{JJS}}{|x_{12}|^{4-\Delta_S}}  \times \\
&\quad  \times \left( (\Delta_S-2)(z_1\cdot z_2) -  (\Delta_S-4) \frac{(x_{12}\cdot z_1)(x_{12}\cdot z_2)}{|x_{12}|^2} \right)\frac{S(x_2)}{\sqrt A}+\ldots\bigg] \, .\\
\end{split}
\eea

Next, let us consider the contribution to the OPE of the stress tensor $T_{\mu\nu}$. Following \cite{Dymarsky:2017xzb}, we can write this term as
\be\label{eq:JJTOPE}
J(x_1) \times J(x_2)  \sim  \frac{C_J}{(4\pi)^2}  \frac{3}{32\pi} \frac{(4\pi)^2}{C_T}  \left(t_1(x_{12},z_1,z_2)^{\alpha\beta}+12 \,\gamma \, t_2(x_{12},z_1,z_2)^{\alpha\beta}\right) T_{\alpha\beta}(x_2)\,,
\ee
where 
\bea
\begin{split}
t_1(x_{12},z_1,z_2)^{\alpha'\beta'}(x) &=& z_1^\mu z_2^\nu P^{\alpha'\beta'}_{\alpha\beta}(6 \hat{x}_{(\mu}\delta_{\nu)}^{\alpha}\hat{x}^\beta
+ 2\delta^\alpha_\mu\delta^\beta_\nu+3\hat{x}_\mu\hat{x}_\nu\hat{x}^\alpha\hat{x}^\beta-5\delta_{\mu\nu}\hat{x}^\alpha\hat{x}^\beta)\,,\\
t_2(x_{12},z_1,z_2)^{\alpha'\beta'}(x) &=& z_1^\mu z_2^\nu P^{\alpha'\beta'}_{\alpha\beta} (2\hat{x}_{(\mu}\delta_{\nu)}^{\alpha}\hat{x}^\beta
- 2\delta^\alpha_\mu\delta^\beta_\nu-3\hat{x}_\mu\hat{x}_\nu\hat{x}^\alpha\hat{x}^\beta-3 \delta_{\mu\nu}\hat{x}^\alpha\hat{x}^\beta )\,,
\end{split}
\eea
and we explicitly introduced the projector on traceless symmetric indices
\be
P^{\alpha'\beta'}_{\alpha\beta} = \frac12 \left(\delta_\alpha^{\alpha'}\delta_\beta^{\beta'} + \delta_\beta^{\alpha'}\delta_\alpha^{\beta'} -\frac23 \eta_{\alpha\beta}\eta^{\alpha'\beta'} \right)\,.
\ee
In order to compute the conductivity we need to take the Fourier transform at point $x_{1,2}$ of the expressions \eref{eq:JJSOPE} and \eref{eq:JJTOPE}. Using standard formulas, see for instance appendix B of \cite{Benvenuti:2019ujm}, we obtain:
\bea
&&\!\!\!\!\!\!\!\int d^3x_1 d^3x_2 e^{i p_1\cdot x_1} e^{i p_2\cdot x_2} J(x_1,z_1) \times J(x_2,z_2)  \sim \nonumber \\ 
&& \sim\left(z_1^\mu \widetilde I_{\mu\nu}(p_1)  z_2^\nu\right) \left(- |p_1| \frac{\pi^3 C_J }4 \delta^3(p_1+ p_2) - \frac{\lambda_{JJS}}{4\pi |p_1|^{\Delta_S-1}}\frac{\Gamma(\Delta_S+1)\sin\left(\frac{\pi\Delta_S}2\right)}{2-\Delta_S} \frac{\widetilde S(p_2)}{\sqrt{A}} \right)+\nonumber\\
&&\qquad +\frac{C_J}{C_T}  \frac{1}{|p|}  \left(\widetilde t_1(p_1,z_1,z_2)_{\alpha\beta}+12\gamma \widetilde t_2(p_1,z_1,z_2)_{\alpha\beta}\right)\widetilde T^{\alpha\beta}(p_1+p_2) +\ldots \,,
\eea
where 
\bea
 \widetilde t_1(p_1,z_1,z_2)_{\alpha\beta} &=& 3 z_1^\mu z_2^\nu \left(\eta _{\alpha  \nu } \eta _{\beta  \mu
   }+\eta _{\alpha  \mu } \eta _{\beta
    \nu }-\eta _{\alpha  \beta } \eta
   _{\mu  \nu }+\hat{p}_{\mu } \hat{p}_{\nu } \eta
   _{\alpha  \beta }-\hat{p}_{\beta } \hat{p}_{\nu
   } \eta _{\alpha  \mu }\right.\nonumber\\
    &&\left.-\hat{p}_{\alpha }
   \hat{p}_{\nu } \eta _{\beta  \mu
   }-\hat{p}_{\beta } \hat{p}_{\mu } \eta _{\alpha
    \nu }-\hat{p}_{\alpha } \hat{p}_{\mu } \eta
   _{\beta  \nu }+\hat{p}_{\alpha } \hat{p}_{\beta
   } \eta _{\mu  \nu }+\hat{p}_{\alpha }
   \hat{p}_{\beta } \hat{p}_{\mu } \hat{p}_{\nu }\right)     \,,\\
 \widetilde t_2(p,z_1,z_2)_{\alpha\beta} &=& \left(z_1^\mu \widetilde I_{\mu\nu}(p)  z_2^\nu \right)  \left(\eta_{\alpha\beta} - 3\hat{p}_{\alpha} \hat{p}_\beta\right)\  \, ,\nonumber
  \eea
  with $ \widetilde I_{\mu\nu}(p) = \left(\eta_{\mu\nu} - \hat{p}_{\mu} \hat{p}_\nu\right)$ and $\hat{p}_\mu =p_\mu /|p|$.
By choosing the polarizations  along the $2$nd direction $z_i=(0,1,0)$, the momenta
\be
p_1 = w\,,\qquad p_2 = -w + p \,,\qquad w=(\Omega,0,0)\,,
\ee
and taking the expectation value of the previous expression at finite temperature, we obtain:
\bea
\!\!\!\!\!\!\!\!\!\!\!\!\!\!\!\!\!\!\!\! &&\< J_2(-w) J_2(w+p) \>_T \sim (2\pi^3)\delta^3(p) |\Omega| \times \nonumber\\
\!\!\!\!\!\!\!\!\!\!\!\!\!\!\!\!\!\!\!\!&&\qquad \times\left(- \frac{C_J}{32} - \frac{C_J \lambda_{JJS}}{4\pi} \frac{\Gamma(\Delta_S+1)\sin\left(\frac{\pi\Delta_S}2\right)}{2-\Delta_S} \Upsilon^{-1} \left(\frac{T}{|\Omega|}\right)^{\Delta_S}- 72\frac{C_J \gamma }{C_T} \frac{\Omega^2}{|\Omega|^2}H_{xx} \left(\frac{T}{|\Omega|}\right)^3 \ldots \right)\,,
\eea
where we defined
\be
\< S(0)\>_T  = B T^{\Delta_S}\, , \qquad  \Upsilon = \frac{\sqrt{A}}B\,, \qquad \< T_{22}(0)\>_T =\<T_{33}(0)\>_T =-\frac12 \<T_{11}(0)\>_T =  H_{xx} T^{3}\,.
\ee
Finally, given the relation \cite{Katz:2014rla},
\be
\frac{\sigma(iw)}{\sigma_Q} (2\pi)^3 \delta^3(p) = -\frac1{|w|} \< J_2(-w) J_2(w+p) \>_T \,,
\ee
we obtain equation \eref{eq:conductivity_us} shown in the main text.

%% file: appendices/appendices.tex

\section{Three point functions}
\label{app:threepoints}
\subsection{Scalar-scalar OPE}
We start by normalizing the OPE of two scalars $\Ocal_1\times \Ocal_2$ such that
\be
\Ocal(x,z)\Ocal_1(0) \sim 
\l_{12\Ocal}\, \frac{(-x\cdot z)^\ell}
{(x^2)^\frac{\D+\D_{12}+\ell}{2}} 
\Ocal_2(0) \, ,
\label{OPE}
\ee
where $z_{\m}$ is a null polarization vector.
The operator $\Ocal$ is a parity even operator of spin $\ell$ and $\l$  is the  OPE coefficient. 
 The symbol $\sim$ means that we consider only one primary operator exchange, in this case $\Ocal_2$. We are therefore omitting the contribution of all the other primaries and all the descendants in the OPE of $\Ocal\times \Ocal_1$.
 We use this normalization for both the cases $\Ocal_1=\Ocal_2=\phi$ and $\Ocal_1=\phi, \Ocal_2=\phib$. 
Notice however that for $\phi\times \phi$, the equality of the operators forces $\ell$ to be even.
\subsection{Current-scalar OPE}
\label{app:JsOPE}
 We normalize the scalar current OPE $J\times \phi$ as
  \be
\label{Vector_scalar_OPE}
\begin{array}{l}
\Ocal_{\ell,+}(x,z) J(0,z_1)\sim  \dfrac{1}{\sqrt{a_\ell} }\;
 \dfrac{\phi(0)}{(x^2)^{\a}} \,\l_{J \phi \Ocal_+} \ \sum_{p=1}^2\omega_p \, t_+^{(p)}(x,z,z_1) \ ,
\\
\Ocal_{\ell,-}(x,z) J(0,z_1)\sim \dfrac{1}{\sqrt{b_\ell}} 
\; \dfrac{\phi(0)}{(x^2)^{\a}} \, \l_{J \phi \Ocal_-}  \ t_-(x,z,z_1)
\ ,
\end{array}
\ee
where $\Ocal_{\ell,\pm}$ is a parity even/odd operator of spin $\ell$ and charge one,
 and $\l$  are  OPE coefficients and $\a\equiv \frac{\D+\D_J-\D_{\phi}+\ell+1}{2}$. 
 The coefficients $a_\ell$ and $b_\ell$ are defined to match the conventions of \cite{Costa:2016xah},
\be
 a_\ell \equiv \frac{(-2)^{\ell-1} (d/2)_{\ell-1}}{\ell^2   (d-1)_{\ell-1}} \, , 
 \qquad
  \qquad
  b_\ell \equiv  \frac{ a_\ell}{-2 \ell (d+\ell-3)}  \, .
\ee
The leading OPE tensor structures are defined as follows
\begin{align}
\label{tpl}
\begin{split}
 t_+^{(1)}(x,z,z_1)&=\displaystyle (x \cdot z)^\ell (x \cdot z_1) \ , \\ 
t_+^{(2)}(x,z,z_1)&=\displaystyle (x \cdot z)^{\ell-1} x^2 (z \cdot z_1) \ , \\
t_-(x,z,z_1)&= |x| (x \cdot z)^{\ell-1}  \epsilon(x,z_1,z) \ . \\
\end{split}
\end{align}
When $\ell=0$ only $ t_+^{(1)}$ survives.
 In \eqref{Vector_scalar_OPE} a single combination of $ t_+^{(p)}$ is used. This is written in terms of the vector $\omega =\{2 (\alpha -1),-2 \alpha +d+\ell\}/\ell $, determined by imposing conservation of $J$.
 Finally we set $d=3$ in all formulae above.

\subsection{Current-current OPE}
The normalization of the current-current OPE is as follows
\be \label{OPEwithSPIN}
\Ocal^\pm(x,z)J_1(0,z_1)\sim \frac{J_2(0,\partial_{z_2})}{(x^2)^{\a_\pm}} \sum_{q}
c_{12\Ocal^\pm}^{(q)} \; t^{[JJ]\,(q)}_{\pm}( x,z,z_1,z_2)  \ ,
\ee
where $\a_+\equiv \frac{1}{2}+  \a_-\equiv \frac{\D+\D_1-\D_2+\ell+2}{2}$.
In  (\ref{OPEwithSPIN}) $q$ runs from one to five for parity even operators. The correspondent OPE structures take the form
\begin{align}
\begin{split}
t_{\ell+}^{[JJ]\,(1)}( x,z,z_1,z_2)&\equiv (x\cdot z)^{\ell} (z_1 \cdot z_2) x^2\ ,\\
t_{\ell+}^{[JJ]\,(2)}( x,z,z_1,z_2)&\equiv (x\cdot z)^{\ell} (x\cdot z_1) (x \cdot z_2)\ ,\\
t_{\ell+}^{[JJ]\,(3)}( x,z,z_1,z_2)&\equiv (x\cdot z)^{\ell-1} (z\cdot z_1) (x \cdot z_2) x^2\ ,\\
t_{\ell+}^{[JJ]\,(4)}( x,z,z_1,z_2)&\equiv (x\cdot z)^{\ell-1}(z\cdot z_2) (x \cdot z_1) x^2\ ,\\
t_{\ell+}^{[JJ]\,(5)}( x,z,z_1,z_2)&\equiv (x\cdot z)^{\ell-2} (z\cdot z_1) (z \cdot z_2) x^4\ .
\end{split}
\label{t_five_allowed}
\end{align}
By imposing equality and conservation of the currents $J_i$ we find only two linearly independent structures
\be
\label{ConservedEqualBasis+}
\sum_{p=1}^5 (m_+)_{\tilde{p} p} \  t^{[JJ]\,(p)}_{\ell +}(x,z,z_1,z_2) \ , \qquad (\tilde{p}=1,2) \ ,
\ee
where
\be
\label{m+}
m_+=\left(
\begin{array}{ccccc}
 (2-\Delta ) (\ell+\Delta ) & (\Delta -\ell) (\ell+\Delta ) & 2 \ell (\Delta -2) & 0 & -\ell (\Delta -2) \\
 \ell-\Delta +2 & 0 & -\ell+\Delta -2 & \Delta -\ell & \ell-\Delta +1 \\
\end{array}
\right) \ .
\ee

For parity odd operators there are four possible tensor structures 
\be \label{structuresParityOdd}
\begin{array}{l} 
t^{[JJ]\,(1)}_{\ell\, -}= \e(x,z_1,z_2)  (x \cdot z)^{\ell}\\
t^{[JJ]\,(2)}_{\ell\, -}=\e(x,z,z_1) (x \cdot z_2) (x \cdot z)^{\ell-1}\\
t^{[JJ]\,(3)}_{\ell\, -}= \e(x,z,z_2) (x \cdot z_1) (x \cdot z)^{\ell-1}\\
t^{[JJ]\,(4)}_{\ell\, -}=[\e(x,z,z_1) (z \cdot z_2)+\e(x,z,z_2) (z \cdot z_1)] (x \cdot z)^{\ell-2} x^2 \ .
\end{array}
\ee
For conserved equal currents we obtain just one structure which takes a different form for $\ell$ even or odd,
\be
\label{ConservedEqualBasis-}
\sum_{p} \; (m_{-})_{p} \;  t^{[JJ]\,(p)}_{\ell-}(x,z,z_1,z_2) \ ,
\ee
with
\be
\label{m-}
m_{-}= 
\left\{
\begin{array}{l l}
(\Delta -3,\ell,\ell,0)   & \  \ell \mbox{ even,} \\
(0,\Delta -\ell-3,\Delta +\ell+1,1-\Delta)  & \ \ell>1, \mbox{ odd.}
\end{array}
\right.
\ee
When  $\ell=1$ there are no allowed tensor structures, while for $\ell=0$ there is one.

\newpage
\section{Conformal Blocks}
\subsection{$ JJ \phi \phib$}
\label{app:CBJJss}
The blocks for $ JJ \phi \phib$ are computed using the recurrence relation \eqref{recurrence_rel}.
We first consider correlators of two vectors $V_1,V_2$ and two scalars $\phi_1,\phi_2$, we finally restrict to the equal, conserved case. In order to use  \eqref{recurrence_rel}, we have to compute the coefficients $R_{A}$ and the functions $h_{\infty}$. 

As described in \cite{Penedones:2015aga}, $R_{A}$ are obtained as the product of three terms,
\be
\label{def:RA}
(R_{A})_{p p'}=(M_A^{(L)})_{p p'} \, Q_A \, M_A^{(R)} \, ,
\ee 
where $Q_A$ and $M_A$ are respectively related to the two- and three-point functions with primary-descendants operators.
For our case the three terms in \eqref{def:RA} were already computed in the literature. 
Indeed $Q_A$ and $M_A^{(R)}$ are the same as for the scalar blocks of \cite{Penedones:2015aga}, while  $(M_A^{(L)})_{p p'}$ are the same of  \cite{Dymarsky:2017xzb}.
Thus, the only missing computation is that of $h^{(p)}_{\infty \, \ell, s}$.
These functions are obtained by solving the Casimir differential equation at leading order in large $\D$ \cite{Kos:2013tga, Penedones:2015aga}.
The Casimir equation mixes the five structures  resulting in a system of 5 coupled differential equations. We introduce the ansatz
\begin{equation}
h_s^{(s')}(r,\eta)\equiv \cA(r,\eta) P_s^{(s')}(r,\eta) \, ,
\qquad 
\cA(r,\eta)\equiv\frac{\left(1-r^2\right)^{1-h} 
}{\sqrt{r^2-2 \eta  r+1} \left(r^2+2 \eta  r+1\right)^{3/2}}
\, .
\end{equation}
The resulting differential equations for $P_s^{(s')}(r,\eta)$ are then easily solved using Mathematica. The solution is given by
\begin{equation}
	P=\left(
	\begin{array}{ccccc}
	\left(r^2-1\right)^2 \left(2 r \eta+A_3\right) & 0 & 0 & 0 & 0 \\
	0 & A_1 A_3^2 & -2 r^2 \eta A_1 A_3 & -2 r^2 \eta A_1 A_3 & 4 r^4 \eta^2 A_1 \\
	0 & -2 r A_1 A_3 & -A_1 A_2 A_3 & 4 r^3 \eta A_1 & 2 r^2 \eta A_1 A_2 \\
	0 & -2 r A_1 A_3 & 4 r^3 \eta A_1 & -A_1 A_2 A_3 & 2 r^2 \eta A_1 A_2 \\
	0 & 4 r^2 A_1 & 2 r A_1 A_2 & 2 r A_1 A_2 & A_1 A_2^2 \\
	\end{array}
	\right) \, ,
\end{equation}
with $A_1=(1+r^2-2 r \eta), A_2=(-1+r^2-2 r \eta), A_3=(1+r^2)$. Hence the functions $h_\infty$ can be written as a linear combination of the five $h_s^{(s')}$ as follows
\begin{equation}
h_{\infty \ell +,s}^{(p)}(r,\eta)=\sum_{s'}^5 h_s^{(s')}(r,\eta) f_{\ell +, s'}^{(p)}(\eta) \ ,
\end{equation}
where the functions $f$ are constants of integration that can be fixed by imposing the correct initial conditions $f^{(p)}_{\ell +, s}(\eta)=h_{\infty \ell +,s}^{(p)}(0,\eta)$. 
We then determine these constants by studying the OPE limit $x_2 \to x_1$, $x_4 \to x_3$ of the four point function \cite{Penedones:2015aga},
\begin{equation}
 \lim_{\substack{ x_2\to x_1 \\ x_4\to x_3}} \sum_{s=1}^{43} f^{(p)}_{\ell\pm,s}(\eta) Q_s= \frac{
 t^{(p)}_{\ell +}( \hat x_{12}, I(x_{24})\cdot D_z ,I(x_{12}) \cdot  z_1, z_2) (-x_{34} \cdot z)^\ell 
}{\ell! (h-1)_\ell}
  \ .\label{smallr:eq.forf}
\end{equation}
Here $(-x_{34}\cdot z)^\ell $ comes from the scalar OPE \eqref{OPE} and $t^{(p)}_{\ell +}$ are the OPE structures defined in \eqref{t_five_allowed}. Here we also introduced the differential operator $D^\m_z \equiv \left(d/2-1+z\cdot \partial_z \right) \partial_z^\m -\frac{1}{2} z^\m  
\partial_z^2$ and the reflection matrix $I(x)^{\m \n}=\d^{\m \n}-2 x^\m x^\n/x^2$.
Finally,  the conserved blocks for $J J\phi \phib$ are obtained from the contraction $(m_+)_{p q}h^{(q)}$, where $m_+$ is defined in \eqref{m+}.
\subsection{$ J \phi J \phib$}
\label{app:CBJsJs}
As we mention in section \ref{sec:CBs}, we compute the conformal blocks of $ J \phi J \phib$ by using the an improved version of the ancillary file of \cite{Costa:2016xah}.
The code produces a single block for the parity-odd exchanges and  four blocks $g^{(p\rq{},q\rq{})}_{\D \ell +,s}$ (for $p,q=1,2$) for the parity-even exchanges. 
To be consistent with the OPE basis defined in appendix \ref{app:JsOPE}, we write the parity-even conserved block as follows,
\be
g_{\D \, \ell \, +,s}=\sum_{p\rq{},q\rq{}=1}^2  (\tilde \w^{(L)})_{ p' } (\tilde \w^{(R)})_{ q' }  g^{(p\rq{},q\rq{})}_{\D \, \ell \, +,s} \, ,
\ee
where
\be
\tilde \w^{(L)}=\left\{\ell (\ell+1),\Delta -\Delta _{\phi }\right\}\ ,
\qquad 
\tilde \w^{(R)}=\left\{-\ell (\ell+1),\Delta -\Delta _{\phi }\right\} \ .
\ee

\subsection{$\phi JJ \phib$}
\label{app:CBsJJs}
The conformal blocks for the $\phi JJ \phib$ can be obtained from the ones of $J \phi J \phib$ by using crossing symmetry $1\leftrightarrow2$.
Indeed it is easy to see that the functions $ \hat{h}_ {s}$ of \eqref{def:hat} are related to the $ \hat{f}_ {s}$ as follows,
\be
\label{fhtohh}
\begin{array}{l}
 \hat{h}_ {1}(u,v)=-v^{\Delta _{\phi }+\Delta _J+\frac{1}{2}} \left[\hat{f}_{2}\left(\frac{u}{v},\frac{1}{v}\right)+\sqrt{u} \left(\hat{f}_{1}\left(\frac{u}{v},\frac{1}{v}\right)+\hat{f}_{3}\left(\frac{u}{v},\frac{1}{v}\right)\right)\right] \, , \vspace{0.03 cm}\\
 \hat{h}_ {2}(u,v)=\frac{1}{2} v^{\Delta _{\phi }+\Delta _J} \left[(u+v+1) \hat{f}_{1}\left(\frac{u}{v},\frac{1}{v}\right)+2 \sqrt{u} \hat{f}_{2}\left(\frac{u}{v},\frac{1}{v}\right)+(u+v-1) \hat{f}_{3}\left(\frac{u}{v},\frac{1}{v}\right)\right] \, , \vspace{0.03 cm}\\
 \hat{h}_ {3}(u,v)=\frac{1}{2} v^{\Delta _{\phi }+\Delta _J} \left[(-u+v-1) \hat{f}_{1}\left(\frac{u}{v},\frac{1}{v}\right)-2 \sqrt{u} \hat{f}_{2}\left(\frac{u}{v},\frac{1}{v}\right)+(-u+v+1) \hat{f}_{3}\left(\frac{u}{v},\frac{1}{v}\right)\right] \, , \vspace{0.03 cm}\\
 \hat{h}_ {4}(u,v)=v^{\Delta _{\phi }+\Delta _J} \hat{f}_{4}\left(\frac{u}{v},\frac{1}{v}\right) \, , \vspace{0.09 cm}\\
 \hat{h}_ {5}(u,v)=-v^{\Delta _{\phi }+\Delta _J+\frac{1}{2}} \hat{f}_{5}\left(\frac{u}{v},\frac{1}{v}\right) \, .\\
\end{array}
\ee
In terms of radial coordinates the equations above relate $\hat{h}_ {s}(r,\eta)$ to $\hat{f}_ {s}(r,-\eta)$. Therefore, by means of \eqref{fhtohh}, we can reconstruct $\hat{h}_ {s}(r,\eta)$ to some order $O(r^n)$ by knowing $\hat{f}_ {s}(r,-\eta)$ to the same order. In particular, since  $\hat{f}_ {s}$ is evaluated at $-\eta$, its complete dependence in $\eta$ has to be known at that order. 
This would require extra computations. Indeed, we only need some derivatives at $\eta=1$ of the $J \phi J \phib$ blocks and their full dependence in $\eta$ was not computed.
For this reason, instead of using \eqref{fhtohh}, we built the $\phi JJ \phib$ blocks using the  differential operators of \cite{Costa:2011dw}. 
The final conserved blocks are then put in a basis compatible with  \eqref{fhtohh}. This allows to have the same definition for the OPE coefficients that multiply the $JsJs$ and the $sJJs$ blocks.
\subsection{Conformal block decomposition}
As an example,  let us compute the conformal blocks decomposition of $\phi JJ \phib$ and $J \phi J \phib$ for the theory of a free complex boson. We use the unit normalized current $J^\m \equiv \frac{- i}{\sqrt{2}} (\phi \partial^\m \phib-\phib \partial^\m \phi)$ and compute the correlators by Wick contractions,
\begin{align}
& f_s(u,v)=  \left\{\frac{1}{2} u^{3/4} \left(\sqrt{u} \left(\frac{1}{\sqrt{v}}+2\right)+1\right),0,\frac{1}{2} \sqrt[4]{u} \sqrt{v},\frac{u^{7/4}}{2 v},0\right\} \, , \\
&h_s(u,v) = \left\{\frac{u^{3/4} \left(\sqrt{u} \left(\sqrt{v}+2\right)+\sqrt{v}\right)}{2 v^2},-\frac{u^{3/4}}{2 v},\frac{\sqrt[4]{u}}{2 \sqrt{v}},\frac{u^{5/4}+u^{7/4}}{2 v^{3/2}},-\frac{u^{3/4}}{2 v}\right\} \, .
\end{align}
Here the functions $f_s$ and $h_s$ are the ones defined in  \eqref{JsJs_Tensor_structures} and \eqref{sJJs_Tensor_structures}.
Because of the normalization explained in appendix \ref{app:CBsJJs}, the conformal block decomposition of the functions $f_s$ and $h_s$ give the same OPE coefficients $p_\Ocal$. These are exemplified in the  tables below for $\Ocal$ being either a parity even or odd operator.
\begin{table*}[h!]
\centering
\as{1.2}
\begin{tabular}{@{}l | ccccccccc@{}}
\toprule
$ \Delta ,\ell  $  &  $ \frac{1}{2},0 $  &  $ \frac{5}{2},1 $  &  $ \frac{7}{2},2 $  &  $ \frac{9}{2},3 $  &  $
   \frac{11}{2},4 $  &  $ \frac{13}{2},5 $  &  $ \frac{15}{2},6 $  &  $ \frac{17}{2},7 $  &  $ \frac{19}{2},8 $\\
   \midrule
 $p_{\Delta  \ell +} $  &  $ \frac{1}{2} $  &  $ \frac{3}{8} $  &  $ -\frac{1}{42} $  &  $ \frac{1}{1056} $  &  $ -\frac{12}{25025} $  &  $ \frac{137}{4534920} $  &  $
   -\frac{367}{24025386} $  &  $ \frac{4859}{3893984640} $  &  $ -\frac{5669}{9546570900} $\\
   \bottomrule
\end{tabular}
\caption{OPE coefficients for parity even operators in the conformal block decomposition of $J\phi J \phib$ and $\phi J J \phib$. }
\label{table:OPEs1}
\end{table*}
\begin{table*}[h!]
\centering
\as{1.2}
\begin{tabular}{@{}l |cccccccc@{}}
\toprule
$ \Delta ,\ell  $  &  $  \frac{9}{2},2  $  &  $  \frac{11}{2},3  $  &  $  \frac{13}{2},4  $  &  $  \frac{15}{2},5  $  &  $
    \frac{17}{2},6  $  &  $  \frac{19}{2},7  $  &  $  \frac{21}{2},8  $  &  $  \frac{23}{2},9  $\\
   \midrule
  $p_{\Delta  \ell +} $  &  $ \frac{1}{15} $  &  $ -\frac{1}{182} $  &  $ \frac{1}{510} $  &  $
   -\frac{107}{373065} $  &  $ \frac{17}{198835} $  &  $ -\frac{193}{12606300} $  &  $ \frac{2969}{695987820} $  &  $ -\frac{1319}{1564192575} $\\
  \bottomrule
\end{tabular}
\caption{OPE coefficients for parity odd operators in the conformal block decomposition of $J\phi J \phib$ and $\phi J J \phib$. }
\label{table:OPEs2}
\end{table*}

\section{Vectors for the bootstrap equations}
\label{vectors_bootstrap}
In this appendix we detail the form of the $23$-dimensional vectors in \eqref{BOOTSTRAP_EQNS}.
The vector $V^{Q=0}_{\D,\ell,+}$ takes a different form for $\ell=0$ and $\ell>0$ (even),
\be
\begin{array}{ll}
\left(V^{Q=0}_{\D,\ell,+ \atop \ell  \textrm{ even}} \right)_1 &=F ^{[-]\phi \bar \phi\phi \bar \phi}_{\Ocal_{+}}(u,v)  \left(
\begin{array}{ccc}
1& 0&0\\
0&0 &0\\
0&0&0
\end{array}
\right)
\, , \\
\left(V^{Q=0}_{\D,\ell,+ \atop \ell  \textrm{ even}} \right)_2 &=F^{[+]\bar \phi \phi\phi \bar \phi}_{\Ocal_{+}}(u,v)\left(
\begin{array}{ccc}
 1 & 0&0\\
0&0 &0\\
0&0&0
\end{array}
\right)
\, , \\
\left(V^{Q=0}_{\D,\ell,+ \atop \ell  \textrm{ even}} \right)_3 &=F^{[-]\bar \phi \phi\phi \bar \phi}_{\Ocal_{+}}(u,v) \left(
\begin{array}{ccc}
1& 0&0\\
0&0 &0\\
0&0&0
\end{array}
\right)
\, , \\
\left(V^{Q=0}_{\D,\ell,+ \atop \ell  \textrm{ even}} \right)_7 &=\mathcal{S}^{+}_1(u,v)
\, , \\
\left(V^{Q=0}_{\D,\ell,+ \atop \ell  \textrm{ even}} \right)_8 &=\mathcal{S}^{-}_1(u,v)
\, , \\
\left(V^{Q=0}_{\D,\ell,+ \atop \ell  \textrm{ even}} \right)_9 &=\mathcal{S}^{+}_2(u,v) 
\, , \\
\left(V^{Q=0}_{\D,\ell,+ \atop \ell  \textrm{ even}} \right)_{10} &=\mathcal{S}^{-}_2(u,v)
\, , \\
\left(V^{Q=0}_{\D,\ell,+ \atop \ell  \textrm{ even}} \right)_{11} &=\mathcal{S}^{+}_3(u,u)
\, , \\
\left(V^{Q=0}_{\D,\ell,+ \atop \ell  \textrm{ even}} \right)_{12} &=
\mathcal{S}^{+}_4(u,u)
\, , \\
\left(V^{Q=0}_{\D,\ell,+ \atop \ell  \textrm{ even}} \right)_{13} &=\mathcal{R}^{-}_{13}(u,v)
\, , \\
\left(V^{Q=0}_{\D,\ell,+ \atop \ell  \textrm{ even}} \right)_{14} &=
\mathcal{R}^{-}_{15}(u,v)
\, , \\
\left(V^{Q=0}_{\D,\ell,+ \atop \ell  \textrm{ even}} \right)_{15} &=
\mathcal{R}^{-}_{16}(u,v)
\, , \\
\left(V^{Q=0}_{\D,\ell,+ \atop \ell  \textrm{ even}} \right)_{16} &=
\mathcal{R}^{-}_{17}(u,v)
\, , \\
\left(V^{Q=0}_{\D,\ell,+ \atop \ell  \textrm{ even}} \right)_{17} &=\mathcal{R}^{+}_7(u,v)
\, , \\
\left(V^{Q=0}_{\D,\ell,+ \atop \ell  \textrm{ even}} \right)_{18} &=\mathcal{R}^{+}_1(u,u)
\, , \\
\left(V^{Q=0}_{\D,\ell,+ \atop \ell  \textrm{ even}} \right)_{19} &=\mathcal{R}^{+}_2(u,u)
\, , \\
\left(V^{Q=0}_{\D,\ell,+ \atop \ell  \textrm{ even}} \right)_{20} &=\mathcal{R}^{+}_4(u,u)
\, , \\
\left(V^{Q=0}_{\D,\ell,+ \atop \ell  \textrm{ even}} \right)_{21} &=\mathcal{R}^{+}_5(u,u)
\, , \\
\left(V^{Q=0}_{\D,\ell,+ \atop \ell  \textrm{ even}} \right)_{22} &=
\mathcal{R}^{+}_6(u,u)
\, , \\
\left(V^{Q=0}_{\D,\ell,+ \atop \ell  \textrm{ even}} \right)_{23} &=
\mathcal{R}^{+}_3(\frac{1}{4},\frac{1}{4})
\, , \\
 \left(V^{Q=0}_{\D,\ell,+ \atop \ell  \textrm{ even}} \right)_{i} &= 0 \, , \qquad \textrm{i=4,5,6,}
\end{array}
\qquad
\begin{array}{ll}
\left(V^{Q=0}_{\D,0,+} \right)_1 &=F ^{[-]\phi \bar \phi\phi \bar \phi}_{\Ocal_{+}}(u,v)  \left(
\begin{array}{cc}
1& 0\\
0&0
\end{array}
\right)
\, , \\
\left(V^{Q=0}_{\D,0,+} \right)_2 &=F^{[+]\bar \phi \phi\phi \bar \phi}_{\Ocal_{+}}(u,v)\left(
\begin{array}{ccc}
 1 &0\\
0 &0\\
\end{array}
\right)
\, , \\
\left(V^{Q=0}_{\D,0,+} \right)_3 &=F^{[-]\bar \phi \phi\phi \bar \phi}_{\Ocal_{+}}(u,v) \left(
\begin{array}{ccc}
1& 0\\
0&0\\
\end{array}
\right)
\, , \\
\left(V^{Q=0}_{\D,0,+} \right)_7 &=S^{+}_1(u,v)
\, , \\
\left(V^{Q=0}_{\D,0,+} \right)_8 &=S^{-}_1(u,v)
\, , \\
\left(V^{Q=0}_{\D,0,+} \right)_9 &=S^{+}_2(u,v) %
\, , \\
\left(V^{Q=0}_{\D,0,+} \right)_{10} &=S^{-}_2(u,v)
\, , \\
\left(V^{Q=0}_{\D,0,+} \right)_{11} &=S^{+}_3(u,u)
\, , \\
\left(V^{Q=0}_{\D,0,+} \right)_{12} &=
S^{+}_4(u,u)
\, , \\
\left(V^{Q=0}_{\D,0,+} \right)_{13} &=R^{-}_{13}(u,v)
\, , \\
\left(V^{Q=0}_{\D,0,+} \right)_{14} &=
R^{-}_{15}(u,v)
\, , \\
\left(V^{Q=0}_{\D,0,+} \right)_{15} &=
R^{-}_{16}(u,v)
\, , \\
\left(V^{Q=0}_{\D,0,+} \right)_{16} &=
R^{-}_{17}(u,v)
\, , \\
\left(V^{Q=0}_{\D,0,+} \right)_{17} &=R^{+}_7(u,v)
\, , \\
\left(V^{Q=0}_{\D,0,+} \right)_{18} &=R^{+}_1(u,u)
\, , \\
\left(V^{Q=0}_{\D,0,+} \right)_{19} &=R^{+}_2(u,u)
\, , \\
\left(V^{Q=0}_{\D,0,+} \right)_{20} &=R^{+}_4(u,u)
\, , \\
\left(V^{Q=0}_{\D,0,+} \right)_{21} &=R^{+}_5(u,u)
\, , \\
\left(V^{Q=0}_{\D,0,+} \right)_{22} &=
R^{+}_6(u,u)
\, , \\
\left(V^{Q=0}_{\D,0,+} \right)_{23} &=
R^{+}_3(\frac{1}{4},\frac{1}{4})
\, , \\
 \left(V^{Q=0}_{\D,0,+} \right)_{i} &= 0\, , \qquad \textrm{i=4,5,6.}
\end{array}
\ee
The matrices $\mathcal{R}$ and $\mathcal{S}$ are $3\times 3$ and they can be written in terms of the functions $F$ as follows,
\be
\mathcal{R}^{\pm}_s(u,v)\equiv \left(
\begin{array}{ccc}
0&0&0\\
0& F^{[\pm](1,1)JJJJ}_{\Ocal_{+},s}(u,v) & F^{[\pm](1,2)JJJJ}_{\Ocal_{+},s}(u,v)\\
0& F^{[\pm](2,1)JJJJ}_{\Ocal_{+},s}(u,v) & F^{[\pm](2,2)JJJJ}_{\Ocal_{+},s}(u,v)
\end{array}
\right) \, ,
\ee
\be
\mathcal{S}^{\pm}_s(u,v)\equiv\frac{1}{2}\left(
\begin{array}{ccc}
0&F^{[\pm](1)J J \phi \bar \phi}_{\Ocal_{+},s}(u,v)&  F^{[\pm](2)J J \phi \bar \phi}_{\Ocal_{+},s}(u,v)\\
F^{[\pm](1)J J \phi \bar \phi}_{\Ocal_{+},s}(u,v) &0 &0\\
F^{[\pm](2)J J \phi \bar \phi}_{\Ocal_{+},s}(u,v) &0 &0
\end{array}
\right) \, .
\ee
The matrices $R$ and $S$ are their $2 \times 2$ counterparts,
\be
R^{\pm}_s(u,v)\equiv\left(
\begin{array}{cc}
0&0\\
0 &F^{[\pm](1,1)JJJJ}_{\Ocal_{+},s}(u,v)
\end{array}
\right)
\, ,
\qquad
S^{\pm}_s(u,v)\equiv
\frac{1}{2}F^{[\pm](1)J J \phi \bar \phi}_{\Ocal_{+},s}(u,v)\left(
\begin{array}{cc}
0& 1\\
1&0 
\end{array}
\right) \, .
\ee
All the other vectors do not have any matrix structure,
\be
\!\!\!\!\!\!\!\!\!\!\!\!\!\!\!
\begin{array}{l}
\left(V^{Q=0}_{\D,\ell,+ \atop \ell  \textrm{ odd}}\right)_{1}=-F ^{[-]\phi \bar \phi\phi \bar \phi}_{\Ocal_{+}}(u,v) 
\, , \\
\left(V^{Q=0}_{\D,\ell,+ \atop \ell  \textrm{ odd}}\right)_{2}= F^{[+]\bar \phi \phi\phi \bar \phi}_{\Ocal_{+}}(u,v) 
\, , \\
\left(V^{Q=0}_{\D,\ell,+ \atop \ell  \textrm{ odd}}\right)_{3}= F^{[-]\bar \phi \phi\phi \bar \phi}_{\Ocal_{+}}(u,v) 
\, , \\
\left(V^{Q=0}_{\D,\ell,+ \atop \ell  \textrm{ odd}}\right)_{i}= 0\, ,
\quad  (i\neq 1,2,3)
\end{array}
\qquad \qquad \qquad \quad
\begin{array}{cl}
\left(V^{Q=2}_{\D,\ell,+}\right)_2&=-F^{[+] \phi \phi  \bar \phi \bar \phi }_{\Ocal}(u,v)\, , \\
\left(V^{Q=2}_{\D,\ell,+}\right)_3&=F^{[-] \phi \phi  \bar \phi \bar \phi }_{\Ocal}(u,v)\, , \\
\left(V^{Q=2}_{\D,\ell,+}\right)_i&=0 \, , \quad (i\neq 2,3)\, .
\end{array}
\ee

\be
\begin{array}{l}
\left(V^{Q=0}_{\D,\ell,-} \right)_{13}= F^{[-](1,1)JJJJ}_{\Ocal_{-},13}(u,v)
 \, , \\
\left(V^{Q=0}_{\D,\ell,-} \right)_{14}=F^{[-](1,1)JJJJ}_{\Ocal_{-},15}(u,v)
 \, , \\
\left(V^{Q=0}_{\D,\ell,-} \right)_{15}=F^{[-](1,1)JJJJ}_{\Ocal_{-},16}(u,v)
 \, , \\
\left(V^{Q=0}_{\D,\ell,-} \right)_{16}=F^{[-](1,1)JJJJ}_{\Ocal_{-},17}(u,v)
 \, , \\
\left(V^{Q=0}_{\D,\ell,-} \right)_{17}= F^{[+](1,1)JJJJ}_{\Ocal_{-},7}(u,v)
  \, , \\
\left(V^{Q=0}_{\D,\ell,-} \right)_{18}= F^{[+](1,1)JJJJ}_{\Ocal_{-},1}(u,u)  
 \, , \\
\left(V^{Q=0}_{\D,\ell,-} \right)_{19}= F^{[+](1,1)JJJJ}_{\Ocal_{-},2}(u,u)
  \, , \\
\left(V^{Q=0}_{\D,\ell,-} \right)_{20}=F^{[+](1,1)JJJJ}_{\Ocal_{-},4}(u,u)
  \, , \\
\left(V^{Q=0}_{\D,\ell,-} \right)_{21}=F^{[+](1,1)JJJJ}_{\Ocal_{-},5}(u,u)
  \, , \\
\left(V^{Q=0}_{\D,\ell,-} \right)_{22}=F^{[+](1,1)JJJJ}_{\Ocal_{-},6}(u,u)
  \, , \\
\left(V^{Q=0}_{\D,\ell,-} \right)_{23}= F^{[+](1,1)JJJJ}_{\Ocal_{-},3}(1/4,1/4)
  \, , \\
\left(V^{Q=0}_{\D,\ell,-} \right)_{i}= 0 \, , \qquad \textrm{otherwise} \, ,
 \end{array}
 \qquad
 \qquad
 \begin{array}{cl}
\left(V^{Q=1}_{\D,\ell,\pm}\right)_4 &= \s_\Ocal F^{[-]J\bar \phi J \phi}_{\Ocal_{\pm},1}(u,v)
\, , \\
\left(V^{Q=1}_{\D,\ell,\pm}\right)_5&= \s_\Ocal F^{[-]J\bar \phi J \phi}_{\Ocal_{\pm},2}(u,v)
\, , \\
\left(V^{Q=1}_{\D,\ell,\pm}\right)_6&= \s_\Ocal F^{[+]J\bar \phi J \phi}_{\Ocal_{\pm},3}(u,u)
\, , \\
\left(V^{Q=1}_{\D,\ell,\pm}\right)_7&= -\s_\Ocal F^{[+]\bar \phi J J \phi}_{\Ocal_{\pm},1}(u,v)
\, , \\
\left(V^{Q=1}_{\D,\ell,\pm}\right)_8&= \s_\Ocal F^{[-]\bar \phi J J \phi}_{\Ocal_{\pm},1}(u,v)
\, , \\
\left(V^{Q=1}_{\D,\ell,\pm}\right)_9&=- \s_\Ocal F^{[+]\bar \phi J J \phi}_{\Ocal_{\pm},2}(u,v)
\, , \\
\left(V^{Q=1}_{\D,\ell,\pm}\right)_{10}&= \s_\Ocal F^{[-]\bar \phi J J \phi}_{\Ocal_{\pm},2}(u,v)
\, , \\
\left(V^{Q=1}_{\D,\ell,\pm}\right)_{11}&=- \s_\Ocal F^{[+]\bar \phi J J \phi}_{\Ocal_{\pm},3}(u,u)
  \, , \\
\left(V^{Q=1}_{\D,\ell,\pm}\right)_{12}&= -\s_\Ocal F^{[+]\bar \phi J J \phi}_{\Ocal_{\pm},4}(u,u)
\, , \\
\left(V^{Q=1}_{\D,\ell,\pm}\right)_{i}&=0\, , \qquad \textrm{otherwise.}
\end{array}
\ee
Recall that the sign sigma is defined in \eqref{sigma_sign}.